\documentclass[12pt,preprint]{aastex}

\def\deg{\ifmmode^\circ\else$^\circ$\fi}
\def\kmsMpc{km\,s$^{-1}$Mpc$^{-1}$}
\def\ah{\ifmmode{^\textrm{\scriptsize h}}\else{$^\textrm{\scriptsize h}$}\fi}
\def\am{\ifmmode{^\textrm{\scriptsize m}}\else{$^\textrm{\scriptsize m}$}\fi}
\def\as{\ifmmode{^\textrm{\scriptsize s}}\else{$^\textrm{\scriptsize s}$}\fi}

\newcommand{\Msun}{\ifmmode{\rm M_\odot}\else{M$_\odot$}\fi}
\newcommand{\Zsun}{\ifmmode{\rm Z_\odot}\else{Z$_\odot$}\fi}

\shortauthors{----}
\shorttitle{ALHAMBRA NIR galaxy counts}
\begin{document}

\title{Near-IR Galaxy Counts and Evolution 
from the Wide-Field ALHAMBRA survey\footnote{{Based on observations collected at the
German-Spanish Astronomical Center, Calar
Alto, jointly operated by the Max-Planck-Institut
f\"ur Astronomie Heidelberg and the
Instituto de Astrof\'isica de Andaluc\'ia (CSIC).}}}

            \author{D. Crist\'obal-Hornillos\altaffilmark{1},J.~A.~L. Aguerri\altaffilmark{2},
			M. Moles\altaffilmark{1},
			J. Perea\altaffilmark{1},
			F.~J. Castander\altaffilmark{5},
            T. Broadhurst\altaffilmark{3},
            E. ~J. Alfaro\altaffilmark{1},
            N. Ben\'{\i}tez\altaffilmark{1,11},
            J. Cabrera-Ca\~no\altaffilmark{4},            
            J. Cepa\altaffilmark{2,6},
            M. Cervi\~no\altaffilmark{1},
            A. Fern\'andez-Soto\altaffilmark{12},
            R.~M. Gonz\'alez Delgado\altaffilmark{1},
			C. Husillos\altaffilmark{1},
			L. Infante \altaffilmark{8},
            I. M\'arquez\altaffilmark{1},
            V.~J. Mart\'{\i}nez\altaffilmark{7,9},
            J. Masegosa\altaffilmark{1},
            A. del Olmo\altaffilmark{1},
            F. Prada\altaffilmark{1},
            J.~M. Quintana\altaffilmark{1},
            and S.~F. S\'anchez\altaffilmark{10}
            }
\altaffiltext{1}{Instituto de Astrof\'{\i}sica de Andaluc\'{\i}a, CSIC, Apdo. 3044, E-18080 Granada}
\altaffiltext{2}{Instituto de Astrof\'{\i}sica de Canarias, La Laguna, Spain}
\altaffiltext{3}{School of Physics and Astronomy, Tel Aviv University, Israel}
\altaffiltext{4}{Departamento de F\'{\i}sica At\'omica, Molecular y Nuclear, Facultad de F\'{\i}sica, Universidad de Sevilla, Spain}
\altaffiltext{5}{Institut de Ci\`encies de l'Espai, IEEC-CSIC, Barcelona, Spain}
\altaffiltext{6}{Departamento de Astrof\'{\i}sica, Facultad de F\'{\i}sica, Universidad de la Laguna, Spain}
\altaffiltext{7}{Departament d'Astronom\'{\i}a i Astrof\'{\i}sica, Universitat de Val\`encia, Val\`encia, Spain}
\altaffiltext{8}{Departamento de Astronom\'{\i}a, Pontificia Universidad Cat\'olica, Santiago, Chile}
\altaffiltext{9}{Observatori Astron\`omic de la Universitat de Val\`encia, Val\`encia, Spain}
\altaffiltext{10}{Centro Astron\'omico Hispano-Alem\'an, Almer\'{\i}a, Spain}
\altaffiltext{11}{IFF(CSIC), C/Serrano 113-bis, 28005 Madrid, Spain}
\altaffiltext{12}{Instituto de F\'isica de Cantabria (CSIC), 39005 Santander,Spain}

\email
{dch@iaa.es, jalfonso@iac.es, moles@iaa.es,\\
 jaime@iaa.es, fjc@ieec.fcr.es, emilio@iaa.es, benitez@iaa.es,\\
tjb@wise1.tau.ac.il, jcc-famn@us.es,jcn@iac.es,mcs@iaa.es,\\
alberto.fernandez@uv.es,rosa@iaa.es,cesar@iaa.es,linfante@astro.puc.cl,\\
isabel@iaa.es,vicent.martinez@uv.es,pepa@iaa.es,chony@iaa.es,\\
fprada@iaa.es,quintana@iaa.es,sanchez@caha.es}

\begin{abstract}

The ALHAMBRA survey aims to cover 4 square degrees using a system of
20 contiguous, equal width, medium-band filters spanning the range
3500~$\AA$ to 9700~$\AA$ plus the standard JHKs filters. 
Here we analyze deep near-IR number counts of one of our fields (ALH08) for which we have
 a relatively large area (0.5 square degrees) and faint
photometry (J=22.4, H=21.3 and K=20.0 at the 50\% of recovery efficiency for point-like sources).
We find that the logarithmic
gradient of the galaxy counts undergoes a distinct change to a flatter
slope in each band: from 0.44 at $[17.0,18.5]$ to 0.34 at
$[19.5,22.0]$ for the J band; for the H band 0.46 at $[15.5,18.0]$ to
0.36 at $[19.0,21.0]$, and in Ks the change is from 0.53 in the range
$[15.0,17.0]$ to 0.33 in the interval $[18.0,20.0]$. These
observations together with faint optical counts are used to constrain
models that include density and luminosity evolution of the local
type-dependent luminosity functions. Our models imply a decline in the
space density of evolved early-type galaxies with increasing redshift,
such that only 30$\%$ - 50$\%$ of the bulk of the present day
red-ellipticals was already in place at $z\sim1$.
\end{abstract}


\section{Introduction}

It is well understood that the stellar masses of galaxies are better
examined with near-IR (NIR) observations compared to shorter
wavelengths mainly because the near-IR light is relatively less
affected by recent episodes of star formation and by internal dust
extinction. Moreover, the K-corrections are also smaller and better
constrained in the NIR and hence massive high redshift objects are
relatively prominent in the NIR. Despite this relative insensitivity
to luminosity evolution and the effects of dust, the hope of using the
NIR counts to constrain the cosmological
parameters \citep{1993ApJ...415L...9G,1997ApJ...475..445M,
2001A&A...368..787H,2001MNRAS.323..795M} has not proved feasible
because evolution in the space density of galaxies was soon understood
to be of comparable significance for the NIR counts as the
cosmological curvature \citep{1992Natur.355...55B}.

Disentangling the effects of cosmology from evolution is not
straightforward even in the NIR, and now it has become more
appropriate to turn the question around and make use of the impressive
constraints on the cosmological parameters from WMAP
\citep{2003ApJS..148..175S}, and type Ia supernovae
\citep{1998AJ....116.1009R, 1999ApJ...517..565P}, in order to measure
more carefully the rate of evolution \citep{2001AJ....121..598M,2001A&A...375....1S,
2003ApJ...595...71C,2006ApJ...639..644E}. In addition, imaging in the NIR
has progressed well with fully cryogenic wide-field imagers now
available on several 4m class telescopes. We also have at our disposal
now much better estimates of the luminosity functions of different
classes of galaxies from the large local redshift surveys in particular
the SDSS \citep{2003ApJ...592..819B,2003AJ....125.1682N}, or 2dFGRS \citep{2001MNRAS.326..255C}.

The evolution of the luminosity functions has been addressed making use
of spectroscopic redshift surveys. However, the results of those studies differ due to the
limitations in terms of low statistic, or small fields probed which lead to
uncertainties due to the large scale structure. In \cite{2004A&A...424...23F} a mild (20-30\%)
evolution in the number density of massive objects since $z\sim1$ is found. Also a roughly
constant number density for red galaxies to $z=0.8$ is found in \cite{2002ApJ...571..136I}.
Using COMBO-17 photometric redshift information \citep{2003A&A...401...73W},
and the rest-frame color bimodality at
each redshift, \cite{2004ApJ...608..752B} with a sample of $\sim$25000 galaxies, concluded 
that the stellar mass in red-galaxies has
increase in a factor 2-3 from $z\sim1$ to the present. 
Combining spectroscopic with photometric redshift data
\cite{2006A&A...453..809I} point to an increase in a factor of $\sim2.7$ for the density of 
red bulge-dominated galaxies between $z=1$ and $z=0.6$. \cite{2007ApJ...665..265F} found
a different evolution since $z\sim1$ in number densities in the red and blue galaxy population, being 
constant for the blue galaxies, while the number density of the red galaxies increases in a factor 3. 
Wide field imaging with larger covered area, and greater numbers
selected to uniform faint limits is complementary to the
redshift surveys in examining statistical models proposed for evolution.

In practical terms it is most useful to combine 
faint NIR counts with deep blue counts when examining 
models of evolution to contrast the effects of luminosity and density evolution
which affect these two spectral ranges in different ways. In  
\cite{2000MNRAS.311..707M,2001MNRAS.323..795M} the authors use 
non-evolving models with a higher $\phi^*$ normalization in the B-band
even if this leads to an over-prediction of bright galaxies than is
observed. \cite{2001A&A...368..787H} pointed out that both the optical
and NIR counts present an excess over the no-evolution models, finding
passive evolution models more suitable to match the distributions. The
authors emphasize nevertheless their disappointment with the fact that
in the passive evolution models the faint number counts are dominated
by early-type galaxies, whereas the real data show that spiral and
Sd/Irr galaxies are the main contributors to the faint counts even in
the K-band.

A characteristic feature of the NIR galaxy counts reported in several
works
\citep{1993ApJ...415L...9G,2001A&A...368..787H,2003ApJ...595...71C,
2005A&A...442..423I} is the change of slope at $17 \leq Ks \leq
18$. This distinctive flattening is not observed in the B-band
counts. This effect has been interpreted in terms of a change in the
dominant galaxy population, becoming increasingly dominated
by an intrinsically bluer population
\citep{1993ApJ...415L...9G,2006ApJ...639..644E}. In the model
described in \cite{2003ApJ...595...71C} a delay in the formation of
the bulk of the early-type galaxies to $z_{form}<2$ and the presence
of a dwarf star-forming population are invoked to match the Ks-band
counts. A similar dwarf star-forming population at $z>1$, that is not
present at lower redshift was found compatible in 
\cite{1996Natur.383..236M,2001MNRAS.323..795M} but that work uses
a $q_0=0.5, \Lambda=0.0$ cosmology, requiring some revision.

Alternatively, an increase of $\phi^*$ for late-type galaxies, driven via
mergers, can produce similar results without introducing an ad hoc
population that is unseen in the local LFs
\citep{2006ApJ...639..644E}.  In any case, a low $z_{form}\sim1.5$
for the ellipticals remains necessary to generate a significant
decrease in the number of red galaxies and to account for the
distinctive break in the NIR count slope at Ks$\sim$17.5.

Here we use the NIR data from the first completed ALHAMBRA
field, hereafter ALH08 (details of the project can be found in
\cite{MOLES2008} and http://www.iaa.es/alhambra).  The limiting
magnitudes (at S/N=5 in an aperture diameter of 2$\times$FWHM) reached
in the 3 NIR bands are in mean for the eight frames in ALH8: J=22.6,
H=21.5 and Ks=20.1 with a 0.3 rms (Vega system), and the total area
covered amounts to $\sim 0.5$ square degrees. The completed survey
will cover 8 independent fields with a total area of 4 square
degrees. The ALHAMBRA survey occupies a middle ground in terms of the
product of depth and area in all three standard NIR filters.  The
bright end of the counts is well constrained by our relatively large
area allowing a careful examination of the location and size of the
break in the count-slopes in J,H and Ks at fainter magnitudes. We have
paid special attention to S/G separation, which at the intermediate
magnitude range, is effectively achieved using optical-NIR color
indices by combining our ALH08 NIR data with the corresponding Sloan
DR5 data.

Unless specified otherwise, all magnitudes here are presented in the
Vega system, and the favored cosmological model, with
H$_{0} = 70$\ \kmsMpc, $\Omega_{M} = 0.3$, $\Omega_{\Lambda} = 0.7$
was adopted through this paper.

\section{Observing Strategy and Data Reduction}

\begin{figure}
\epsscale{1.0}
\plotone{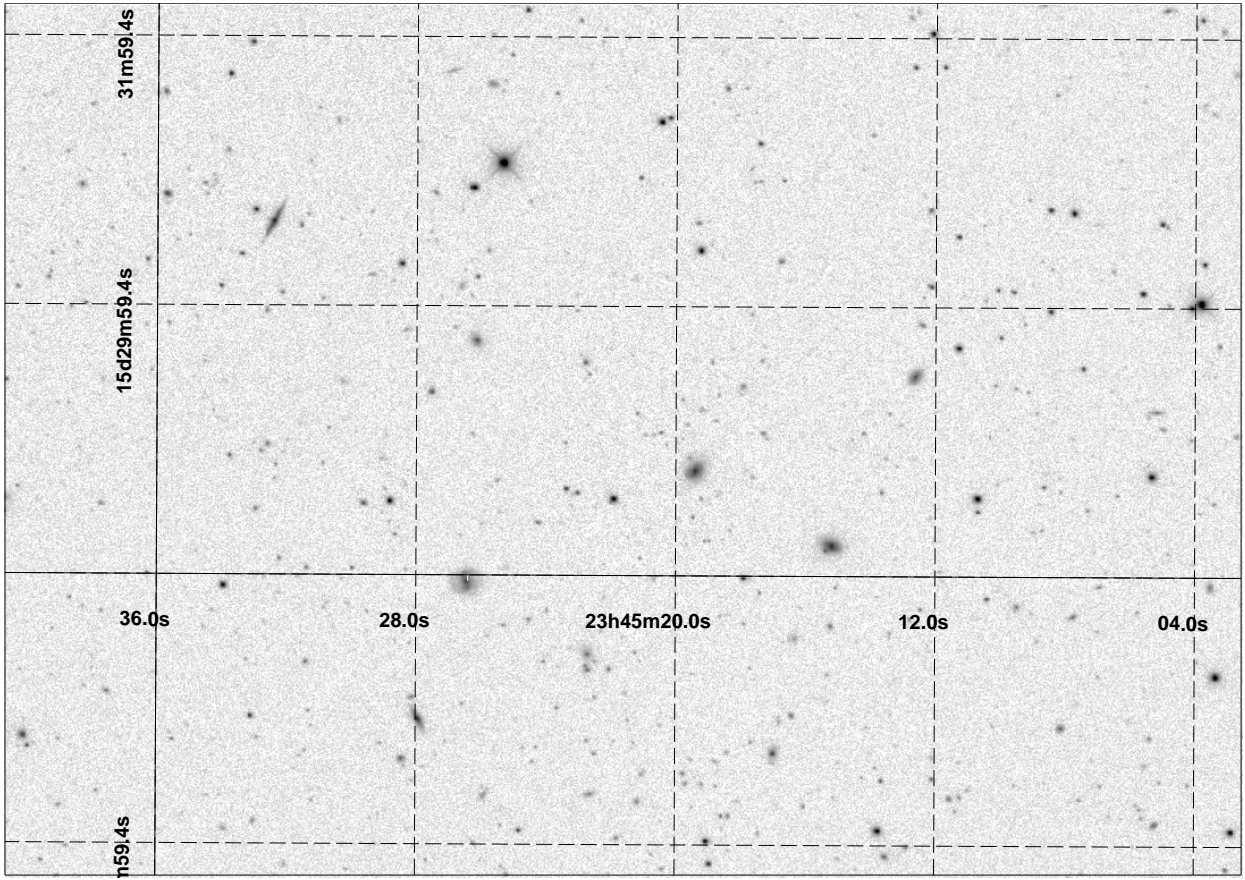}
\caption{ Central part ($\sim9.2^{\prime}\times6.4^{\prime}$) of one of the pointings in the ALH08 field in the Ks band. \label{Fig:f08p11_KS}
}
\end{figure}

The ALHAMBRA survey is collecting data in 23 optical-NIR filters using
the Calar Alto 3.5m telescope with the cameras LAICA for the 20
optical filters, and OMEGA2000 for the NIR, JHKs filters
\citep{MOLES2008,BENITEZ2008}. In this paper we
discuss the galaxy number counts in the J, H and Ks bands computed in
the completed ALH08 pointing. Due to the OMEGA2000 and LAICA geometries, two
parallel strips of $\sim1$ degree $\times 0.25$ degrees are acquired
in each of the eight ALHAMBRA fields. Each of these strips is covered
by four OMEGA2000 pointings. Fig.~\ref{Fig:f08p11_KS} shows the
central part of the processed image for one such pointing in
ALH08.

The OMEGA2000 camera has a focal plane array of type HAWAII-2 by
Rockwell with 2048 x 2048 pixels. The plate scale is 0.45
arcsec/pixel, giving a full field of view of $\sim$236 arcmin$^2$. The JHKs
images were taken using a dither pattern of 20 positions, with single
images time exposures of 80~s in J, 60~s in H and 46~s in Ks band
(obtained using respectively 16, 20 and 23 software co-adds). The
total exposure time was 5~ks in each of the three filters.

Due to the high quantity of raw data images we have implemented a
dedicated reduction pipeline to process the images. This code will
be presented and discussed in a forthcoming technical paper. The use
of the reduction pipeline guaranties the homogeneity of the process,
and allows us to perform a first automatic analysis of the resulting
images and to verify the quality of the products in real time. The 
magnitude at the 50\% of recovery efficiency for point-like sources in
the eight final images corresponds in average to J$=22.4\pm0.24$,
H$=21.3\pm0.14$ and Ks$=20.0\pm0.13$ in the Vega system. The galaxy
number counts have been computed in the high signal to noise region of
the final images, covering a total of $\sim$1600 arcmin$^2$, or 0.45~sq deg.

\subsection{Flat-fielding and sky subtraction}

Firstly, the individual images of each observing run were
dark-subtracted, and flat-fielded by super-flat images constructed
with the science images in each filter. In the case of NIR imaging it
is specially important to remove properly the high sky level that is
changing in short timescales. The sky structure of each image was
removed with the XDIMSUM package
\citep{1995ApJ...450..512S} with a sky image constructed with the
median of the 6 images closest in acquisition time, which in the case
of the J, H, Ks filters correspond to timescales of 480, 360, 276~s
respectively. During this process cosmic ray masks for each
individual image were also created.

SExtractor \citep{1996A&AS..117..393B} was used with the preliminary
sky subtracted images to compute the number of detected objects (above
a given S/N and with ellipticity lower than a certain limit), and to
make a robust estimate of the FWHM of each image. This information together
with the sky level variation was used to automatically remove low S/N
images, lying outside the survey requirements, and also to identify 
exposures presenting problems such as telescope trailing.

\subsection{Ghost and linear pattern masks}

The readout of the detector array produces ghost images coming from
bright stars. As those spurious effects are replicated in all the
readout channels (separated by 128 pixels) of the detector's quadrant
where the real bright source is located, it was possible to mask them
before image combination.

Linear patterns produced by moving objects were located in the images
using two different approaches: i) Objects with high ellipticity and
pixel area were identified from the individual image SExtractor
catalogs. ii) Linear patterns split in multiple spots and
structures were located using the Hough transform. Linear patterns
detected by any of the two methods were masked.

\subsection{Image combination}

Using XDIMSUM, the images were combined masking the cosmic rays,
bad pixels and linear patterns. Those preliminary co-added images were
used to create object and bright cores masks. Object masks were used
to cover the sources in the second iteration of the XDIMSUM sky
subtraction procedure. Bright cores masks operate when
constructing an improved version of the cosmic ray masks.

At this point, the individual images have been dark current corrected,
flat fielded and sky subtracted. Each individual image has an
associated mask containing the bad pixels, cosmic rays, linear
patterns and ghost images.

To perform the final combination of the processed images we used SWARP
\citep{2002ASPC..281..228B}, a code that combines the images,
correcting at the same time the geometrical distortions in the
individual images using the information stored in WCS headers. The
astrometric calibration of each individual image and the update of its
WCS headers was done by the pipeline using an automatic module. During
the SWARP combination the extinction variations among the images were
also corrected. More technical details on the image combination are
given in appendix~\ref{Ap:Combination}.

\subsection{Photometric calibration}

For the photometric calibration we used the 2MASS catalogs
\citep{2003tmc..book.....C}. After having combined all the images of
each pointing, the objects in common with the 2MASS catalogs were
found and those with higher signal to noise ratio selected to compute
a zero point offset.

\begin{figure}
\epsscale{1.0}
\plotone{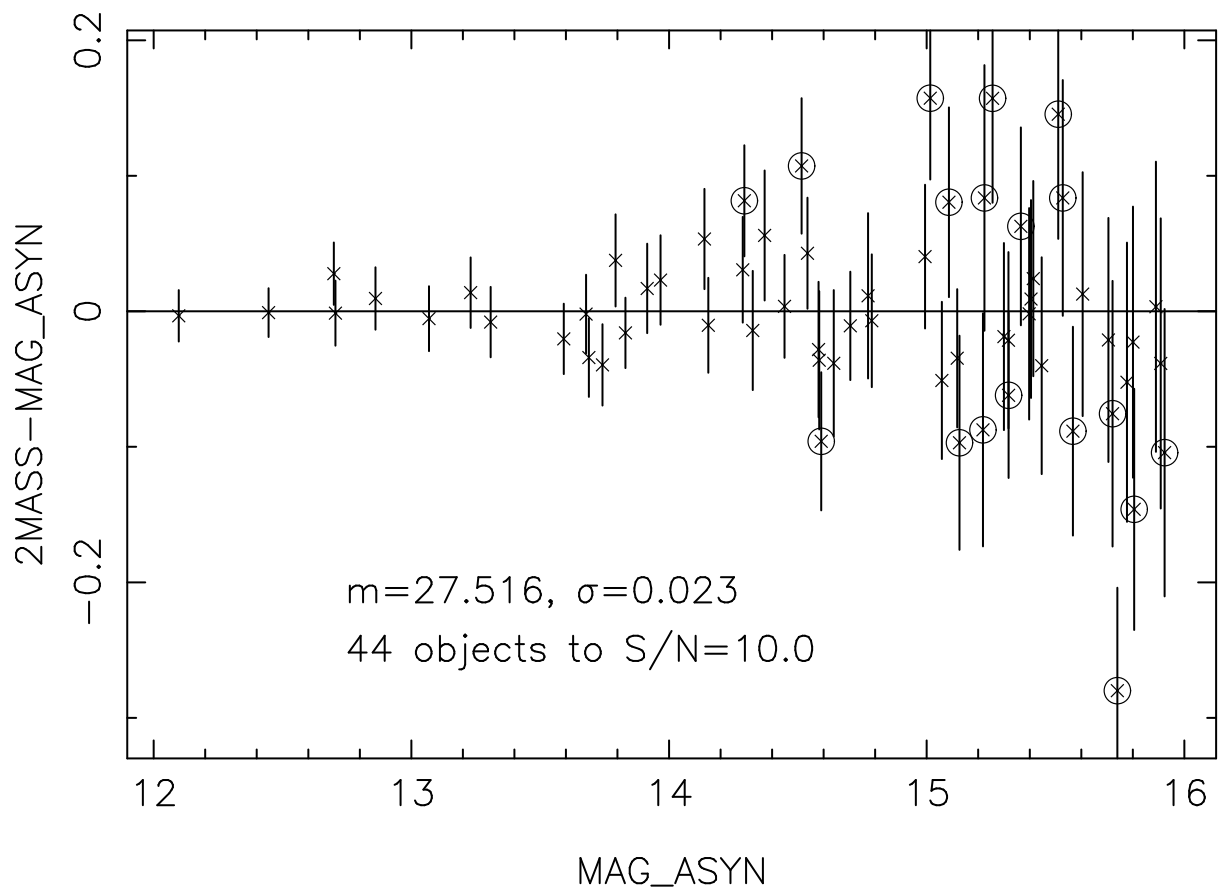}
\caption{ Calibration with 2MASS stars for the one of the pointings included in the ALH08 field in the H filter. Sources marked with a circle
 were eliminated using a 2$\sigma$ reject in the calibration
 fit. Error bars correspond to the 2MASS magnitude error summed in
 quadrature with the SExtractor computed error.}\label{Fig:2mass2}
\end{figure}

\begin{figure}
\epsscale{1.0}
\plotone{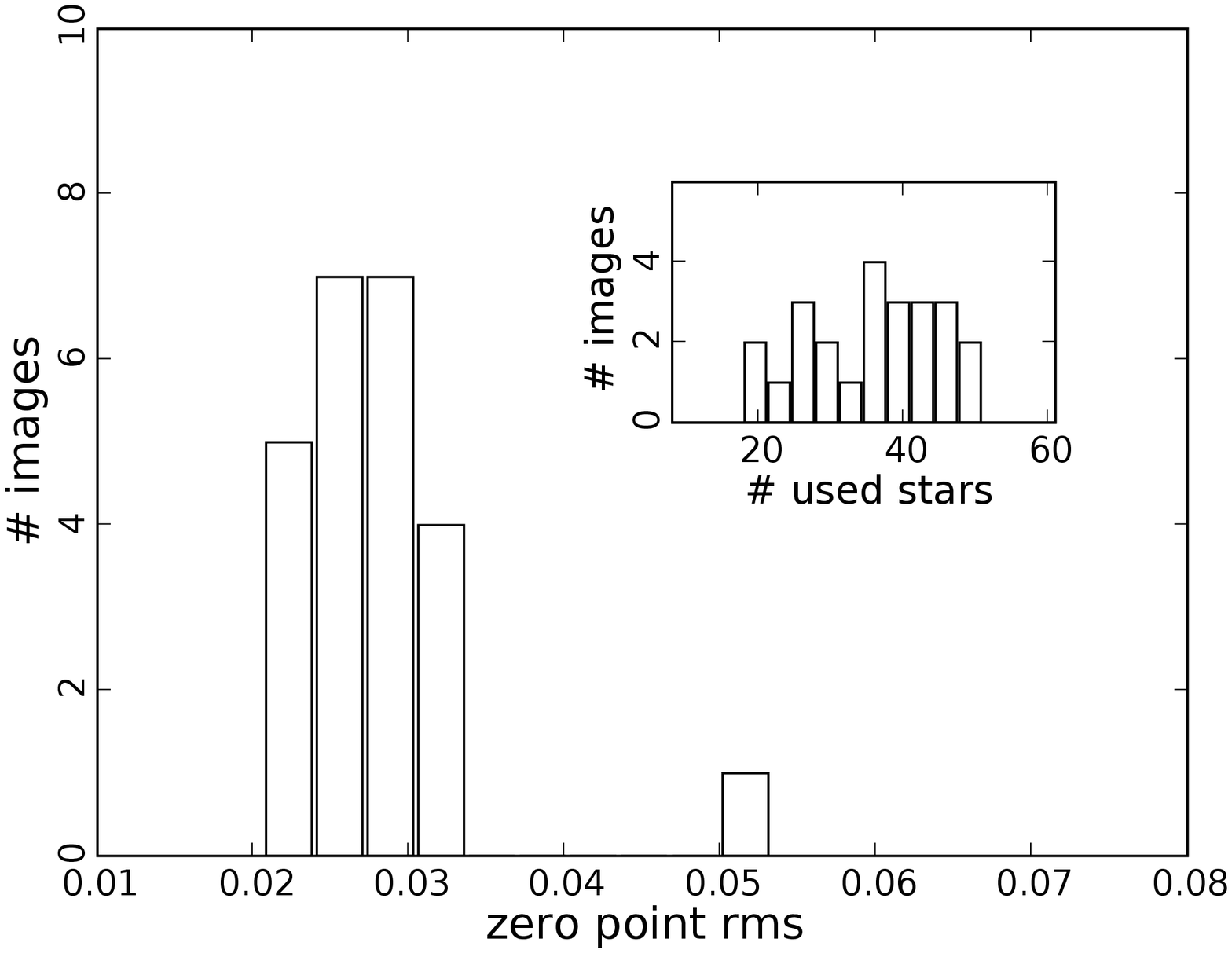}
\caption{ Histogram of the photometric accuracy of the different pointings in the ALH08 field.
In the small panel the histogram of the number of used stars for the
calibration in each pointing is shown.}
\label{Fig:2mass3}
\end{figure}

In Fig.~\ref{Fig:2mass2} an example of the photometric calibration for
one of the pointings in ALH08 is shown. The histogram with the rms in
the photometric calibration for the ALH08 pointings in the three NIR
bands is shown in Fig.~\ref{Fig:2mass3}. The mean value for the rms is
$0.028\pm0.006$ mags, and a mean of 36 calibration sources have been used in each frame.
We have not found any appreciable color related 
trend.

The calibrated images were inspected for the possible presence of ring
patterns that would have been produced by pupil ghosts. We computed
for each band, and for all the sources used in the calibration process of the different final frames, the
difference between our photometry and the 2MASS photometry as a
function of the radial distance to the nominal field center.

\begin{figure}
\epsscale{1.0}
\plotone{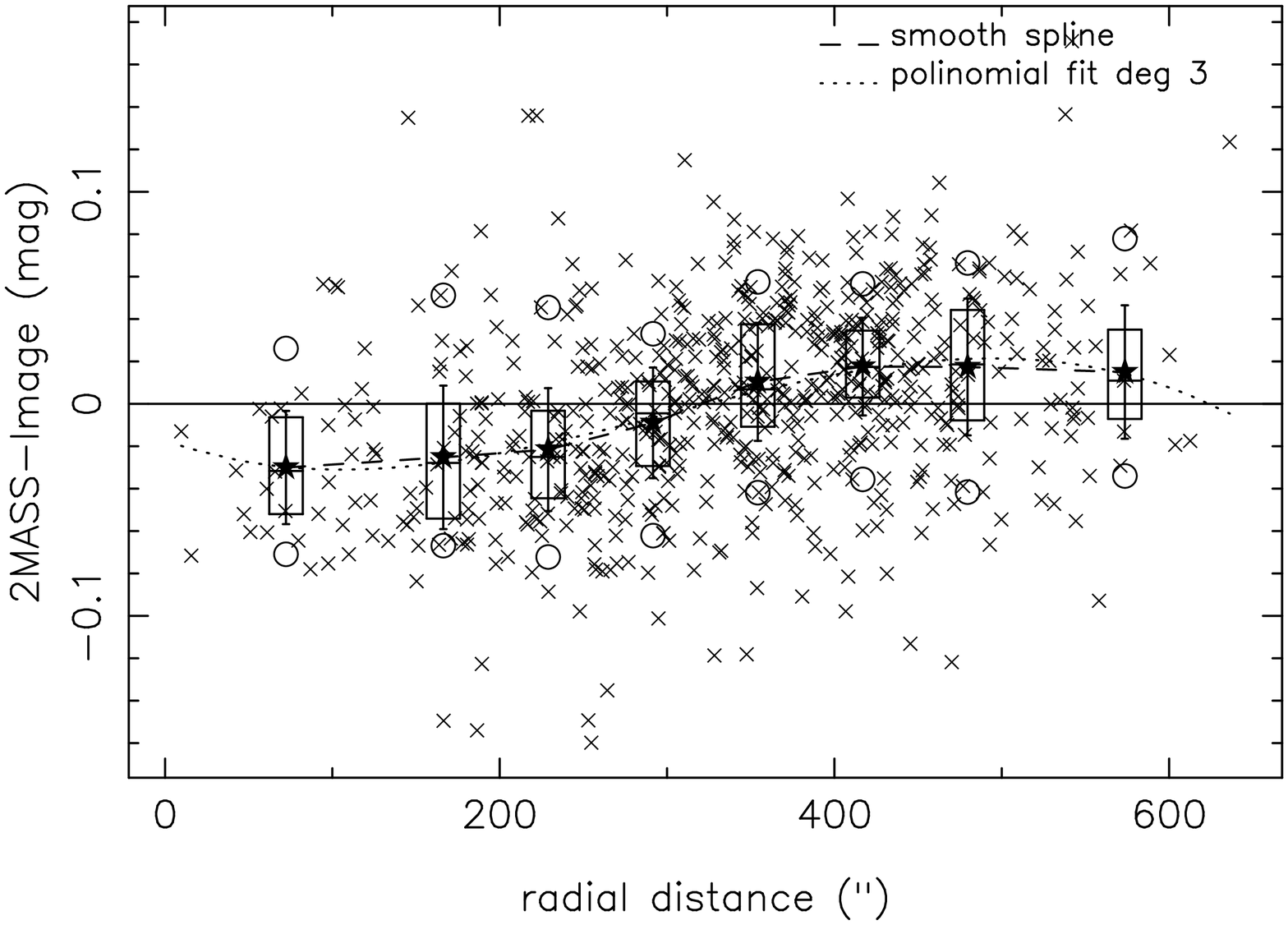}
\caption{Differences between 2MASS and ALHAMBRA photometry 
versus the distance to the nominal pointing center in the J band. 16
pointings have been combined and a binning have been performed in the
x-axis to increase the signal. For each bin the boxplot shows the mean
({\it stars}), the median, 1st and 3rd quartile ({\it boxes}), the
minimum and maximum ({\it
circles}),
values after trimming the 10\% on either side of the sample in each bin. The error bars represent the $1\sigma$ interval.
\label{Fig:radial}}
\end{figure}

\begin{figure}
\epsscale{1.0}
\plotone{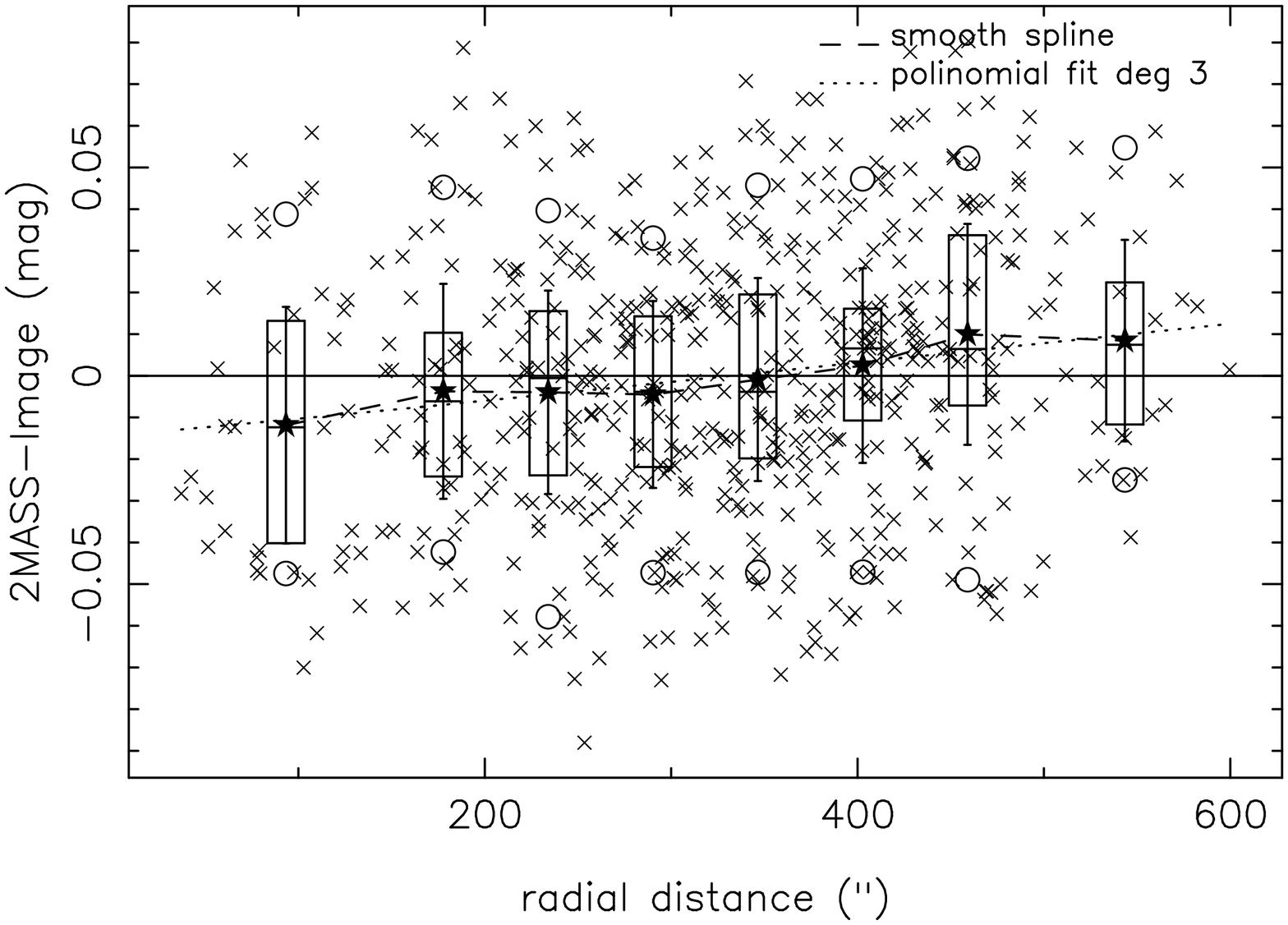}
\caption{Differences between 2MASS and ALHAMBRA photometry versus the distance 
to the nominal pointing center after applying the pupil ghost
correction in the flat-field as explained in the text. Symbols as in
Fig.~\ref{Fig:radial}. Note that the scale in the y-axis has change with respect to Fig.~\ref{Fig:radial}.
\label{Fig:radial2}}
\end{figure}

We found a significant effect, over 0.02 mags only in the J-band
images, as it is shown in Fig.~\ref{Fig:radial}.  This effect was
corrected by fitting the pupil ghost, using the mscpupil task in the
IRAF MSCRED package \citep{2002adaa.conf..309V}, and removing this
pupil image from the flat-fields and individual images.  The final
resulting radial differences are shown for the J band in
Fig.~\ref{Fig:radial2} where it clearly appears that all the
systematics was corrected well below the $1\sigma$ level.

\section{Galaxy Number Counts}

The steps to compute robust counts to the 50\% detection level of the
images are similar to those described in \cite{2003ApJ...595...71C}
and in \cite{2006ApJ...639..644E}. In the next section we detail how
the best set of SExtractor parameters was estimated as a compromise
between optimizing the depth at the 50\% completeness level while
keeping low the number of spurious sources.

\subsection{Completeness corrections}

\begin{figure}
\epsscale{0.8}
\plotone{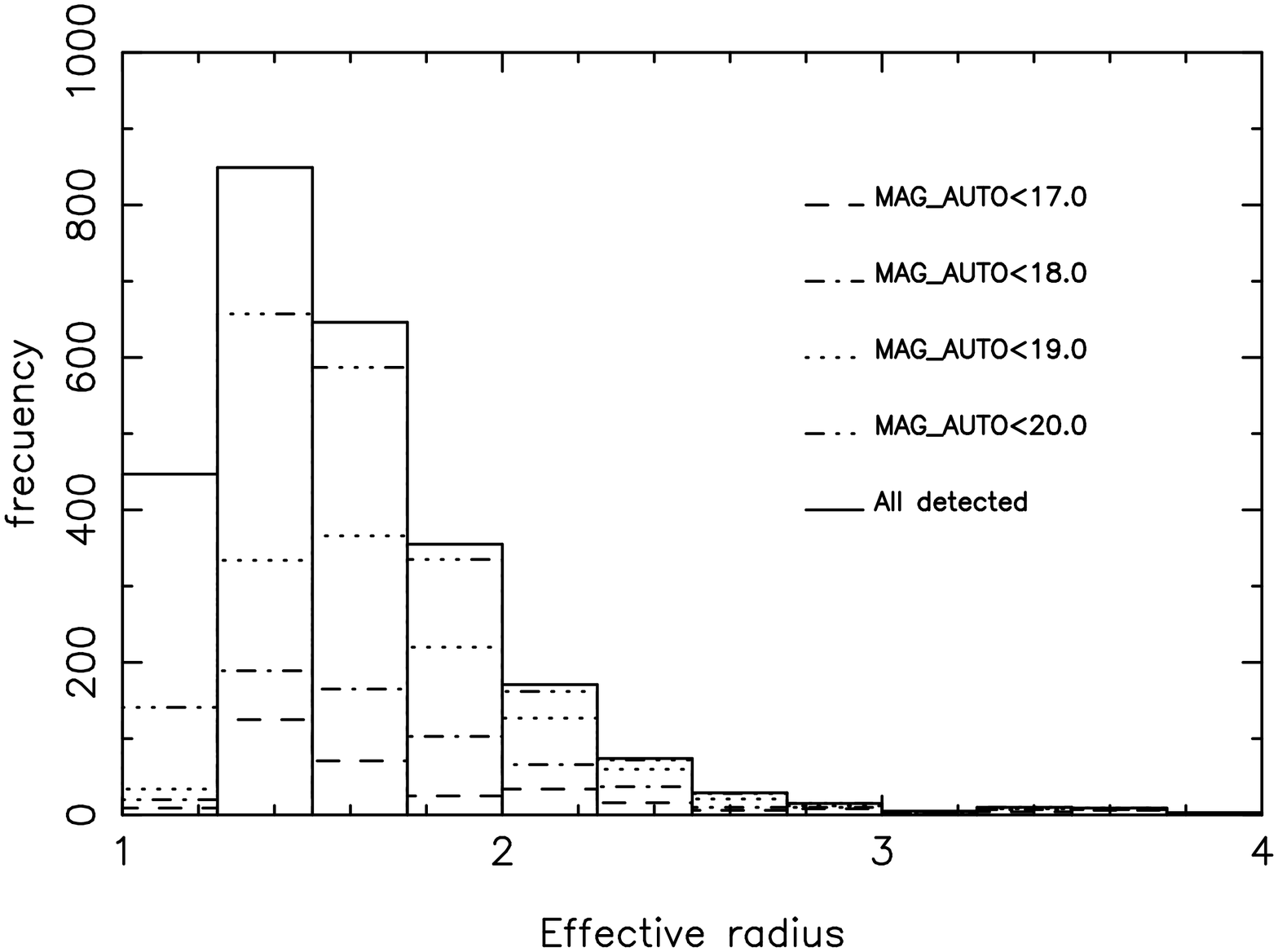}

\plotone{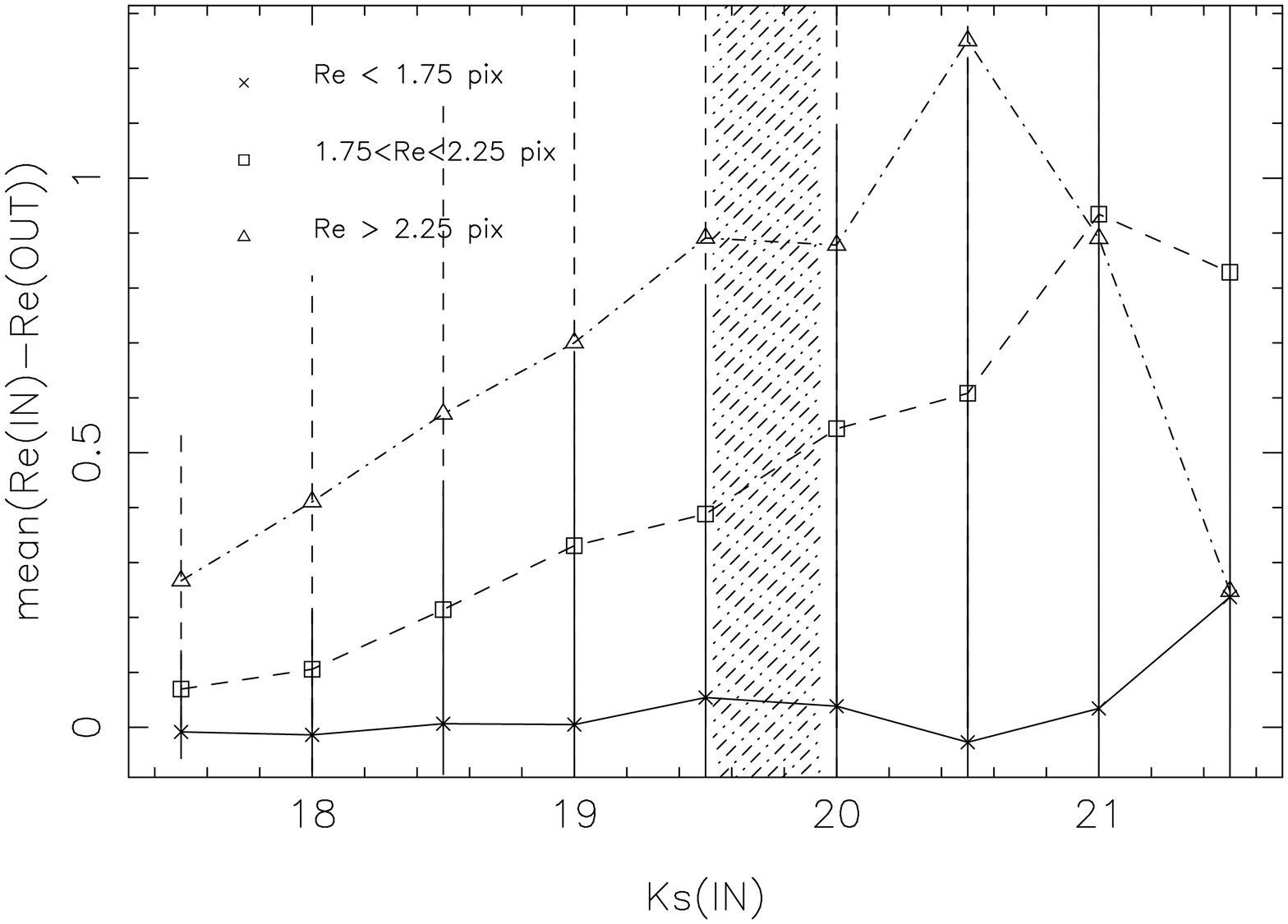}
\caption{ {\it (Top panel)} Histogram of Re for the sources in one of the pointings of the field ALH08
 {\it (Bottom panel)} Differences between Re(in) and Re(out) for the dimmed injected sources. The symbols correspond to $\times$ Re(in)$\le1.75$ pixels, $\square$ $1.75<$Re(in)$\le2.25$ and $\triangle$ Re(in)$>2.25$ pixels. The shaded is the range of magnitudes between the 50\% and the 80\% of detection efficiency. \label{Fig:rerec}} 
\end{figure}

In order to compute the corrections that should be applied to the
faint part of the galaxy counts we have performed a set of Monte Carlo
simulations where real sources from the images were injected back to
the same science image at different positions. The completeness
correction to be applied depends on the surface brightness profile of
the source. To account for this we computed a different correction
function for sources in three effective radius (Re) intervals. Those
Re intervals in pixels for the simulations were chosen from the
histogram of the Fig.~\ref{Fig:rerec} (top panel): $Re\le1.75$, $1.75
< Re \le 2.25$ and $Re > 2.25$. In Fig. \ref{Fig:rerec} (bottom
panel) it is shown how the Re recovered decreases as the magnitude
goes fainter (see below for a description of how these values were
obtained).

\begin{figure}
\epsscale{1.0}
\plotone{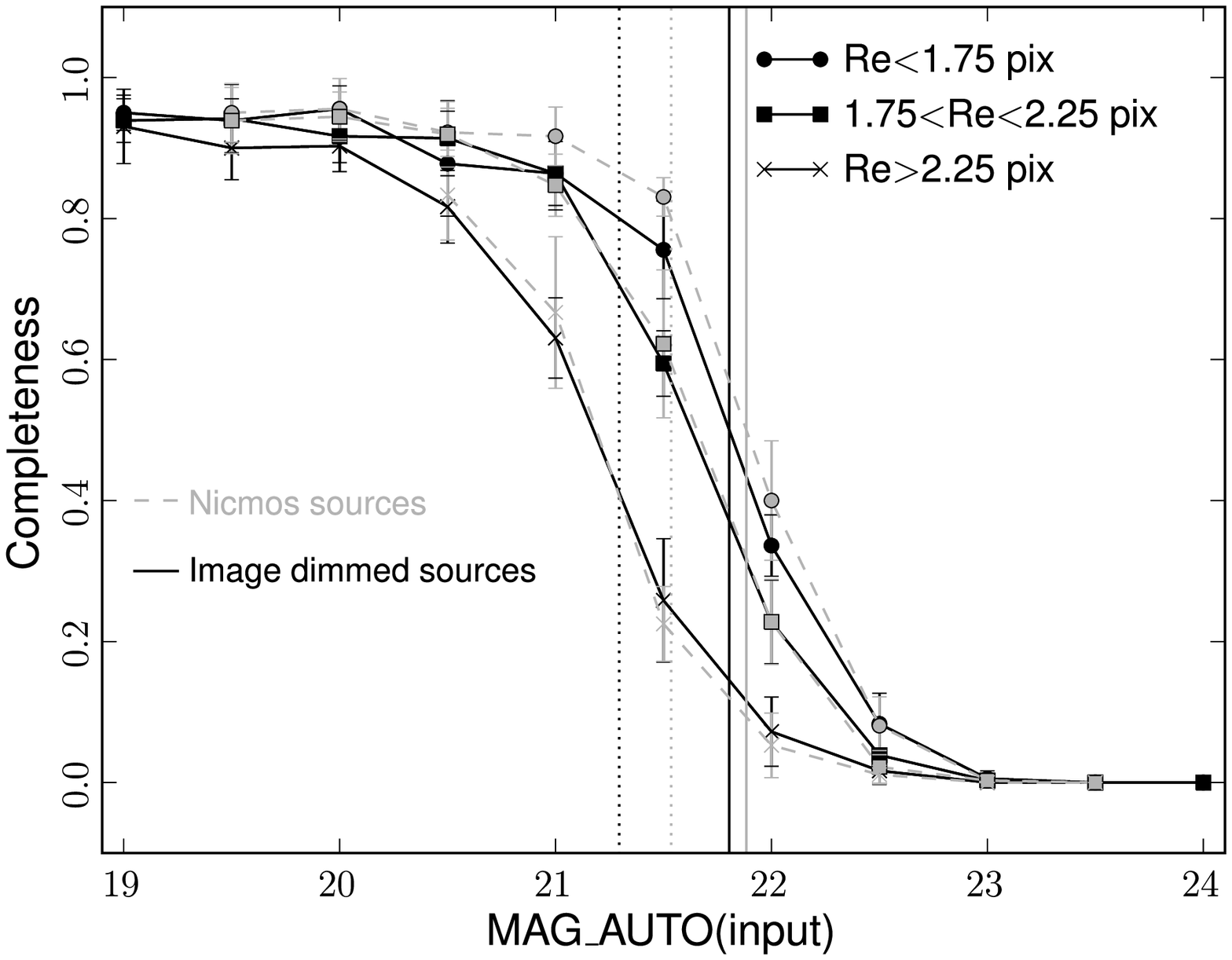}
\caption{Completeness correction for the same pointing of the previous figures in the J band using 0.8$\sigma$ DETECT\_THRESH and the 2.5 pixel gaussian kernel. The vertical dotted (solid) lines are the magnitude at the 50\% (80\%) of detection efficiency.
\label{Fig:completeness}}
\end{figure}

For the simulations, bright sources were selected from the
image in each Re interval. These sources were artificially dimmed to
the 0.5 magnitude bin under study and injected back on the image
randomly. In each iteration 40 sources were simulated, computing the
recovering fraction and the robust mean (trimming the 10\% on either side of the distribution)
of the
SExtractor desired output parameters differences between the dimmed
input sources and the recovered ones. Using 9 of these iterations in
each magnitude bin, we estimated the mean and rms of the recovered
fraction, and SExtractor parameters differences over all the
meaningful magnitude range. In Fig.~\ref{Fig:completeness} the
completeness correction for the three Re ranges is shown for one of
the pointings in the J band.

It can be argued that with that method unrealistic pseudo-artificial
sources could be produced. Whereas this could be true, the goal of
this procedure is to parameterize the recovering efficiency on the
basis of the source and image characteristics without any physical
consideration, which will be implicitly taken into account when
performing the corrections on the real data. As a validation of the
procedure, the same simulations have been performed using real NICMOS
F110W sources from the HDF-S Flanking Fields. The NICMOS images were
resampled and convolved with a suitable Gaussian kernel to match the
pixel scale and FWHM of the OMEGA2000 image under study. Initial
sources were taken in the interval [m-1,m+1] (being m the magnitude
under study), which produces a realistic $mag-R_e$ relation. As can be
seen from the Fig.~\ref{Fig:completeness} the results in the
completeness correction are the same that using dimmed sources from
the image.

\begin{figure}
\epsscale{0.8}
\plotone{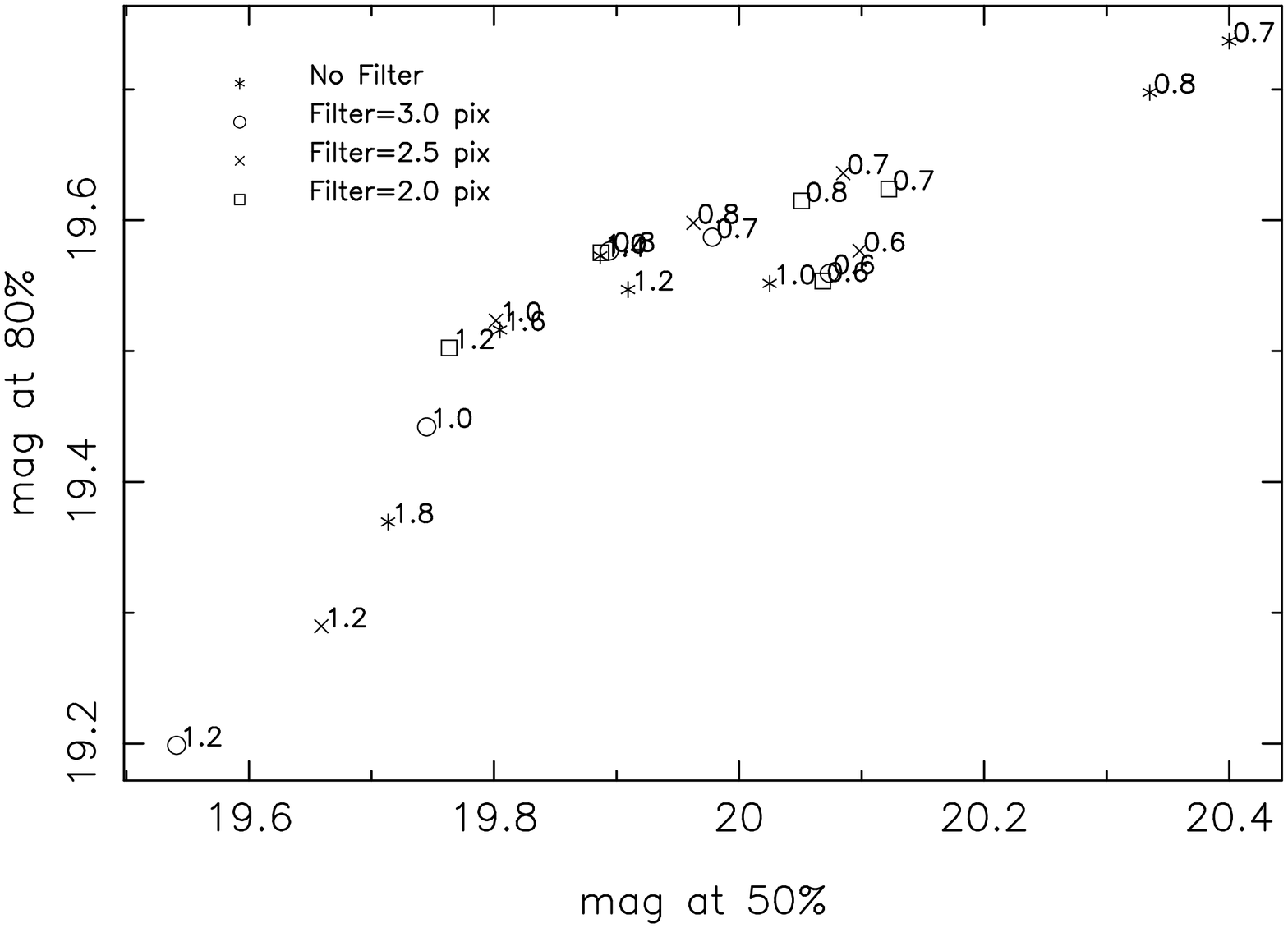}

\plotone{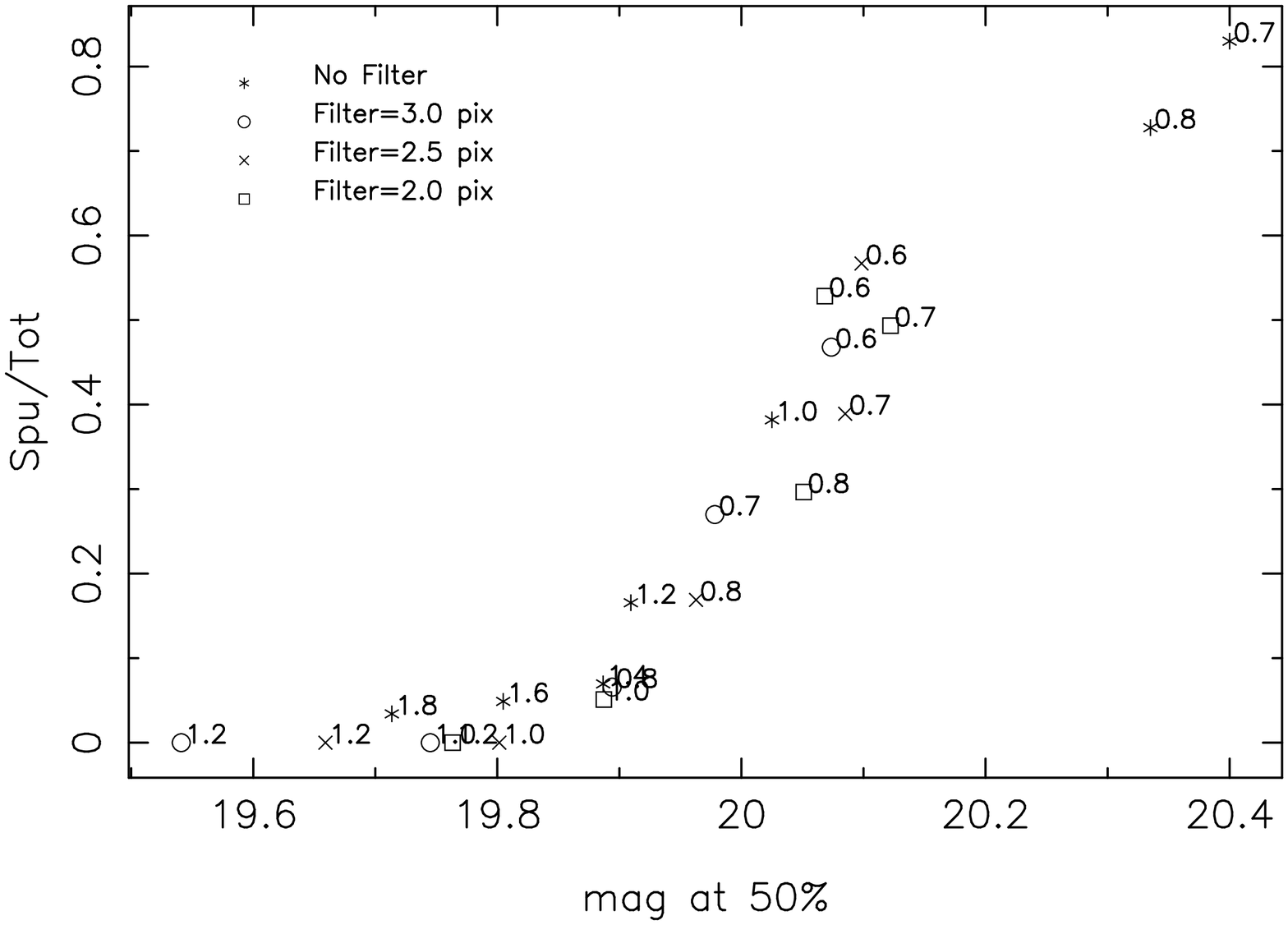}

\caption{ {\it (Top panel)} 50\% vs. 80\% of recovering efficiency for the same pointing as before in Ks band using different SExtractor detection strategies. The size in pixels of the SExtractor Gaussian convolution kernel is given in the labels. The detection thresholds are indicated at the points. {\it (Bottom panel)} 50\% of detection efficiency vs. spurious to total ratio.\label{Fig:profprof}}
\end{figure}

In the top panel of Fig.~\ref{Fig:profprof} we show the depth to the
50\% and 80\% of recovering efficiency, computed using a linear
interpolation in the magnitude vs. efficiency data. The figure
indicates that a decreasing of the detection thresholds, in order to
get a fainter level at the 50\% of recovery efficiency limit, will not
improve by much the limit at the 80\% of completeness, which seems to
reach a plateau. Moreover, as will be seen from the reliability plots,
it produces a significant increase of detected spurious sources, as we
explain below.

\subsection{Detection reliability}

To accurately compute the galaxy number counts it is important to
establish the reliability of the detections at the faint magnitude
end.  To find out the optimum way to evaluate that reliability we have
studied the performance of three different methods. The first approach
was to create artificial sky images with the same rms and background
distribution than the real ones. The ratio of spurious versus real
detections was computed running SExtractor over the science and the
artificial sky images.

We have also inspected the performance of the method used in
\cite{2003ApJ...595...71C}. Basically half exposure time images were
created from two complementary sets of the data. The detections were
performed in the total time image and the source fluxes measured in
the half time images using the SExtractor double image mode for the
same automatic apertures. Those created sources showing a magnitude
difference greater than $3\sigma$ were considered spurious.

The last method and the one that, at the end, produces the best
results, consisted in constructing sky images using a similar
combination procedure that the used to create the science images:
combining the unregistered processed images with subtracted background
using an artificial dither pattern. The major difficulty here was to
remove the extended sources that could bias the sky even when doing
trimmed mean (discarding 20\% of the pixels at each side). We have
confirmed that SExtractor does locate these smooth deviations over the
sky rms when filtering is used. To avoid this we multiplied those sky
images by -1 and used them as real sky images.

\begin{figure}
\epsscale{1.0}
\plotone{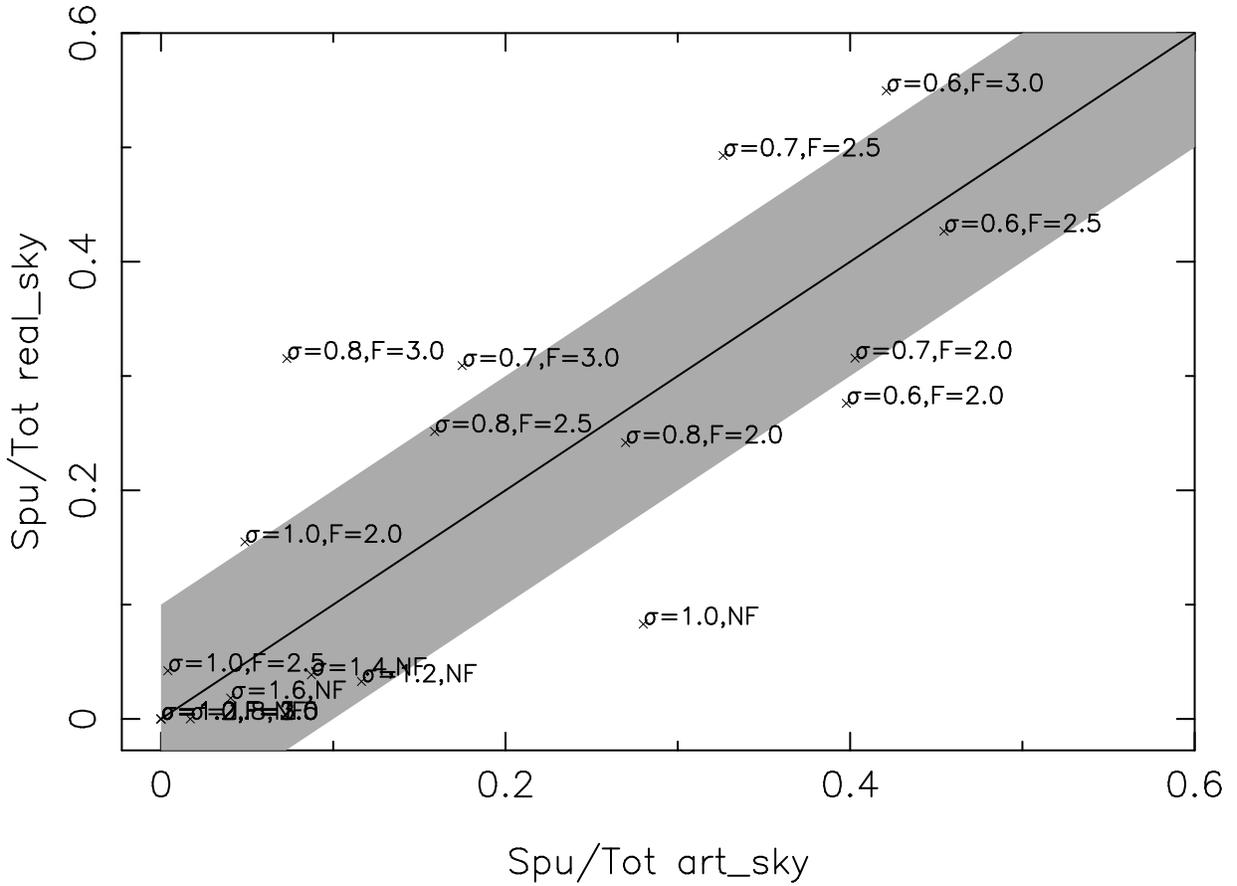}
\caption{Ratio of spurious to total detections when artificial and real sky images 
(see text) were used. Different SExtractor DETECT\_THRESH ($\sigma$)
and FILTER (F) combinations are shown. The grey area corresponds to a
difference less than 0.1.
\label{Fig:spuvsspu}} 
\end{figure}

In the Fig.~\ref{Fig:spuvsspu} we show a comparison between the
spurious rate at the 50\% of detection efficiency computed over the
'real sky' and the artificial sky. A good agreement between both
methods is observed, indicating that the use of artificial images to
compute the ratio of spurious to total detections is adequate. Being
artificial images faster to construct we made use of them to estimate
the detection reliability in each pointing.

\begin{figure}
\epsscale{0.8}
\plotone{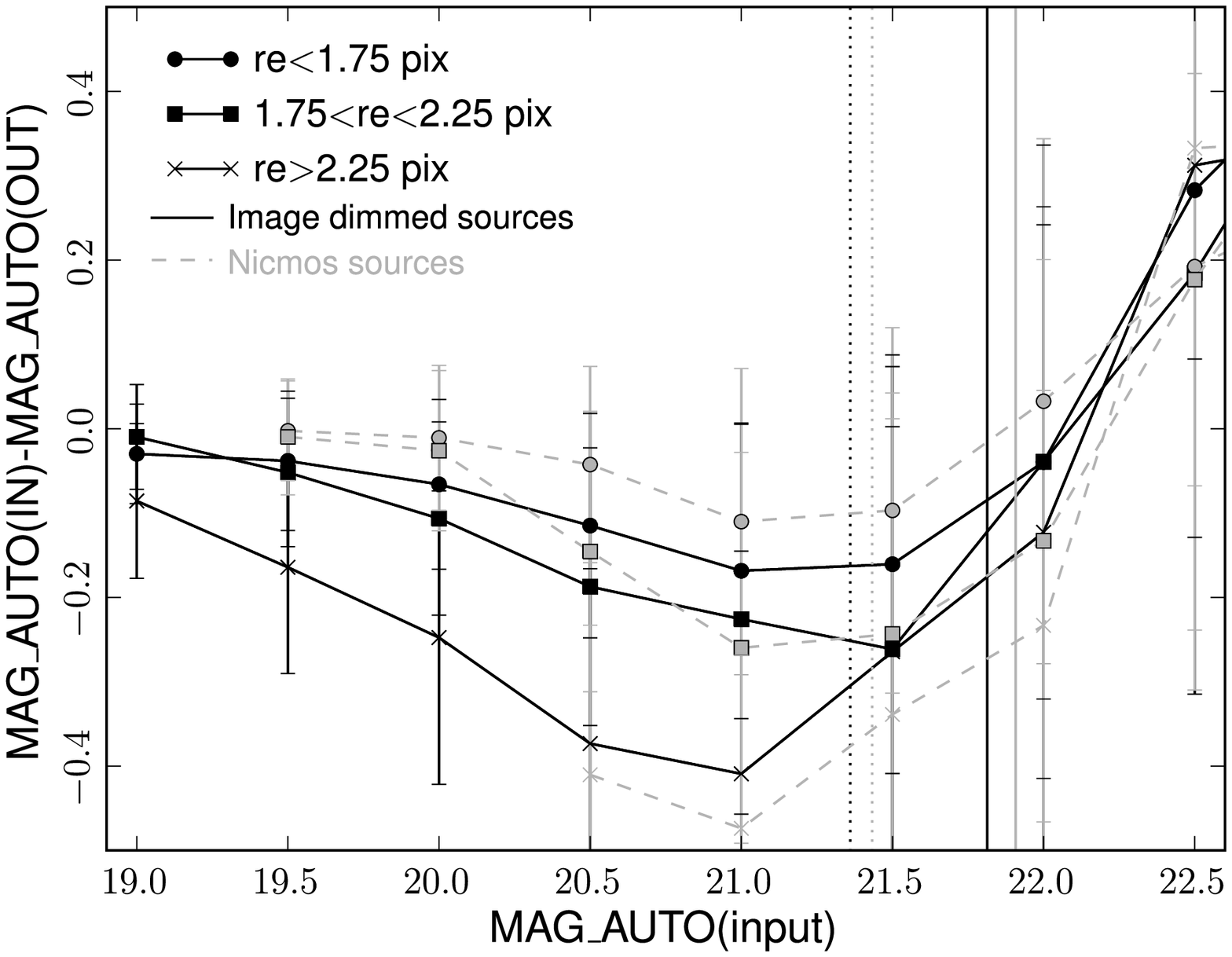}

\plotone{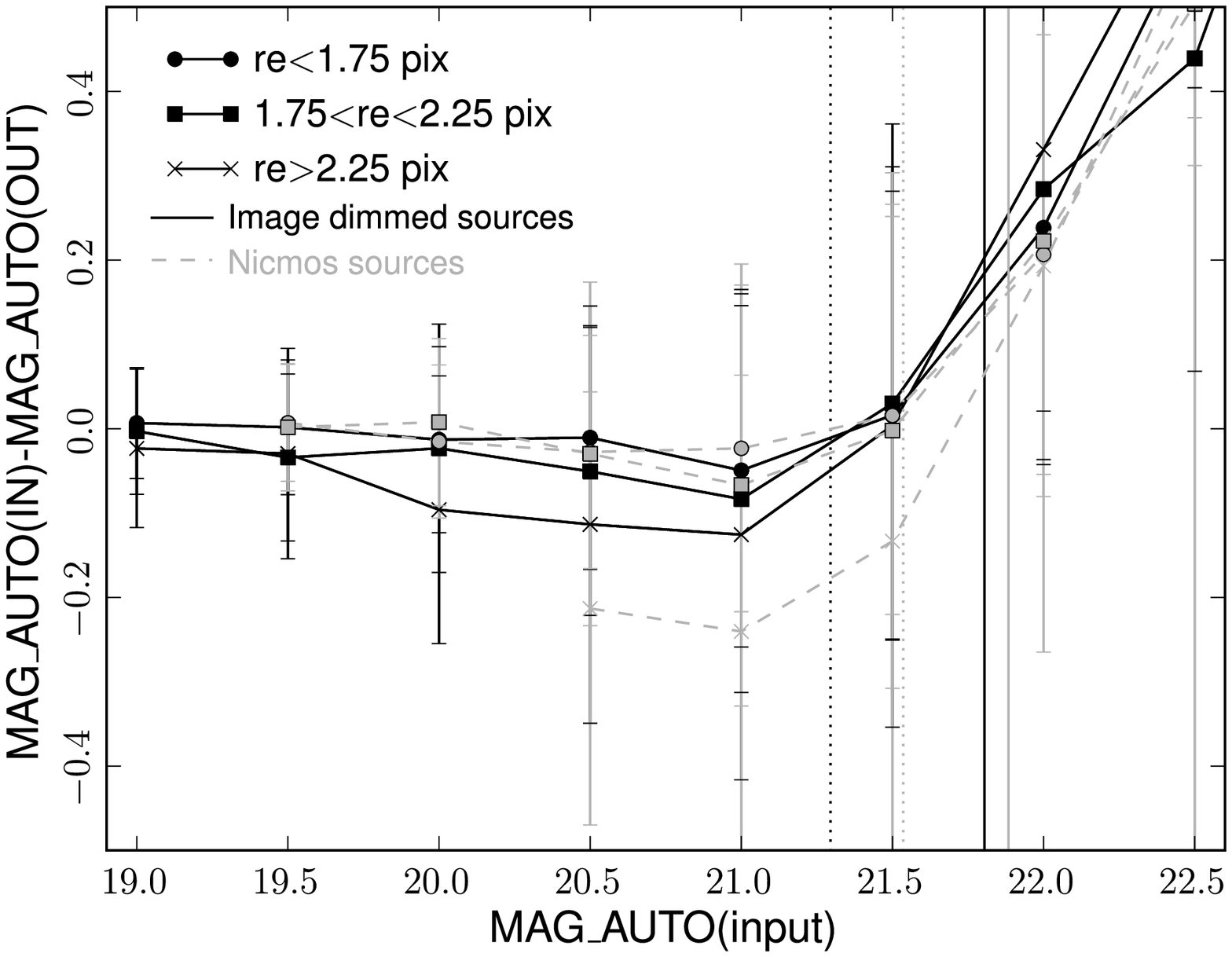}
\caption{Differences between the input and the SExtractor recovered AUTO magnitudes in one of the pointings of ALH08 in the J band for two SExtractor FILTER and DETECT\_THRESH combinations: no filtering and DETECT\_THRESH=1.2 {\it (upper panel)}, and 2.5 pixel filtering and DETECT\_THRESH=0.8 {\it (bottom panel)}. The vertical dotted (solid) lines are the magnitude at the 50\% (80\%) of detection efficiency.
\label{Fig:magrec}} 
\end{figure}

The bottom panel of Fig.~\ref{Fig:profprof} shows the magnitude
reached at the 50\% of detection efficiency versus the spurious to
total ratio at the same magnitude bin. From the figure we can see that
in order to reach a deep 50\% detection limiting magnitude,
maintaining at the same time the number of spurious detections below
20\%, the optimum SExtractor DETECT\_THRESH-FILTER combinations are:
DETECT\_THRESH=1.2 or 1.4 without filtering, or DETECT\_THRESH=0.8
using a filtering with a Gaussian kernel of size similar to the image
FWHM. In the top panel of Fig.~\ref{Fig:profprof} it is shown that
those combinations reach roughly the same magnitude at the 80\% of
recovery efficiency. We have decided to use the latter
filter-DETECT\_THRESH combination because, as can be outlined from the
bottom panel in Fig.~\ref{Fig:magrec}, the differences between the
input and the recovered AUTO magnitudes in the simulations are close
to zero in all the magnitude range up to the magnitude at the 80\% of
recovery efficiency. Close to the magnitude at the 50\% of
completeness the recovered magnitude appears to be $\sim0.1-0.2$ mag
brighter than in the previous bin. This effect could be due to the
fact than only those sources that suffer from noise brightening were
found by SExtractor and used to compute the input-output magnitude
differences, producing a bias towards brighter recovered
objects. Following those results, the number counts in the following
sections will be computed and corrected up to the magnitude of 80\% of
recovery efficiency for point-like sources, avoiding any possible
systematics due to a magnitude shift at the 50\% completeness bin.

\begin{figure}
\plotone{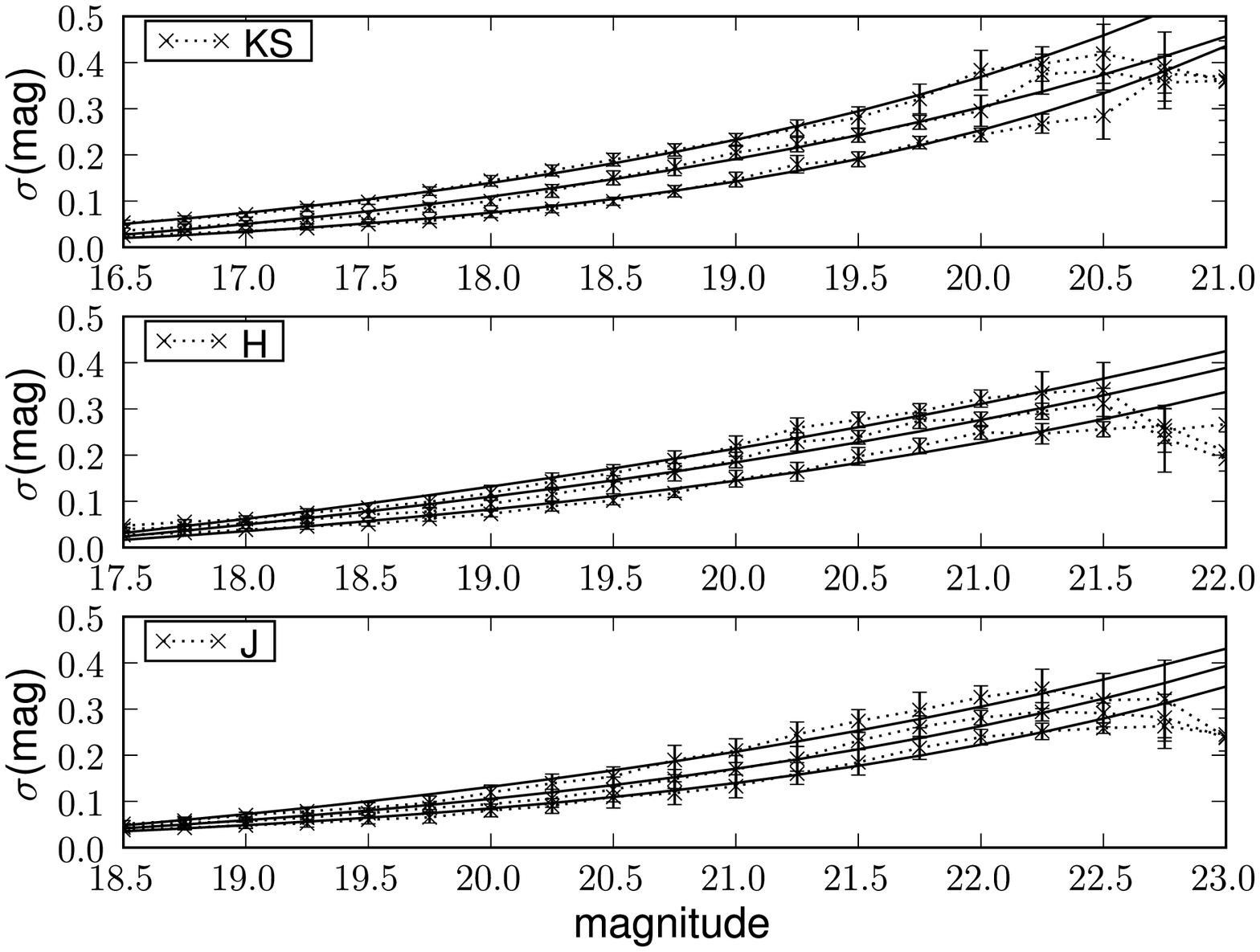}
\caption{Median photometric random errors in magnitudes per magnitude bin ($\sigma_m$) in the three NIR filters for the eight ALH08 pointings {\it (crosses and dotted lines)}, the three lines represent sources with $Re<=1.75$, $1.75<Re<=2.25$, and $Re>2.25$ pixels. The magnitude errors are computed as the rms of the recovered sources magnitudes in each bin of the simulations described in the text. Error bars represent the rms of the computed $\sigma_m$ among the pointings. Exponential grow fit ($\sigma(m)= \sigma_0 + a\exp{[b(m-m_0)]}$) to the magnitude errors in each band  {\it (solid lines)}.
\label{Fig:MagError}} 
\end{figure}

As pointed before, the photometry of the sources was obtained
using MAG\_AUTO. Simulations indicate no significant differences
between simulated and recovered magnitudes (at the 80\%
recovery efficiency) for the used values of the SExtractor
parameters FILTER and DETECT\_THRESH.
Using the same kind of simulations we computed the rms of the recovered  MAG\_AUTO values
in each bin to characterize the photometric error, the results are shown in Fig.~\ref{Fig:MagError}. 

\subsection{Star-galaxy separation}

\begin{figure}
\epsscale{1.0}
\plotone{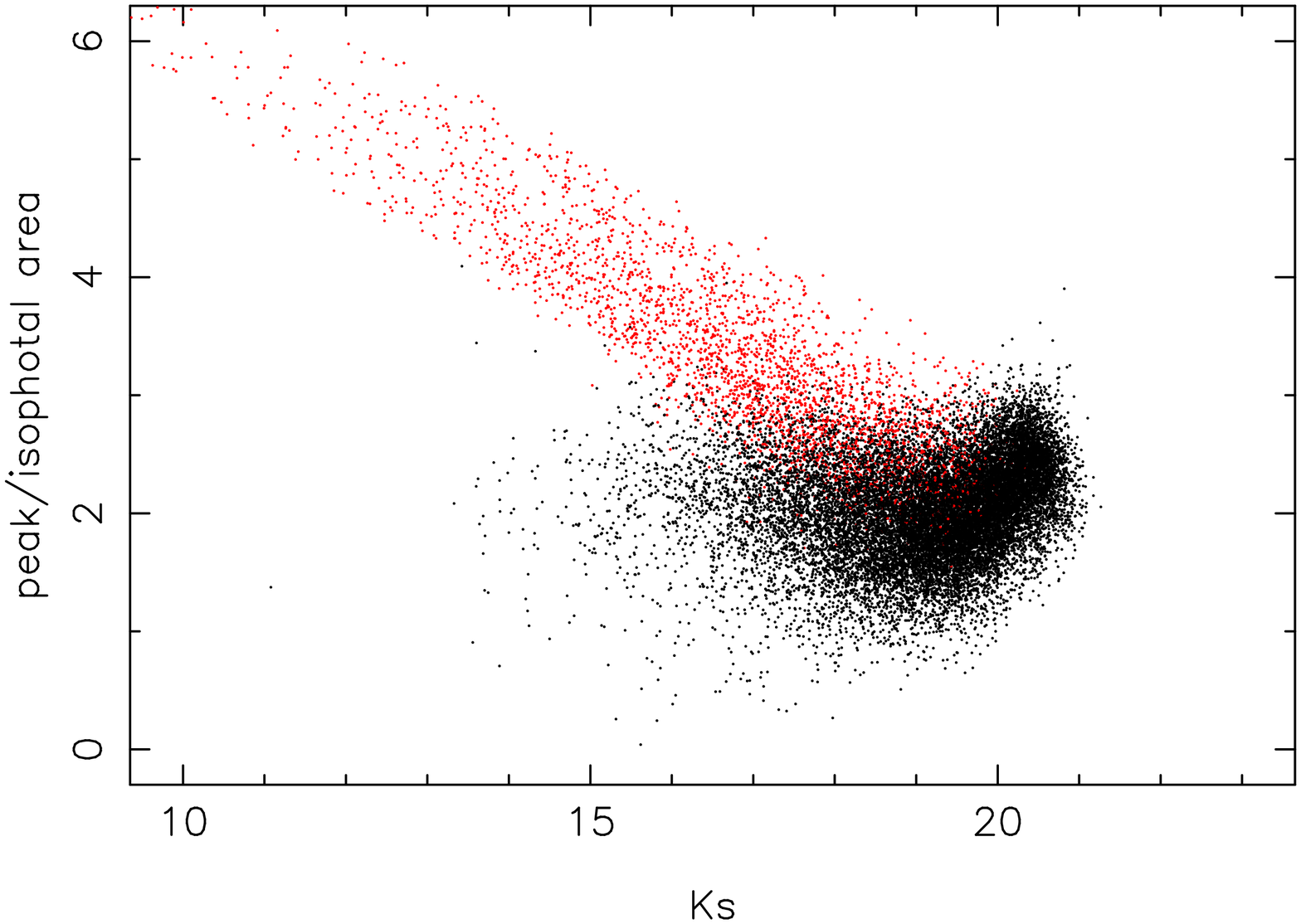}
\caption{\label{Fig:peakiso} Peak/ISOPHOTAL\_AREA vs. Ks diagram. 
The red dots are object for which the SExtractor stellarity index is $>$0.8.} 
\end{figure}

In \cite{2003ApJ...595...71C} it was shown that fainter than Ks=17.0
the correction due to stars is $<0.06$ dex. However, given the lower
Galactic latitude of the ALH08 field, a higher number of contaminating
stars is expected. A correct star subtraction is relevant in the
intermediate magnitude range. At bright magnitudes stars can be easily
separated from galaxies using a compactness criteria. In contrast, at
intermediate magnitudes the star/galaxy separation is more demanding
because many galaxies are barely resolved with small apparent sizes
compared with the FWHM and pixel resolution, as can be seen from the
diagram shown in Fig.~\ref{Fig:peakiso}.

\begin{figure}
\epsscale{0.8}
\plotone{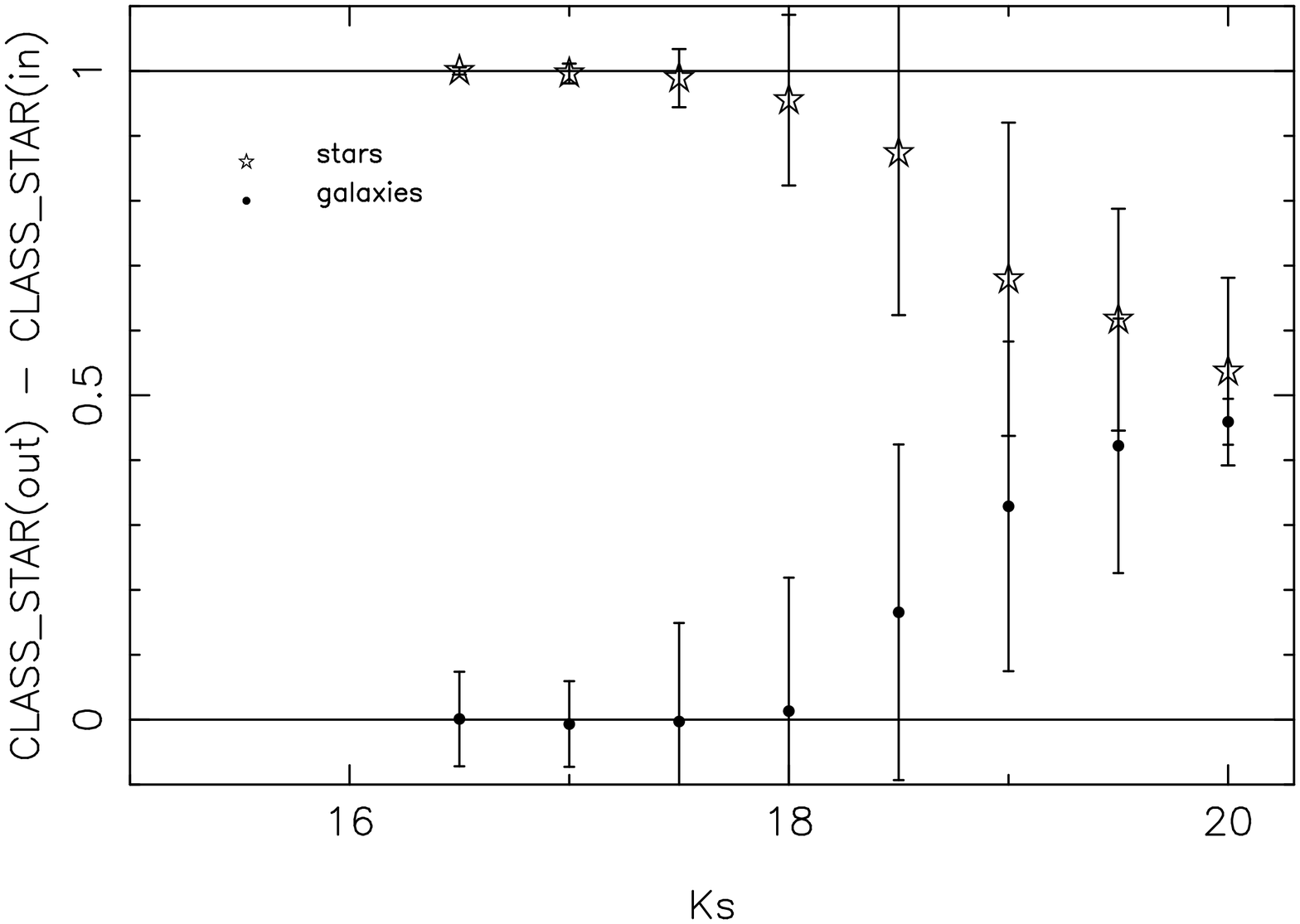}

\plotone{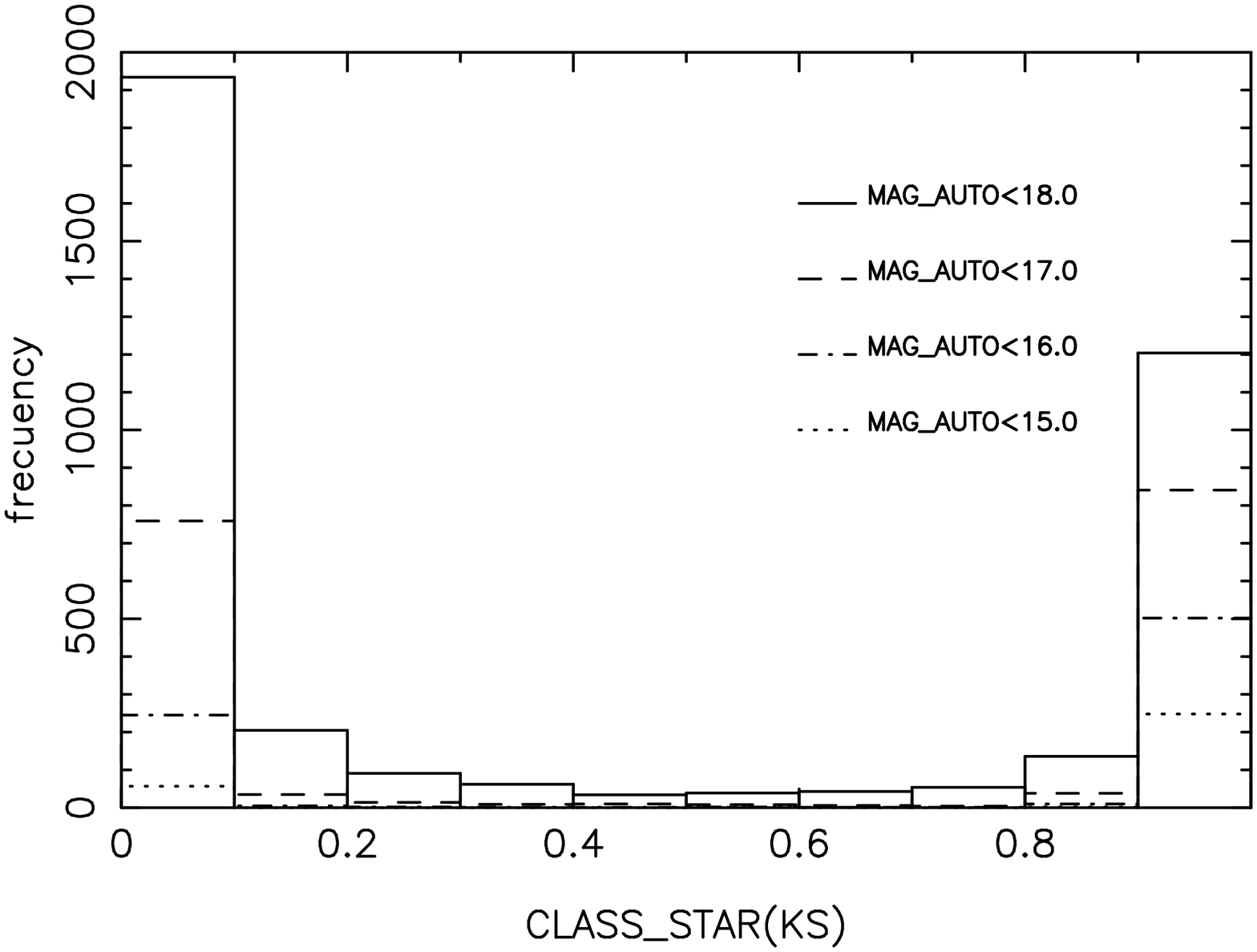}
\caption{\label{Fig:classstar1} {\it (Top panel)} CLASS\_STAR input-output differences as a function of input artificially dimmed magnitude for a set of well defined stars (CLASS\_STAR(in) $\sim 1$) and galaxies (CLASS\_STAR(in) $\sim 0$). In the case of stars 1-(CLASS\_STAR(out)-CLASS\_STAR(in)) is plotted. {\it (Bottom panel)} Histogram of CLASS\_STAR for different ranges in the Ks magnitudes.}

\end{figure}

The viability of star/galaxy separation using the SExtractor
neural network has also been analyzed. For this purpose, using the
same Monte Carlo method explained before, bright stars and galaxies
from the images have been artificially dimmed to each magnitude bin
and their SExtractor stellarity parameter recovered. In
Fig.~\ref{Fig:classstar1} it clearly appears how, in one final frame
with FWHM=1.1, the input and recovered CLASS\_STAR differences are
less than 0.1 up to Ks=18.0. But, in the next bin Ks$>18.5$, the
CLASS\_STAR of the dimmed stars and galaxies could lead to some
misclassification. Nevertheless, as can be seen from the histogram in
the bottom panel of Fig.~\ref{Fig:classstar1}, a non negligible number
of objects start to populate the range from 0.4 to 0.8 in CLASS\_STAR
at magnitudes fainter than 17 in Ks, and the selection of the
CLASS\_STAR cut off might bias the star counts estimates.

\begin{figure}
\epsscale{1.0}
\plotone{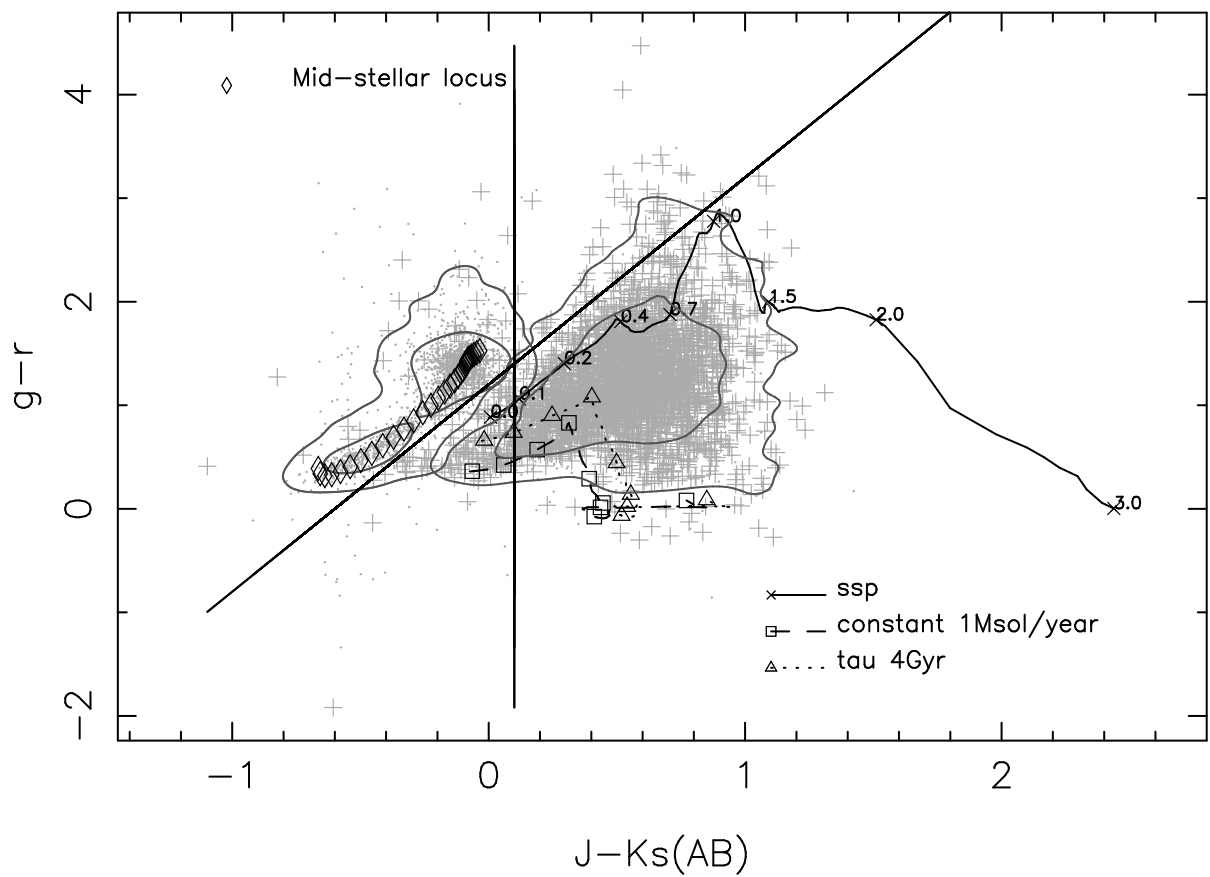}
\caption{g-r vs J-Ks plot used to perform the star-galaxy 
separation. All objects in ALH08 with $S/N>5$ are plotted. The line
separating the two classes is $g-r = 1.2+2.0\cdot (J-Ks)_{AB}$. The
plotted symbols correspond to the Sloan classification {\it dots} for
representing stars and the {\it crosses} for galaxies. Redshift-color
tracks for three galaxy models constructed using the
\cite{2003MNRAS.344.1000B} code are shown. The redshift values are
displayed over the Single Stellar Population (SSP) track, the marks
over the other 2 curves have the same redshift spacing. The {\it
diamonds} represent the mid stellar locus given in
\cite{2007AJ....134.2398C}. Two contour lines containing the 95\% and
the 68\% are displayed for the galaxies and stars using the Sloan
classification.}
\label{Fig:sgsep} 
\end{figure}

\begin{figure}
\epsscale{0.8}
\plotone{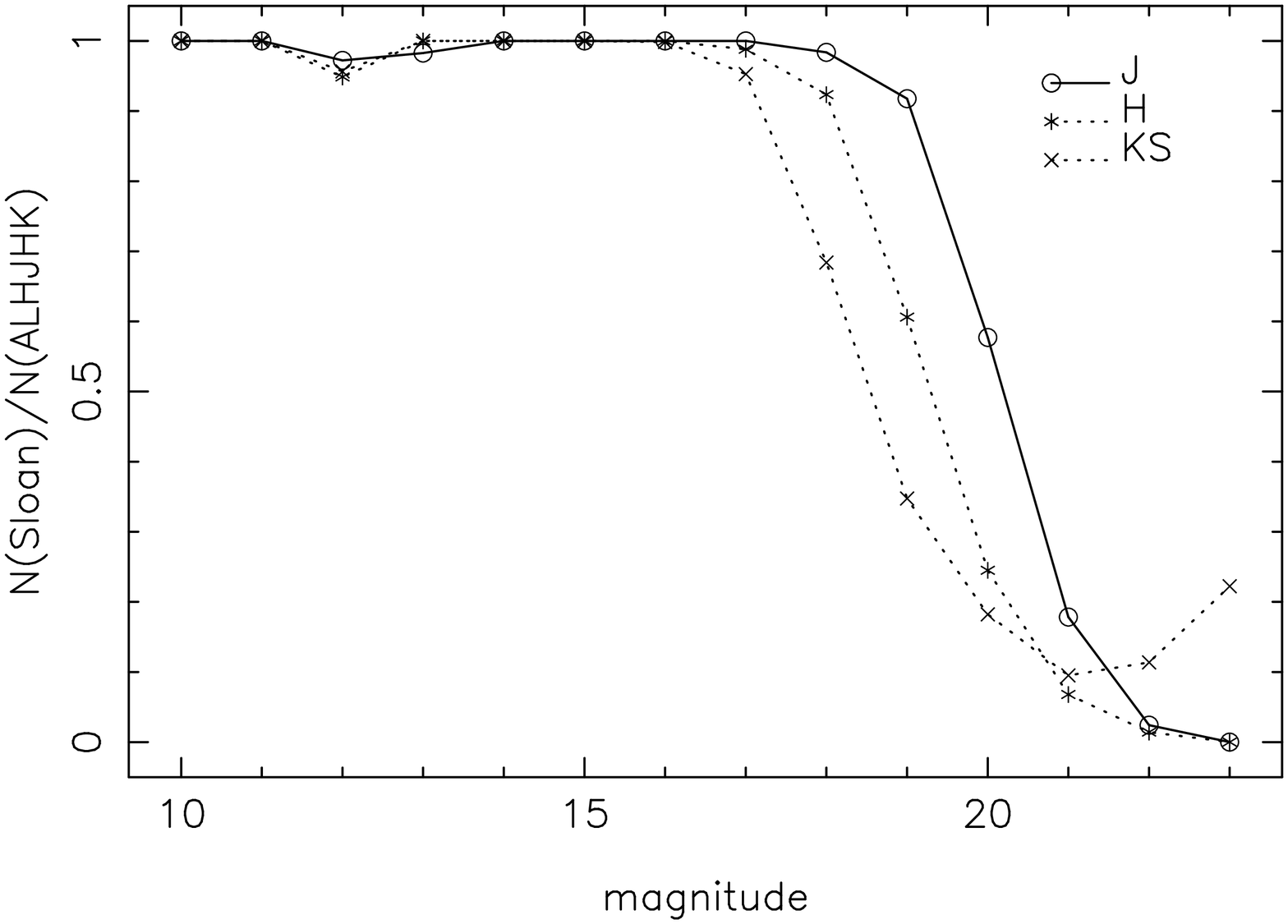}

\plotone{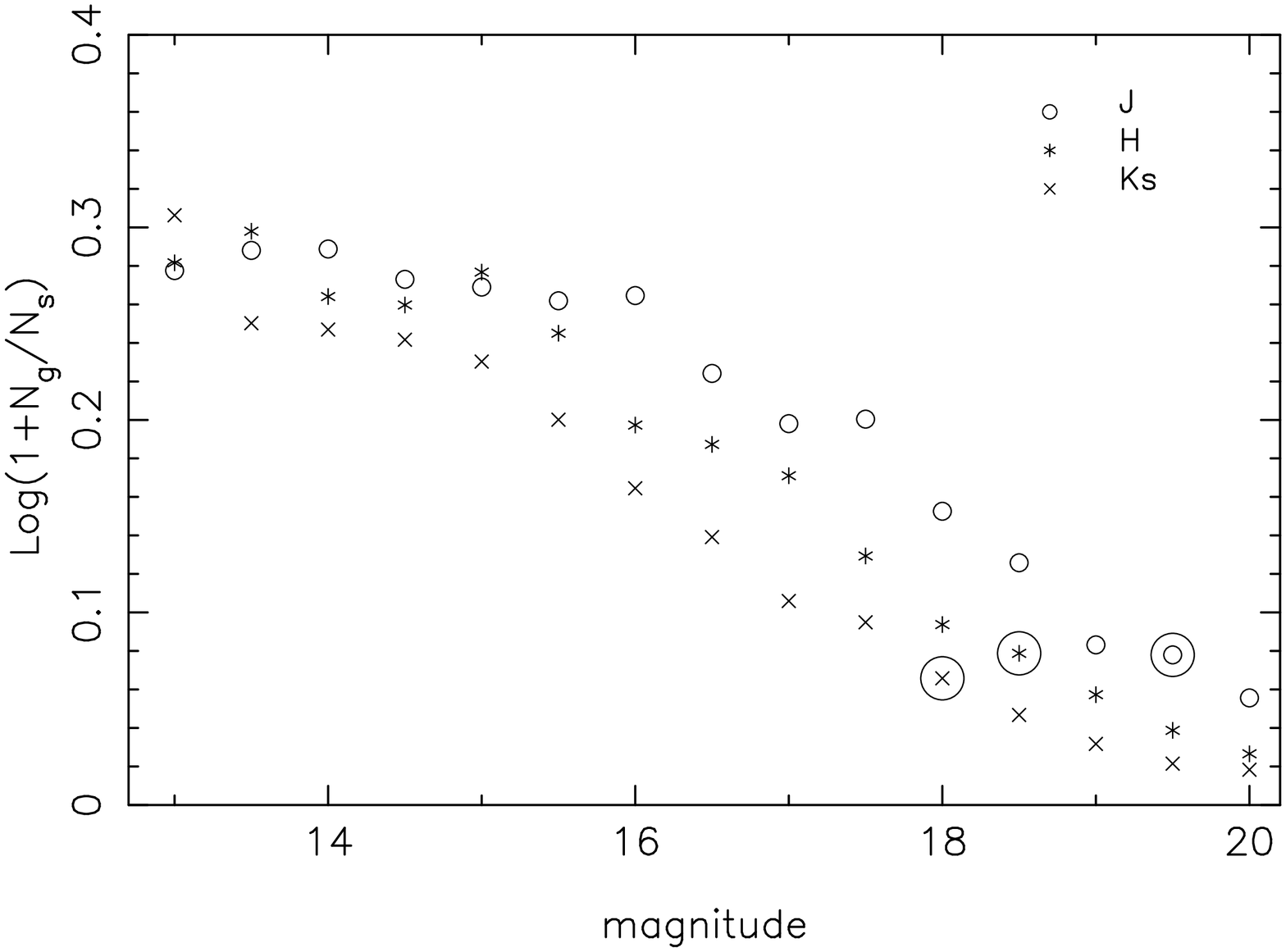}
\caption{\label{Fig:classstar} {\it (Top panel)} The ratio of sources identified in the Alhambra 
field for which there is a counterpart in the Sloan SDSS DR5 catalog. {\it
(Bottom panel)} The correction subtracted from log(N)
galaxy counts due to stars. The big open circle indicates the point up
to which the number of stars can be estimated from the Sloan-Alhambra
colors (see text), the fainter star counts assume a flat extrapolation
from this value.}
\end{figure}

\begin{figure}
\epsscale{1.0}
\plotone{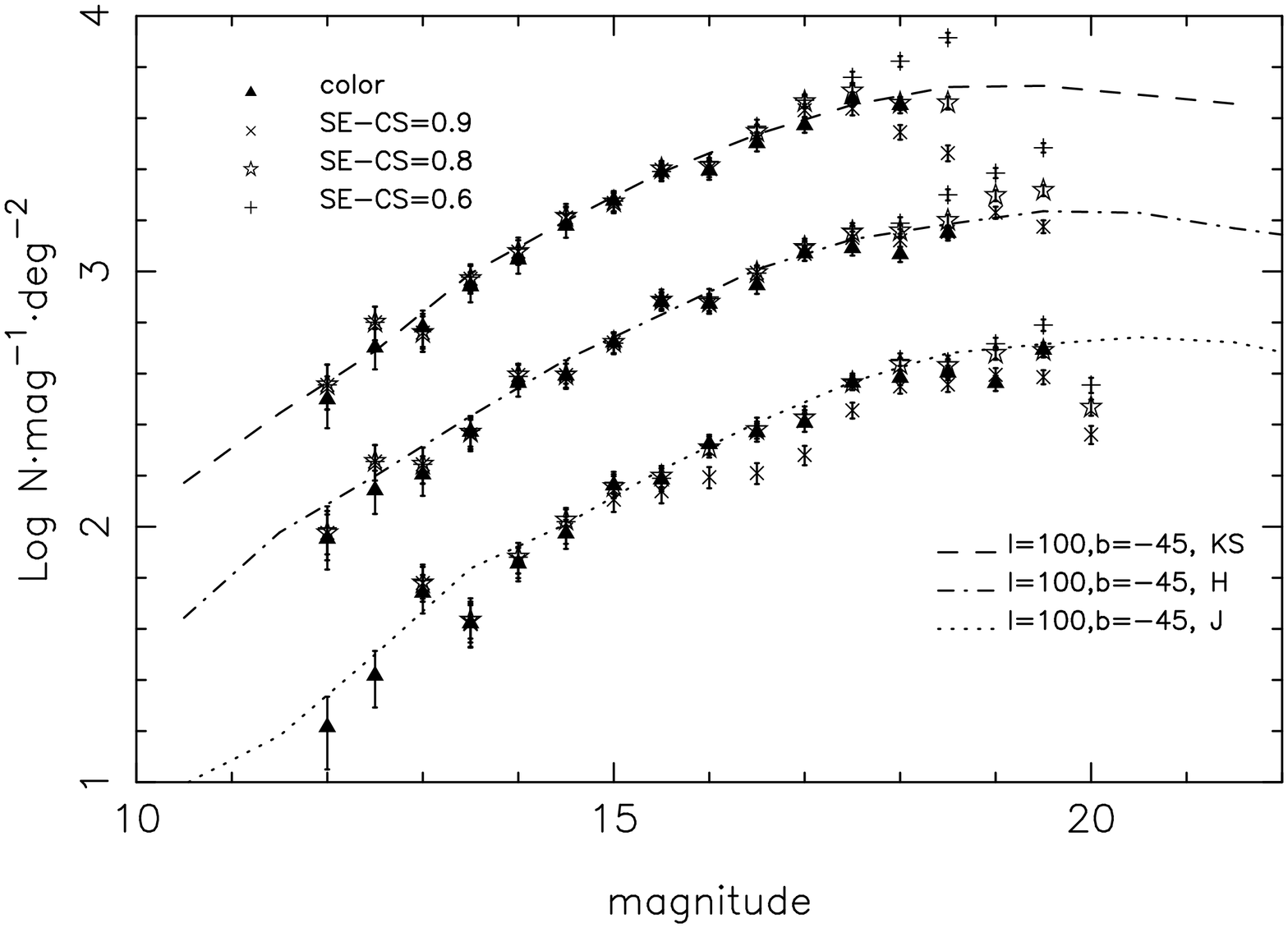}
\caption{\label{Fig:starcor} Star counts for the ALH08 field compared with the \cite{1996AA...305..125R} Galactic count model. The J and Ks data have been plotted with an offset of -0.5 and 0.5 dex in the y-axis.} 
\end{figure}

An alternative way to perform the star/galaxy separation makes use
simply of color-color diagrams.
\cite{1997ApJ...476...12H} have established a reliable
star/galaxy separation using the B-I vs I-K colors. Here we make use
of the SDSS DR5 data for the ALH08 field to proceed with this
separation using the g-r vs J-Ks colors shown in
Fig.~\ref{Fig:sgsep}. The star counts are corrected by the ratio
of the Sloan/Alhambra completeness factors in the corresponding filter
shown in Fig.~\ref{Fig:classstar}. The star counts using
Sloan-Alhambra colors were computed to magnitude 19.5,18.5 and 18.0
respectively in the J,H and Ks band, where the Sloan/Alhambra
completeness is $>$0.5. For the fainter star counts we used a 0-slope
extrapolation where the models of star counts in the Galaxy are
flatter (see Fig.~\ref{Fig:starcor}). In Fig.~\ref{Fig:classstar} bottom
panel, the correction to the $\log(N)$ galaxy counts is presented,
showing that fainter than the magnitudes up to where the color-color separation
could be performed the correction in the three bands is $<0.08$ dex.

To check the validity of the star counts computed using the
color-color approach, in Fig.~\ref{Fig:starcor} those are compared
with the \cite{1996AA...305..125R} Galactic star counts models. The
lines correspond to star counts in the Galaxy for galactic coordinates
(l=100,b=-45), very close to the coordinates of the Alhambra-08 field
(l=99, b=-44). From this figure it appears that it is possible to
accurately remove the stars from the galaxy counts using the
color-color diagram. This was the method we have finally used to do
the star/galaxy separation. 

Finally is noteworthy to mention that, star/galaxy separation 
will be done more accurately when the photometry in the 23 Alhambra bands is
available, as the system will provide a sort of low resolution spectroscopy for each object.

\section{Results}

The corrected galaxy number counts have been computed for the ALH08
field in the three standard NIR filters in a consistent way in the
sense that they have been estimated following the same scheme for the
three bands. The number count data, computed and corrected up to the
magnitude of 80\% of recovery efficiency for point-like sources, are
presented in the Tabs.~\ref{Tab:cJ}, \ref{Tab:cH} and \ref{Tab:cK}.

The error in the number counts for each pointings are the sum in quadrature of the rms
in the estimation of the completeness corrections, with the contribution 
of Poisson noise and galaxy clustering calculated for each magnitude bin 
following eq. \ref{Eq:error} \citep{1997ApJ...476...12H}, that includes the
angular correlation function. 

\begin{equation}\label{Eq:error}
\sigma_i^2= N_i(m)+5.3\left(\frac{r_0}{r_{\star}}\right)\Omega_i^{(1-\gamma)/2}N_i^2(m)
\end{equation}

being $N_i$ the raw counts in the pointing, $r_0=7.1$ Mpc, $\gamma=1.77$
and $5\log(r_{\star})=m-M_{\star}-25$. $M_{\star}$ 
is set to -22.95,-23.69 and -23.93 for the J, H and Ks filters. The final uncertainty in the
combined counts per square degree and magnitude is given by:

\begin{equation}
\sigma(m)= \frac{\sqrt{\sum\sigma_i^2}}{\Omega\Delta m}
\end{equation}

\begin{deluxetable}{llllll}
\tabletypesize{\small}
\tablewidth{0pt}
\tablecaption{
Corrected galaxy number counts in the J band.\label{Tab:cJ}}
\tablehead{
\colhead{Magnitude}	& \colhead{raw counts}\tablenotemark{a} & \colhead{eff. cor.}\tablenotemark{b} & \colhead{$\log(Nc)$}\tablenotemark{c} & \colhead{$\log(Nm)$}\tablenotemark{d} & \colhead{area} \\ 
& & & $N\cdot mag^{-1}\cdot deg^{-2}$ & $N\cdot mag^{-1}\cdot deg^{-2}$ & $deg^2$\\}
\startdata
13.00 & 43.0 & 1.00 & $1.31^{+0.49}_{-1.31}$ & $1.30^{+0.73}_{-1.30}$ & 0.441 \\
13.50 & 31.0 & 1.00 & $0.92^{+0.73}_{-0.92}$ & $0.95^{+0.96}_{-0.95}$ & 0.441 \\
14.00 & 53.0 & 1.00 & $1.13^{+0.66}_{-1.13}$ & $1.08^{+0.90}_{-1.08}$ & 0.441 \\
14.50 & 75.0 & 1.00 & $1.63^{+0.36}_{-1.63}$ & $1.64^{+0.61}_{-1.64}$ & 0.441 \\
15.00 & 118.0 & 1.00 & $1.88^{+0.28}_{-1.07}$ & $1.89^{+0.49}_{-1.89}$ & 0.441 \\
15.50 & 129.0 & 1.00 & $2.00^{+0.23}_{-0.54}$ & $2.00^{+0.41}_{-2.00}$ & 0.441 \\
16.00 & 175.0 & 1.00 & $2.11^{+0.22}_{-0.47}$ & $2.11^{+0.34}_{-2.11}$ & 0.441 \\
16.50 & 242.0 & 1.00 & $2.55^{+0.10}_{-0.13}$ & $2.56^{+0.24}_{-0.55}$ & 0.441 \\
17.00 & 309.0 & 1.00 & $2.77^{+0.07}_{-0.08}$ & $2.77^{+0.15}_{-0.23}$ & 0.441 \\
17.50 & 436.0 & 1.00 & $2.91^{+0.06}_{-0.07}$ & $2.91^{+0.09}_{-0.12}$ & 0.441 \\
18.00 & 634.0 & 1.00 & $3.22^{+0.04}_{-0.04}$ & $3.22^{+0.11}_{-0.14}$ & 0.441 \\
18.50 & 832.0 & 1.00 & $3.40^{+0.03}_{-0.03}$ & $3.40^{+0.09}_{-0.11}$ & 0.441 \\
19.00 & 1128.0 & 1.07 & $3.63^{+0.02}_{-0.02}$ & $3.64^{+0.06}_{-0.07}$ & 0.441 \\
19.50 & 1615.0 & 1.08 & $3.80^{+0.02}_{-0.02}$ & $3.80^{+0.05}_{-0.05}$ & 0.441 \\
20.00 & 2303.0 & 1.09 & $3.99^{+0.06}_{-0.08}$ & $3.99^{+0.08}_{-0.10}$ & 0.441 \\
20.50 & 3285.0 & 1.12 & $4.18^{+0.04}_{-0.05}$ & $4.18^{+0.06}_{-0.07}$ & 0.441 \\
21.00 & 4455.0 & 1.17 & $4.34^{+0.03}_{-0.03}$ & $4.34^{+0.05}_{-0.05}$ & 0.441 \\
21.50 & 5030.0 & 1.25 & $4.50^{+0.02}_{-0.03}$ & $4.50^{+0.03}_{-0.03}$ & 0.381 \\
22.00 & 1739.0 & 1.52 & $4.67^{+0.03}_{-0.03}$ & $4.67^{+0.01}_{-0.02}$ & 0.109 \\
\enddata
\tablenotetext{a}{ Raw counts including stars.}
\tablenotetext{b}{ Effective correction, defined as $(Nc+Nst)/(2\cdot area\cdot raw)$.}
\tablenotetext{c}{ Corrected galaxy counts,errors corresponds to the Poissonian and galaxy clustering uncertainty plus error in completeness added in quadrature.}
\tablenotetext{d}{ Mean of the 8 pointings and its rms.}
\end{deluxetable}

\begin{deluxetable}{llllll}
\tabletypesize{\small}
\tablewidth{0pt}
\tablecaption{
Corrected galaxy number counts in the H band.\label{Tab:cH}}
\tablehead{
\colhead{Magnitude}	& \colhead{raw counts}\tablenotemark{a} & \colhead{eff. cor.}\tablenotemark{b} & \colhead{$\log(Nc)$}\tablenotemark{c} & \colhead{$\log(Nm)$}\tablenotemark{d} & \colhead{area} \\ 
& & & $N\cdot mag^{-1}\cdot deg^{-2}$ & $N\cdot mag^{-1}\cdot deg^{-2}$ & $deg^2$\\}
\startdata
12.00 & 22.0 & 1.00 & $0.97^{+0.64}_{-0.97}$ & $0.98^{+0.98}_{-0.98}$ & 0.444 \\
12.50 & 40.0 & 1.00 & $1.62^{+0.30}_{-1.62}$ & $1.60^{+0.47}_{-1.60}$ & 0.444 \\
13.00 & 39.0 & 1.00 & $1.19^{+0.56}_{-1.19}$ & $1.19^{+0.79}_{-1.19}$ & 0.444 \\
13.50 & 53.0 & 1.00 & $0.52^{+1.19}_{-0.52}$ & $0.55^{+1.44}_{-0.55}$ & 0.444 \\
14.00 & 97.0 & 1.00 & $1.85^{+0.28}_{-0.95}$ & $1.86^{+0.49}_{-1.86}$ & 0.444 \\
14.50 & 106.0 & 1.00 & $1.94^{+0.24}_{-0.61}$ & $1.94^{+0.54}_{-1.94}$ & 0.444 \\
15.00 & 131.0 & 1.00 & $1.81^{+0.33}_{-1.81}$ & $1.81^{+0.59}_{-1.81}$ & 0.444 \\
15.50 & 223.0 & 1.00 & $2.39^{+0.14}_{-0.21}$ & $2.39^{+0.25}_{-0.63}$ & 0.444 \\
16.00 & 288.0 & 1.00 & $2.74^{+0.07}_{-0.09}$ & $2.74^{+0.16}_{-0.25}$ & 0.444 \\
16.50 & 363.0 & 1.00 & $2.88^{+0.06}_{-0.07}$ & $2.88^{+0.12}_{-0.17}$ & 0.444 \\
17.00 & 541.0 & 1.00 & $3.10^{+0.04}_{-0.05}$ & $3.10^{+0.06}_{-0.07}$ & 0.444 \\
17.50 & 788.0 & 1.00 & $3.37^{+0.03}_{-0.03}$ & $3.37^{+0.10}_{-0.14}$ & 0.444 \\
18.00 & 1074.0 & 1.00 & $3.56^{+0.02}_{-0.02}$ & $3.57^{+0.06}_{-0.07}$ & 0.444 \\
18.50 & 1512.0 & 1.05 & $3.76^{+0.02}_{-0.02}$ & $3.76^{+0.06}_{-0.07}$ & 0.444 \\
19.00 & 2110.0 & 1.05 & $3.94^{+0.07}_{-0.08}$ & $3.94^{+0.08}_{-0.10}$ & 0.444 \\
19.50 & 3138.0 & 1.07 & $4.14^{+0.04}_{-0.05}$ & $4.14^{+0.05}_{-0.06}$ & 0.444 \\
20.00 & 4456.0 & 1.12 & $4.32^{+0.03}_{-0.03}$ & $4.32^{+0.04}_{-0.05}$ & 0.444 \\
20.50 & 5879.0 & 1.23 & $4.49^{+0.02}_{-0.02}$ & $4.49^{+0.04}_{-0.05}$ & 0.444 \\
21.00 & 2578.0 & 1.48 & $4.65^{+0.02}_{-0.02}$ & $4.65^{+0.02}_{-0.02}$ & 0.165 \\
\enddata
\tablenotetext{a}{ Raw counts including stars.}
\tablenotetext{b}{ Effective correction, defined as $(Nc+Nst)/(2\cdot area\cdot raw)$.}
\tablenotetext{c}{ Corrected galaxy counts,errors corresponds to the Poissonian and galaxy clustering uncertainty plus error in completeness added in quadrature.}
\tablenotetext{d}{ Mean of the 8 pointings and its rms.}
\end{deluxetable}

\begin{deluxetable}{llllll}
\tabletypesize{\small}
\tablewidth{0pt}
\tablecaption{
Corrected galaxy number counts in the Ks band.\label{Tab:cK}}
\tablehead{
\colhead{Magnitude}	& \colhead{raw counts}\tablenotemark{a} & \colhead{eff. cor.}\tablenotemark{b} & \colhead{$\log(Nc)$}\tablenotemark{c} & \colhead{$\log(Nm)$}\tablenotemark{d} & \colhead{area} \\ 
& & & $N\cdot mag^{-1}\cdot deg^{-2}$ & $N\cdot mag^{-1}\cdot deg^{-2}$ & $deg^2$\\}
\startdata
12.00 & 25.0 & 1.00 & $1.14^{+0.54}_{-1.14}$ & $1.13^{+0.74}_{-1.13}$ & 0.440 \\
12.50 & 44.0 & 1.00 & $1.61^{+0.32}_{-1.61}$ & $1.60^{+0.52}_{-1.60}$ & 0.440 \\
13.00 & 41.0 & 1.00 & $...$ & $...$ & 0.440 \\
13.50 & 78.0 & 1.00 & $1.89^{+0.24}_{-0.55}$ & $1.90^{+0.32}_{-1.90}$ & 0.440 \\
14.00 & 101.0 & 1.00 & $2.03^{+0.20}_{-0.39}$ & $2.03^{+0.47}_{-2.03}$ & 0.440 \\
14.50 & 141.0 & 1.00 & $2.21^{+0.16}_{-0.26}$ & $2.22^{+0.32}_{-2.22}$ & 0.440 \\
15.00 & 188.0 & 1.00 & $2.41^{+0.12}_{-0.17}$ & $2.41^{+0.26}_{-0.73}$ & 0.440 \\
15.50 & 291.0 & 1.00 & $2.74^{+0.07}_{-0.09}$ & $2.74^{+0.14}_{-0.21}$ & 0.440 \\
16.00 & 375.0 & 1.00 & $2.96^{+0.05}_{-0.06}$ & $2.96^{+0.11}_{-0.15}$ & 0.440 \\
16.50 & 585.0 & 1.00 & $3.22^{+0.03}_{-0.04}$ & $3.22^{+0.08}_{-0.09}$ & 0.440 \\
17.00 & 940.0 & 1.00 & $3.49^{+0.02}_{-0.02}$ & $3.49^{+0.06}_{-0.07}$ & 0.440 \\
17.50 & 1307.0 & 1.03 & $3.67^{+0.02}_{-0.02}$ & $3.67^{+0.05}_{-0.06}$ & 0.440 \\
18.00 & 1836.0 & 1.04 & $3.86^{+0.01}_{-0.02}$ & $3.86^{+0.06}_{-0.07}$ & 0.440 \\
18.50 & 2614.0 & 1.05 & $4.04^{+0.05}_{-0.06}$ & $4.04^{+0.07}_{-0.08}$ & 0.440 \\
19.00 & 3698.0 & 1.11 & $4.24^{+0.04}_{-0.04}$ & $4.24^{+0.06}_{-0.07}$ & 0.440 \\
19.50 & 4477.0 & 1.37 & $4.42^{+0.03}_{-0.03}$ & $4.42^{+0.04}_{-0.04}$ & 0.440 \\
20.00 & 574.0 & 1.55 & $4.50^{+0.04}_{-0.05}$ & $4.50^{+0.02}_{-0.02}$ & 0.054 \\
\enddata
\tablenotetext{a}{ Raw counts including stars.}
\tablenotetext{b}{ Effective correction, defined as $(Nc+Nst)/(2\cdot area\cdot raw)$.}
\tablenotetext{c}{ Corrected galaxy counts,errors corresponds to the Poissonian and galaxy clustering uncertainty plus error in completeness added in quadrature.}
\tablenotetext{d}{ Mean of the 8 pointings and its rms.}
\end{deluxetable}

\begin{figure}
\epsscale{1.0}
\plotone{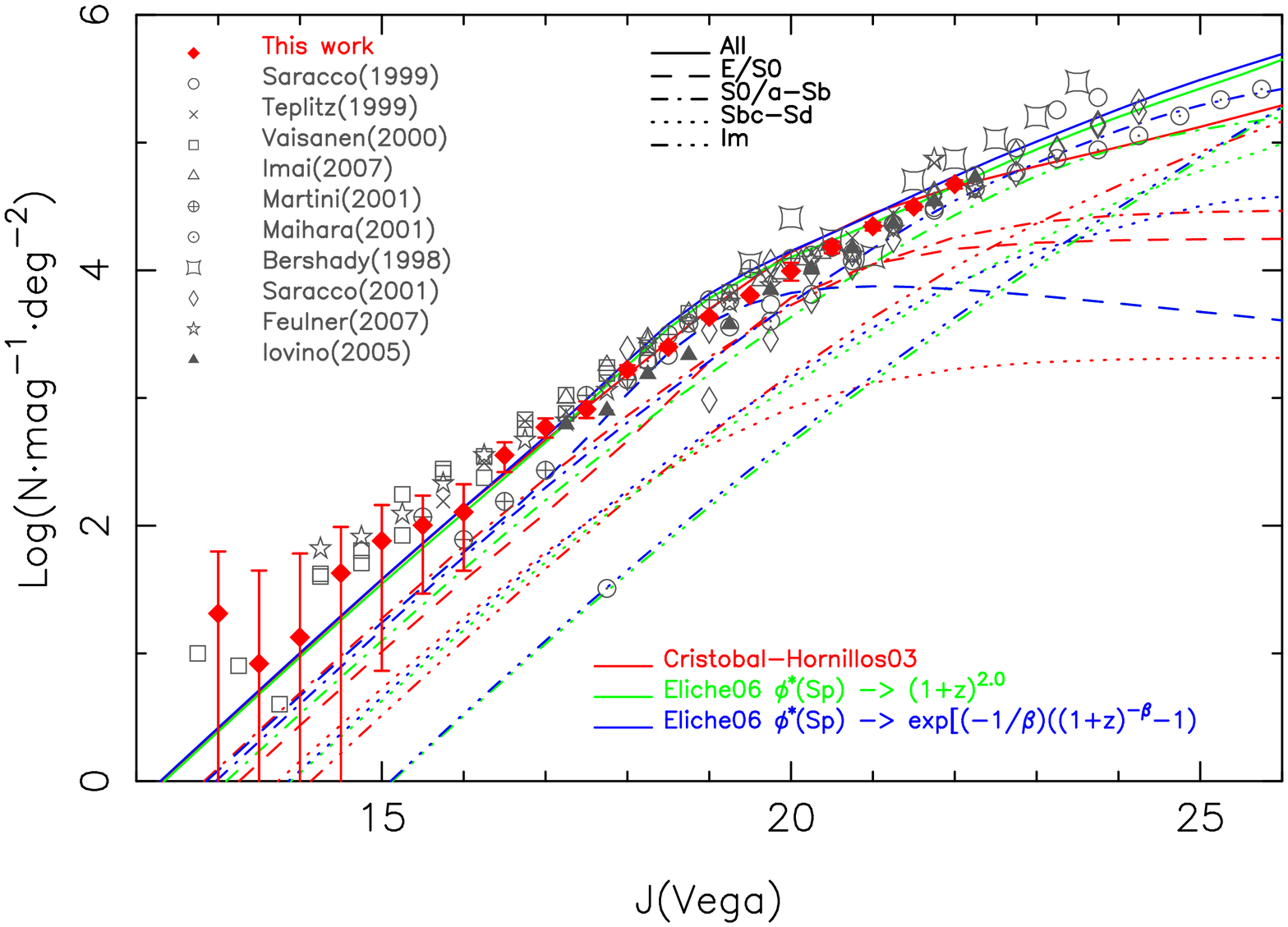}
\caption{\label{Fig:countsJ} Galaxy number counts in the J filter compared 
with data from other surveys. The lines correspond to the number
counts models in \cite{2003ApJ...595...71C} and
\cite{2006ApJ...639..644E} described in the text.}
\end{figure}

\begin{figure}
\epsscale{1.0}
\plotone{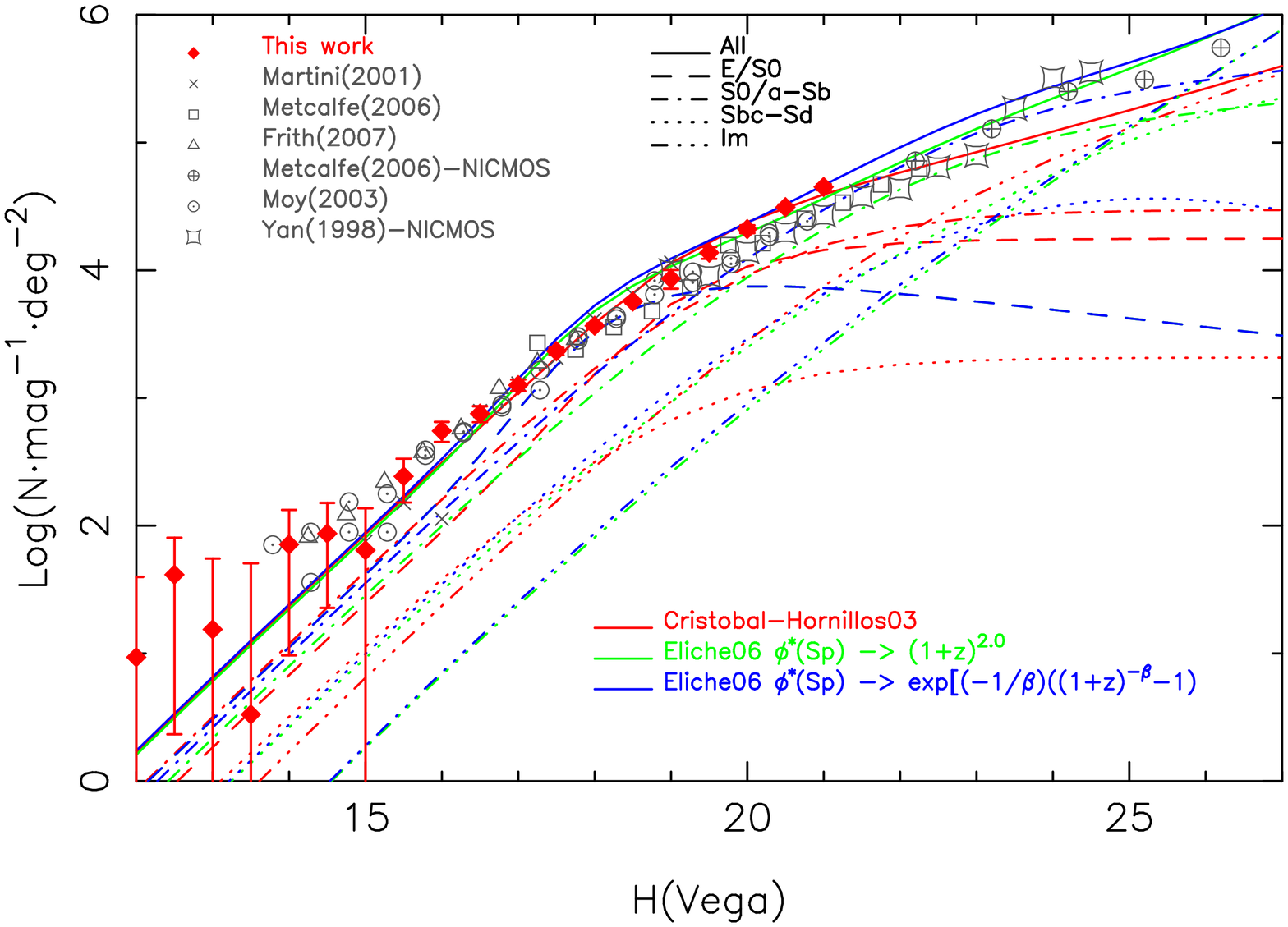}
\caption{\label{Fig:countsH} Galaxy number counts in the H filter compared 
with data from other surveys and the models in
\cite{2003ApJ...595...71C} and \cite{2006ApJ...639..644E} described in
the text.}
\end{figure}

\begin{figure}
\epsscale{1.0}
\plotone{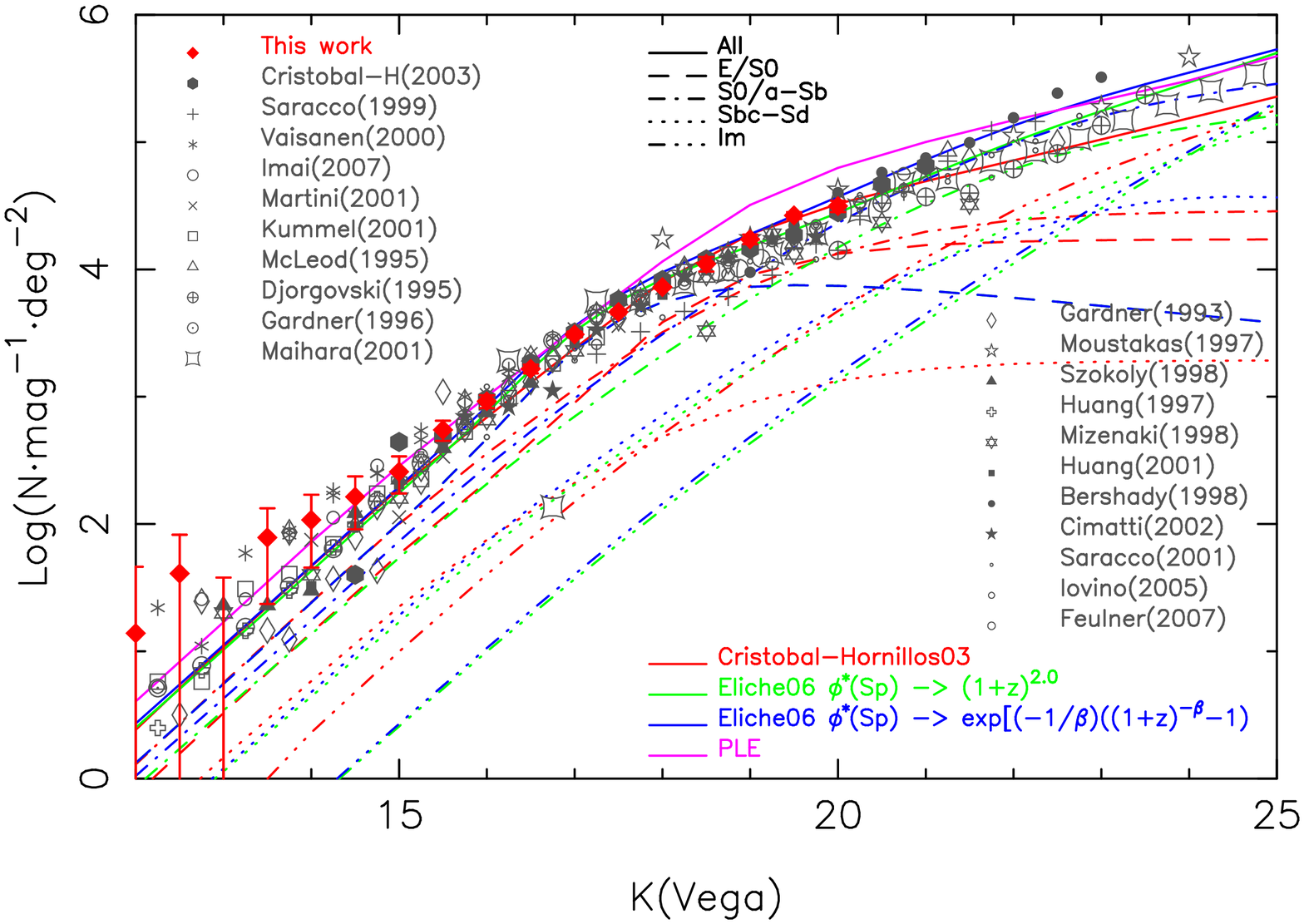}
\caption{\label{Fig:countsKS} Galaxy number counts in the Ks filter compared with 
data from the literature and the galaxy counts models in
\cite{2003ApJ...595...71C} \cite{2006ApJ...639..644E} described in the
text. Also it is shown a model where only the passive evolution of
stellar populations is considered (PLE).}
\end{figure}

In the Figs.~\ref{Fig:countsJ}, \ref{Fig:countsH}, and
\ref{Fig:countsKS} the Alhambra counts in the J,H, and Ks bands
are plotted together with the computed number counts
from other surveys. 

Let us first point out the general aspect of those
counts, leaving more detailed considerations for the next section.
Our corrected J band galaxy counts are in overall
agreement with those computed in other works 
\citep{1999ApJ...520..469T,2007MNRAS.378..429F}, the bright end of the
results presented by \cite{2001PASJ...53...25M}, and the ELAIS-N2 data
from \cite{2000ApJ...540..593V}. Nevertheless, our counts are lower
than those given by \cite{1998ApJ...505...50B} at magnitudes $J>21.5$.

The Alhambra counts in the H band are in good correspondence with the
published data by \cite{2001AJ....121..598M, 2006MNRAS.371.1601F} and
with the bright part of the data from \cite{2003A&A...403..493M}, after applying an
offset of -0.215 mag. This offset was calculated following the -0.065
calibration difference among Las Campanas Infrared Survey (LCIRS,
\cite{2002ApJ...570...54C}) reported in \cite{2003A&A...403..493M},
and the systematic of -0.28 magnitude difference between LCIRS and
2MASS magnitudes reported in \cite{2006MNRAS.371.1601F}. However, the
faint end ($H>20$) of our data is significantly above the faint number
count data from \cite{1998ApJ...503L..19Y} and
\cite{2006MNRAS.370.1257M}, obtained from NICMOS observations.

Regarding the Ks filter, for which there are numerous number counts
studies, our galaxy counts are in good agreement with most of the
published data, as can be appreciated in the Fig.~\ref{Fig:countsKS}.

\subsection{The measured slopes}

\begin{figure}
\epsscale{0.5}
\plotone{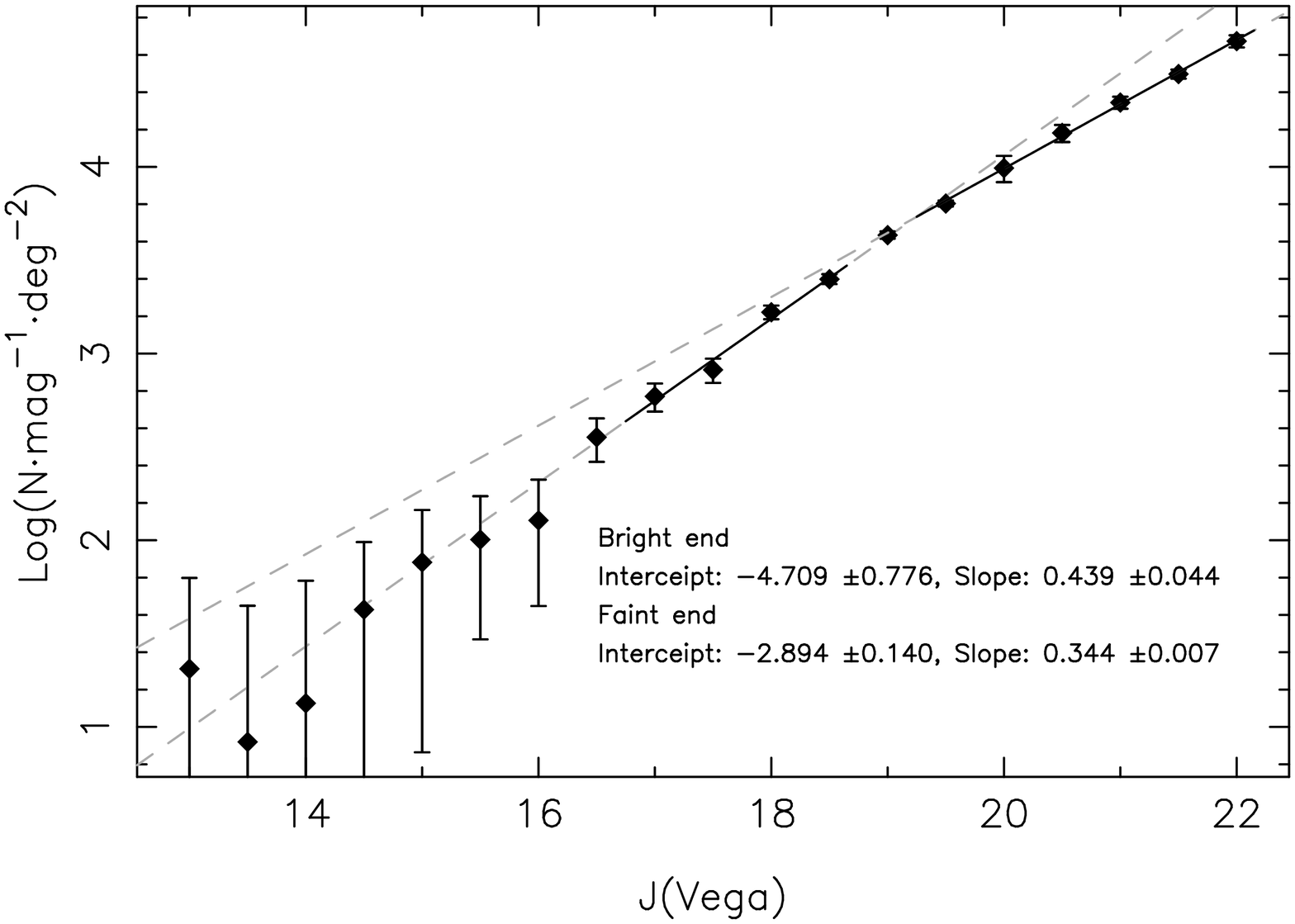}
\vspace{0.5cm}

\plotone{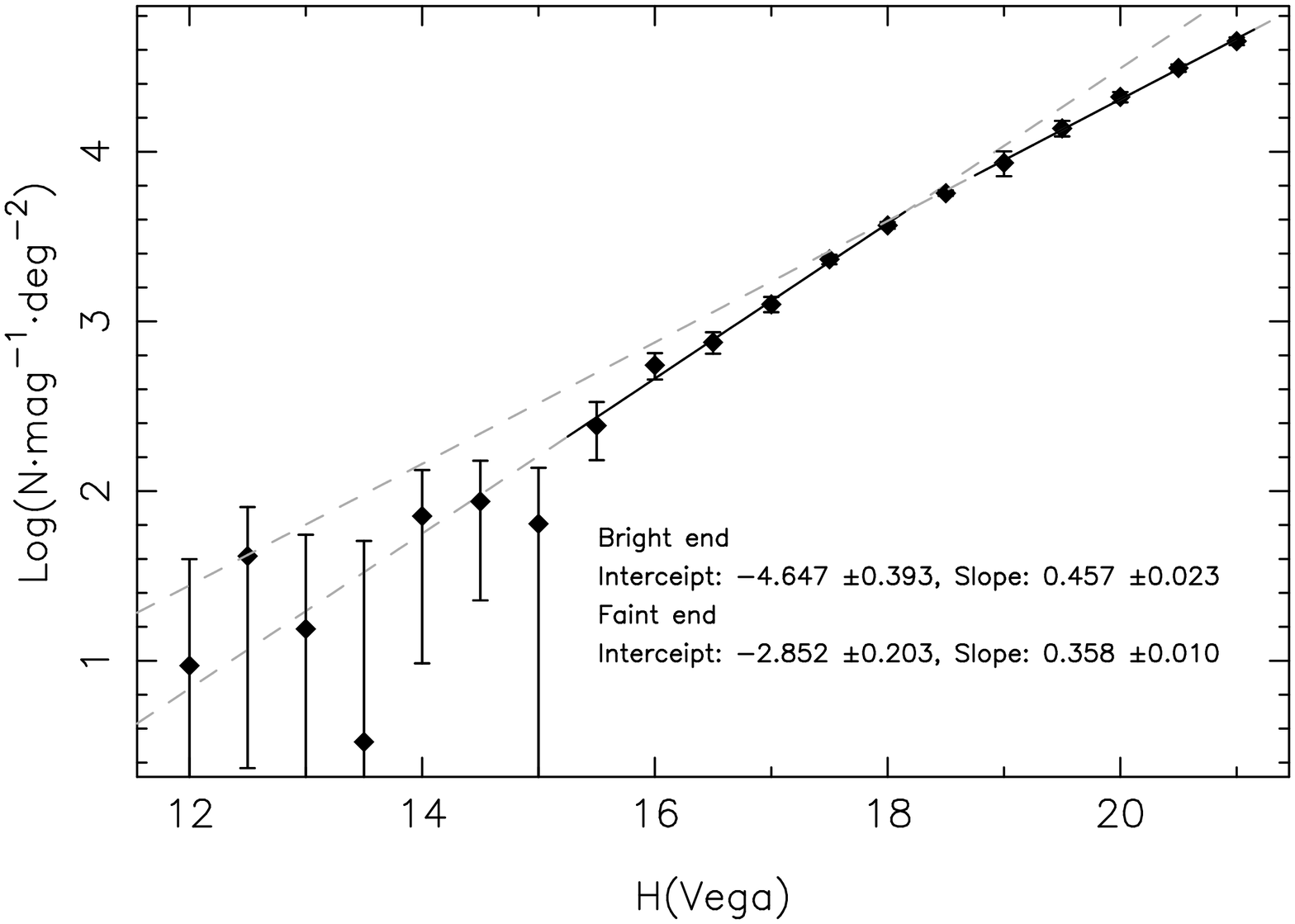}
\vspace{0.5cm}

\plotone{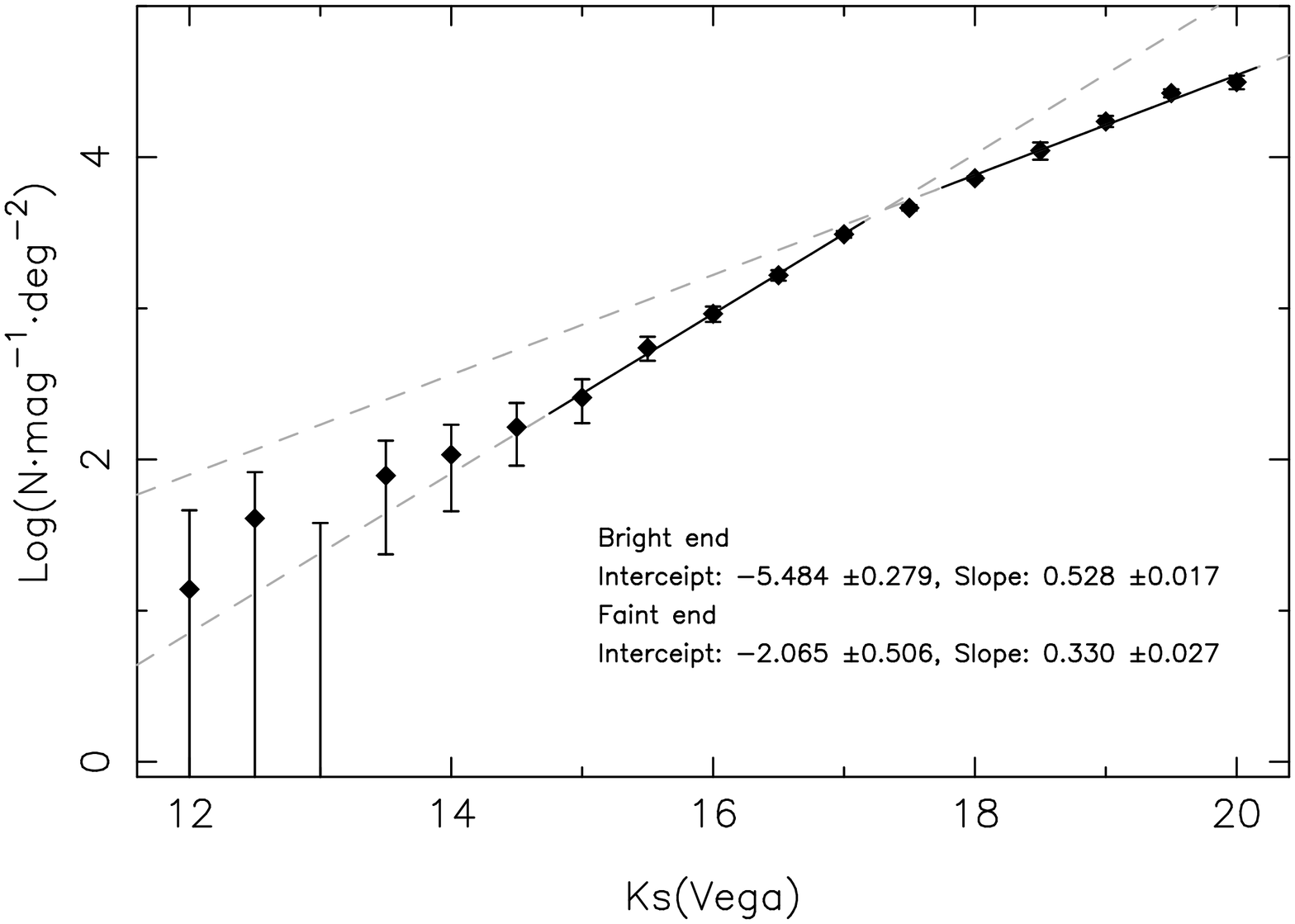}
\caption{\label{Fig:slopes1} Galaxy number count bright and faint slopes found in 
this work. Error bars are the Poissonian and galaxy clustering uncertainty added in quadrature with
the rms in the estimation of the completeness corrections.}
\end{figure}

We have found, as in previous studies, that the slope of the galaxy
number counts displays a clear change at Ks$\sim$17.3 (see
Fig.~\ref{Fig:slopes1}). However, in the present work we also have
found this change of slope in the J and H band galaxy counts at
J$\sim$19.1 and H$\sim$ 18.2.

The slopes of the bright end and faint end counts in the
different bands were measured using the least squares method. The
values that we have calculated and the magnitude ranges used for the
fits are listed in Tab.~\ref{Tab:slopes}. The results show that the
faint slopes for the three NIR filters are the same within $2\sigma$,
whereas the bright slope is steeper in the Ks band.

As is described in \cite{2004A&A...424...73T} the fact that the photometric error increases at fainter magnitudes, and that the 
differential galaxy number counts also rises towards lower fluxes, lead to a stepper observed slope at the faint end. 
This effect is related with the Eddington bias \citep{1913MNRAS..74....5E}.
To investigate if the computed magnitude errors could bias our slope estimates we have done the following study.
First, in each filter we use the exponential grow function fit to the magnitude errors (Fig.~\ref{Fig:MagError}).
Then we extended our computed number counts one magnitude fainter using the corresponding slope in Tab.~\ref{Tab:slopes},
simulating a Gaussian decay for fainter magnitudes. The parameter $\sigma$ for the Gaussian is the value of the exponential
fit to the error values for point-like sources at one magnitude fainter that the last bin given in 
Tabs.~\ref{Tab:cJ}, \ref{Tab:cH} and \ref{Tab:cK} ($\sigma=0.35,0.41$ and 0.44  at respectively J=23.0, H=22.0 and Ks=21.0).  
Using the fitted magnitude errors for point-like objects ($\sigma(m)$), and assuming that real distribution is close to
the extended number counts ($N(m)$), we simulate the bias produced by the photometric error on the observed counts, 
by convolving the $N(m)$ with $\sigma(m)$ using the eq.~\ref{Eq:convol}. The results indicate an increase
in the slope less than 0.015 in the case of J and H filters, and 0.03 in the Ks band, meaning that the original distributions
would have a faint end slope of 0.33,0.34 and 0.30 in the J,H, and Ks filters in the ranges given in Tab.~\ref{Tab:slopes}.

\begin{equation}\label{Eq:convol}
N(m_{obs})=\int N(m)\frac{1}{\sqrt{2\pi \sigma}} e^{\frac{(m-m_{obs})^2}{2\sigma^2(m)}}dm 
\end{equation}

The increase of the slope is not observed when we simulate this
bias directly over the observed count distribution.
This is due to  the fact that fainter than the 80\% completeness magnitude bin there is a fast decrease of the number
of detected sources, and that in this bin the typical photometric error ($\sigma_m=0.23-0.26$) for the dominant
point-like sources, makes that the bias do not significantly affect to the previous bins.
In this case the slopes after applying the bias are lower than the computed
from the observed number counts ($0.33 \pm 0.01$, $0.33 \pm 0.02$, and $0.30 \pm 0.02$ in J,H and Ks).
This result would suggest that the original distribution slopes at the faint end would be higher than
the ones given in Tab.~\ref{Tab:slopes}. We have used the combined results from this paragraph and the previous
one to increase the uncertainty in the slope, leaving the values computed directly from the observed number counts as
satisfactory estimates of the real distribution slope at the faint end.

\begin{deluxetable}{lllll}
\tabletypesize{\small}
\tablewidth{0pt}
\tablecaption{
Measured slopes in the J,H and Ks filters\label{Tab:slopes}}
\tablehead{
\colhead{Filter}& \colhead{Bright range} & \colhead{Bright slope} & \colhead{Faint range} & \colhead{Faint slope}}
\startdata
J &  [17.0,18.5] &  0.44$\pm$0.04 & [19.5,22.0] & 0.34$\pm$0.01 \\
H &  [15.5,18.0] &  0.46$\pm$0.02 & [19.0,21.0] & 0.36$\pm$0.02 \\
Ks & [15.0,17.0] &  0.53$\pm$0.02 & [18.0,20.0] & 0.33$\pm$0.03 \\  

\enddata
\end{deluxetable}

Our results for the Ks band are in good agreement with other 
K-band surveys which also report a similar change of slope in the galaxy counts
in the range 17.0-18.0 \citep{2000A&A...361..535D,2001A&A...368..787H,
2003ApJ...595...71C,2005A&A...442..423I}. At the bright part our
slopes are close to the values measured in
\cite{2000A&A...353..867K,2001AJ....121..598M,2003ApJ...595...71C,2005A&A...442..423I}
(see Tab.~\ref{Tab:surveysK}), while other references point to a
steeper bright slope
\citep{1993ApJ...415L...9G,1996MNRAS.282L...1G,1997ApJ...476...12H,2001A&A...368..787H}. This
could be due to the fact that the last authors could extend the
power-law fit to brighter magnitudes due to their larger surveyed
area, whereas our fit is closer to the magnitude where the break is
found which would lead to a decrease in the slope if the break
transition is smooth.  In the faint part of the Ks counts a
slope of $0.33$ found in this work is in agreement with
\cite{1998ApJ...505...50B,2001A&A...368..787H,2007AJ....133.2418I}. The
value of the faint slope from the ALHAMBRA data is however steeper
than the value reported in some surveys covering smaller areas, where
the fitted range extends to fainter magnitudes,
\cite{1993ApJ...415L...9G,1997ApJ...475..445M,2001PASJ...53...25M}. Here
the differences might be due to cosmic variance. A larger than
$2\sigma$ disagreement is found with the faint slope of
\cite{2003ApJ...595...71C} fitted in the range [17.5,19.5], although
their value increase to 0.29 when the fit interval is extended to
Ks=21.0. \cite{2000A&A...353..867K} give a higher value for the faint
slope, however due to the brighter limiting magnitude of their number
counts they could established a break or the beginning of a roll-over
in the interval Ks=[16.5,17.0].

\begin{deluxetable}{llllllll}
\tabletypesize{\scriptsize}
\tableheadfrac{0.01}
\tablewidth{0pt}
\tablecaption{
Characteristics of the surveys in the K band \label{Tab:surveysK}}
\tablehead{
\colhead{Reference}	& \colhead{Surveyed area} & \colhead{limit magnitude} & \colhead{bright range} & \colhead{bright slope} & \colhead{faint range} & \colhead{faint slope} &\colhead{filter}\\ 
  &	 \colhead{sqarcmin}	 \\
}
\startdata
Gardner93	&5688	 &	14.5\tablenotemark{c}   	 &	 [10.0,16.0]  	& 0.67		   		& [18.0,23.0]	&	0.23	   	 	&   K'  	\\
Gardner93	&582	 &   	16.75\tablenotemark{c} &   ---	   		& 	---		   		& 	---			&	---			 	&    ---    \\
Gardner93   &167.7	 &   18.75\tablenotemark{c}  	 &   ---	   		& 	---		   		& 	---			&	---			 	&    ---    \\
Gardner93   &16.5    &	22.5    	 &   ---	   		& 	---		   		& 	---			&	---			 	&    ---    \\
Glazebrook94 & 552   &  16.5\tablenotemark{c} &  ---	   		& 	---		   		& 	---			&	---			 	&    K \\
Djorgovski95 & 3     &  23.5\tablenotemark{c} & ---    & --- &  [20.0,23.5]  & 0.32$\pm$0.02 & K \\
McLeod95 & 22.5,2.0 & 19.5,21.25\tablenotemark{c} &  ---	   		& 	---		   		& 	---			&	---			 	&    Ks \\
Gardner96   &35424   &	15.75\tablenotemark{c}  	 &   $<$16.0	   	& 0.63$\pm$0.01    	&  ---        	& 	---   		 	&    K	    \\
Moustakas97 &2.0     	 &   24.0\tablenotemark{c}		 &    ---       		& 	---         		& [18.0,23.0]  	&	0.23$\pm$0.02   &    K      \\
Huang97     &29628   & 	16.0\tablenotemark{c}	 & [12.0,16.0]    		&  0.689$\pm$0.013 	&	---	    & 	---	         	&    K'     \\
Minezaki98  &181,2.21&    19.1,21.2\tablenotemark{d}  	 &   ---	   		& ---			   	& [18.25,18-75]	&	0.28		 	&    K'     \\
Bershady98  &1.5	 &   	24.00\tablenotemark{b} &   	---	   		& ---			   	& $>$18.5		&	0.36		 	&    K      \\
Szokoly98   & 2185   &   16.5\tablenotemark{c} &  [14.5,16.5] & 0.50$\pm0.03$ & ---& ---& Ks \\
Saracco99	&20		 &   22.25\tablenotemark{c}	 &   ---		   		& ---			  	& [17.25,22.5]	&	0.38		 	&    Ks     \\
V{\"a}is{\"a}nen00	&3492,2088	 &   16.75,17.75\tablenotemark{c}    	 &   [15.0,18.0]   	& 0.40-0.45  	&  ---        	& 	---			 	&    K      \\
Martini01	&180,51	 &       17.0,18.0\tablenotemark{c} &   [14.0,18.0] 	& 0.54		   		&  ---			&	---			 	&    K	    \\
Daddi00 &  701,447 & 18.5,19.0\tablenotemark{c} &  [14.0,17.5] & 0.53$\pm0.02$ & $>$17.5 & 0.32$\pm0.02$ & Ks \\
K{\"u}mmel00	&3348 	 &   17.25\tablenotemark{c}	 &   [10.5,17.0] 	&	0.56$\pm$0.01 	& [16.5,17.5] 	&	0.41  		 	&    K	    \\
Maihara01   &4		 &   25.25\tablenotemark{c}	 &    ---	   		& 	---         		& $>$20.1	    &	0.23	  	 	&    K'     \\
Saracco01   &13.6   &    22.75\tablenotemark{c}	 &    ---      		&   ---		   		& $>$19			&	0.28		 	&    Ks     \\
Huang01		&720	 &   	19.5\tablenotemark{c}    &   $<$16.5	   	& 0.64		   		& $>$17			&	0.36		 	&    K'     \\
Cimatti02	& 52	& 19.75\tablenotemark{c} &   	---   		& 	---	   		& 	---		&	---		 	&    Ks    \\
Cristobal03 &180,50	 &      20.0,21.0\tablenotemark{b}    &   [15.5,17.5] 	&  0.54		   		& [17.5,19.5]  	&	0.25  		 	&    Ks	    \\
Iovino05    & 414    &      20.75\tablenotemark{c} &   [15.75,18.0]&     0.47$\pm$0.23     & [18.0,21.25]  &   0.29$\pm$0.08   &    Ks       \\
Elston06 & 25560 & 19.2\tablenotemark{b} &  ---	   		& 	---		   		& 	---			&	---			 	&    Ks \\
Imai07		&750,306	 &   	18.625,19.375\tablenotemark{c} &   $<$18.00	   	& 0.32$\pm$0.06	   	& $>$19.00	  	&	0.32$\pm$0.06   & 	 Ks	    \\
Feulner07	&925	 &   	20.75\tablenotemark{c} &   	---	   		& ---	 			   	&	---	 		&	---	 		 	&    K      \\
This work   & 1584,194  &19.5,20.0\tablenotemark{c} & [15.0,17.0] &  0.53$\pm$0.02 & [18.0,20.0] & 0.33$\pm$0.03 & Ks\\ 

\enddata

\tablenotetext{a}{ 3$\sigma$ limit for point sources}
\tablenotetext{b}{ 50\% efficiency for point objects}
\tablenotetext{c}{ the latest magnitude bin in the number counts}
\tablenotetext{d}{ 80\% efficiency for point objects}
\end{deluxetable}

\begin{deluxetable}{llllllll}
\tabletypesize{\scriptsize}
\tableheadfrac{0.01}
\tablewidth{0pt}
\tablecaption{
Characteristics of the surveys in the J band \label{Tab:surveysJ}}
\tablehead{
\colhead{Reference}	& \colhead{Surveyed area} & \colhead{limit magnitude} & \colhead{bright range} & \colhead{bright slope} & \colhead{faint range} & \colhead{faint slope} &\colhead{filter}\\ 
  &	 \colhead{sqarcmin}	 \\
}
\startdata

Bershady98 & 1.5	&	24.5\tablenotemark{b}  &    ---		& ---			& 	$>$19.5	 &	0.35		&  J	\\
Teplitz99 & 180	&	21.75\tablenotemark{c} &		---		& ---			&	---	 &   --- &  J    \\
Saracco99  & 20     &   23.75\tablenotemark{c} &   	---		& ---			&	[18.0,24.0]&	0.36	    &  J    \\
V{\"a}is{\"a}nen00 & 2520,1275	&   18.25,19.25\tablenotemark{c} &  [17.0,19.5]	& 0.40-0.45	& 	---		 &   ---			&   J   \\
Martini01  & 180,27    &   18.5,20.5\tablenotemark{c}  &  [16.0,20.5]  &   0.54		&	---		 &   ---		    &   J   \\
Maihara01  &	4	&	26.25\tablenotemark{c} &		---		& 	---		&	[21.1,25.1]&	0.23		&   J   \\
Saracco01 & 13.6 &   24.25\tablenotemark{c} &  ---			& ---			&   $>$20	 &       0.34	&   J   \\
Iovino05   & 391    &   22.25\tablenotemark{c} &  [17.25,22.25] & 0.39$\pm0.06$ & ---       & --- & J \\ 
Feulner07  & 925   	&   22.25\tablenotemark{c} &   ---	     & ---	 			&---	 	 &   ---	 		&   J   \\
Imai07	   & 	750,306	 &   	19.625,20.375\tablenotemark{c} &  [17.0,19.5]	& 0.39$\pm0.02$	&	$>$19.5	 &   0.30$\pm0.03$		&   J   \\
This work   & 1588,392  &21.0,22.0\tablenotemark{c} & [17.0,18.5] &  0.44$\pm$0.04 & [19.5,22.0] & 0.34$\pm$0.01 & J\\

\enddata
\tablenotetext{a}{ 3$\sigma$ limit for point sources}
\tablenotetext{b}{ 50\% efficiency for point objects}
\tablenotetext{c}{ the latest magnitude bin in the number counts}
\end{deluxetable}

\begin{deluxetable}{llllllll}
\tabletypesize{\scriptsize}
\tablewidth{0pt}
\tablecaption{
Characteristics of the surveys in the H band \label{Tab:surveysH}}
\tablehead{
\colhead{Reference}	& \colhead{Surveyed area} & \colhead{limit magnitude} & \colhead{bright range} & \colhead{bright slope} & \colhead{faint range} & \colhead{faint slope} &\colhead{filter}\\ 
  &	 \colhead{sqarcmin}	 \\
}
\startdata

Teplitz98   	  &	35.4        &   22.8\tablenotemark{a}			& ---	&		---		&		---	&	---		    		&    F160W 		\\
Yan98		 	  &  8.7,2.9	   &   23.5,24.5\tablenotemark{c}		&  ---   &      ---  		& [20,24.5]	&0.315$\pm$0.02         &    F160W   \\
Thompson99 	  &  0.7	   &   27.4\tablenotemark{b}		& ---	&		---		&		---	&	---		            &    F160W       \\
Martini01    	  &  180,80       &   18.0,19.0\tablenotemark{c}		& $<$19	&		0.47	&	---	&	---		            &    H		\\
Chen02	 		  &  1408	   &   20.8\tablenotemark{d} 		& $<$19 	&	0.45$\pm$0.01 	&	$>$19 	&0.27$\pm$0.01  &	H       \\
Moy03			  &  97.2,619  &   20.5,19.8\tablenotemark{b}	& ---	&		---		&		---	&	---			        &	H	    \\
Metcalfe06  	  &  49        &   22.9\tablenotemark{a}		& ---	&		---		&		---	&	---		            &    H		\\
Metcalfe06-Nicmos &  0.90      &   27.2\tablenotemark{c}		& ---	&		---		&		---	&	---		            &    F160W   \\
Frith06		  &  1080	   &   17.75\tablenotemark{c}		&	---	&		---		&	---		&		---	            &	H       \\
This work   & 1598,594  &20.5,21.0\tablenotemark{c} & [15.5,18.0] &  0.46$\pm$0.02 & [19.0,21.0] & 0.36$\pm$0.01 & H\\
\enddata
\tablenotetext{a}{ 3$\sigma$ limit for point sources}
\tablenotetext{b}{ 50\% efficiency for point objects}
\tablenotetext{c}{ the latest magnitude bin in the number counts}
\tablenotetext{d}{ S/N=5 in a 4$^{\prime\prime}$ diameter aperture}
\end{deluxetable}

In the H and J bands there are fewer works reporting count-slope values. Our
results, given in Tab.~\ref{Tab:slopes}, show similar slopes for the J
and H filter at the bright and faint ends. As can be seen in
Tabs.~\ref{Tab:surveysJ} and \ref{Tab:surveysH} our result at the
bright end are in good agreement with the bright-end slope values to H=19
given in \cite{2001AJ....121..598M} and \cite{2002ApJ...570...54C},
and in the J filter with the results in \cite{2000ApJ...540..593V}
and \cite{2005A&A...442..423I}. At the faint end, only the slope
values in \cite{2001PASJ...53...25M} in the J filter, estimated in an
area of 4 arcsec$^2$, and \cite{2002ApJ...570...54C} have a
significant discrepancy.
 
\section{Comparison with Models of Evolution}

Historically, the galaxy number counts have been used to examine
parameters of the cosmological model and to test different galaxy
evolution scenarios. Now that the cosmological parameters are fixed
using other methods the consequences derived from the galaxy counts
for the galaxy evolution have become more precise. The ALHAMBRA
computed counts in the three NIR bands provide a good dataset which, when
combined with other optical data and independent determinations of the
local luminosity functions, allow evolution to be examined, in
particular the still uncertain question of the formation and
evolutionary history of early type galaxies.

In this section we compare our counts with semi-analytic predictions
from number-count models, following the recipes given in
\cite{1998PASP..110..291G}, which trace back the redshift evolution of the
galaxy Spectral Energy Distribution (SED) of different galaxy
classes. The SEDs have been computed using the codes of
\cite{2003MNRAS.344.1000B}. We apply dust attenuation following
\cite{2006ApJ...639..644E}, $\tau_B=0.6$ that corresponds to
$\tau_V=0.4$ if the attenuation follows a $\propto \lambda^{-2}$ power
law. We apply dust extinction either directly and by 
the same amount to all the galaxies using
\cite{2003MNRAS.344.1000B} codes, following the prescription given in
\cite{2000ApJ...539..718C}, or by using the luminosity dependent
extinction law proposed in \cite{1991ApJ...383L..37W}. The parameters
we use to characterize four different galaxy types are
given in Tab.~\ref{Tab:galparams}. In the LF parameterization for the
different bands, $M^{\star}$ is changed according to the rest-frame
colors of the evolved SED (from $z_f$ to $z=0$), whereas $\alpha$ and
$\phi^{\star}$ are assumed to be the same in all filters.

In the first step we compare the ALHAMBRA NIR counts with the
prediction obtained using the model proposed in
\cite{2003ApJ...595...71C}. The
extinction correction was applied directly to the SEDs, as an entry
parameter in the code described in
\cite{2003MNRAS.344.1000B} using $\tau_V=1.2$ for stars younger than
$10^7$yr , and $\mu=0.3$ as the fraction of it coming from an
ambient contribution which affects the old stars too. The parameters
used in the local luminosity function are $M^{\star}=-24.07$,
$\alpha=-1.00$, $\phi^{\star}=4.94\times 10^{-3}$ calculated in
\cite{1997ApJ...480L..99G} (\cite{2001MNRAS.326..255C} provide the
parameters for the $\Lambda$-cosmology), which were transformed to
take into account the presence of different galactic types in the
local LF adopting the galaxy mixing E/S0=28\%, Sab/Sbc=47\%, Scd=13\%.

The model also adds a dwarf star-forming population, characterized by
an stellar population of age 1Gyr at all redshifts, and a steeper
slope LF ($M^{\star}=-23.12$, $\alpha=-1.5$,
$\phi^{\star}=0.96\times10^{-3}$) given in
\cite{1998PASP..110..291G}. The formation redshifts are $z_f=2.0$ for
the E/S0 and intermediate-type disk galaxies and $z_f=1.0$ for the
Scd, although the formation redshift for the Scd could be $z_f=4.0$ or
higher without modifying the total counts in NIR, that at magnitudes
fainter than Ks=21.5 are dominated by the dwarf star-forming
population. Due to the disappearance of the red-galaxy
population at $z>2.0$ this model reproduces the change of slope in the Ks band
observed in the present data (see Fig.~\ref{Fig:countsKS}).

\begin{figure}
\epsscale{1.0}
\plotone{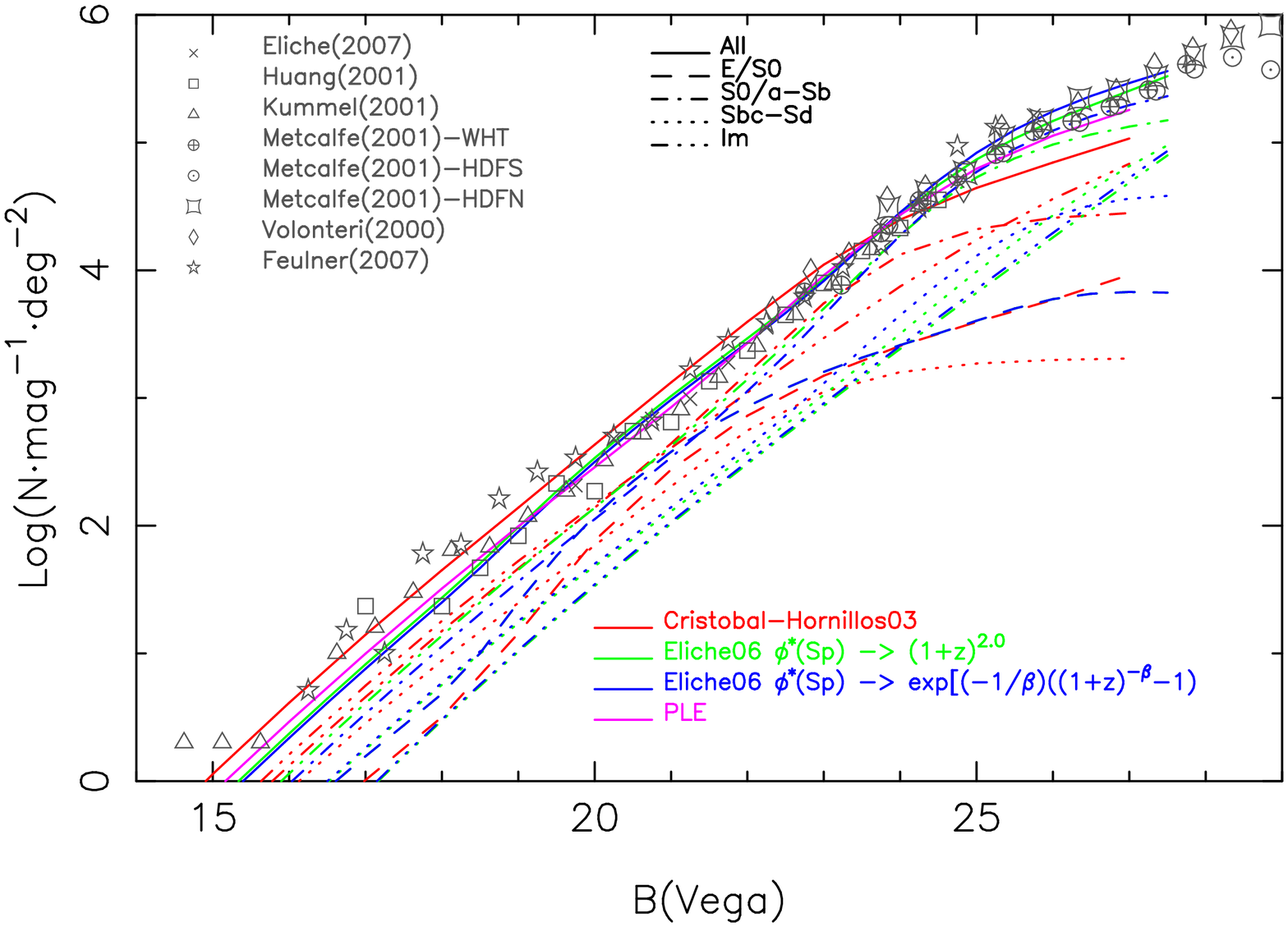}
\caption{\label{Fig:countsB} Galaxy number count in B band taken from the literature. The lines correspond to the number counts predictions from the models in \cite{2003ApJ...595...71C} and \cite{2006ApJ...639..644E}. Also it is shown a model where only the passive evolution of stellar populations is considered (PLE).} 
\end{figure}

However, as it was discussed in \cite{2006ApJ...639..644E}, this model fails
to simultaneously reproduce the blue band counts as can be seen in
Fig.~\ref{Fig:countsB}, where the predicted counts are compared
with some B-band galaxy counts from the literature.

The number counts models that we consider now, below, include some
modifications which aim to simultaneously reproduce the counts in both
the NIR filters and blue filters. For these, the local type
dependent luminosity functions were computed from Sloan data in
\cite{2003AJ....125.1682N}, as shown in Tab.~\ref{Tab:lumfunc}.

\begin{deluxetable}{lcccc}
\tabletypesize{\small}
\tablewidth{0pt}
\tableheadfrac{0.01}
\tablecaption{
Parameters for the SEDs\label{Tab:galparams}}
\tablehead{
\colhead{Galaxy type}& \colhead{functional form} & \colhead{$\tau$} & \colhead{$Z/Z_{\odot}$} & \colhead{IMF}}
\startdata
E/S0 & Single star pop. & --- & 1 & Salpeter\\
early Sp & Exponential &4& 1 & Salpeter\\
late Sp & Exponential & 7& $2/5$ & Salpeter \\
Im & Constant & .. & $1/5$ & Salpeter \\
\enddata
\end{deluxetable}

\begin{deluxetable}{lccc}
\tabletypesize{\small}
\tableheadfrac{0.01}
\tablewidth{0pt}
\tablecaption{
Schechter Parameters for the Luminosity Functions \tablenotemark{a}\label{Tab:lumfunc}}
\tablehead{
\colhead{Galaxy type}& \colhead{$M^{\star}_{AB}(r')$}\tablenotemark{b} & \colhead{$\phi^{\star} \times 10^{-3} Mpc^{-3}$} & \colhead{$\alpha$}}
\startdata
E/S0 & -21.53 		&  1.61& -0.83\\
early Sp & -21.08 	&  3.26& -1.15\\
late Sp & -21.08 	&  1.48& -0.71 \\
Im & -20.78 		&  0.37& -1.90 \\
\enddata
\tablenotetext{a}{ Considering $H_0=70$}
\tablenotetext{b}{ The characteristic galaxy luminosity given in the Sloan $r'$ band in AB system.}
\end{deluxetable}

We consider first the two models proposed in
\cite{2006ApJ...639..644E}. In the first model a $\phi^{\star}$ evolution
$\propto (1+z)^2$, driven via mergers, is considered for the spiral
and irregular galaxies. This evolution in $\phi^{\star}$ is
compensated by the evolution in $M^{\star}$ to conserve the luminosity
density. Is it important to take in mind that these models calculate
the galaxy number counts tracing back the evolution of the stellar
populations to $z=0$, so the intrinsic brightening with $z$ of the
stellar populations must be added to the $M^{\star}$ evolution when is
compared with luminosity functions computed at higher redshifts. The
formation redshift for the majority of the ellipticals and intermediate type
disk galaxies in this model was set to 1.5. In the second model, the low formation redshift for
the early spiral galaxies could be set to $z_f=4$, avoiding at the
same time an unreasonable high number of late type galaxies at high-z, 
using the merger parameterization $\phi^*\propto
\exp{[(-Q/\beta)((1+z)^{-\beta}-1)]}$ given in
\cite{1992Natur.355...55B}. The value of $\beta=1+(2q_0)^{0.6}/2$ was
set to 1.53 using $q_0=-0.55$. A value of $Q=1$ was used as in \cite{2006ApJ...639..644E}.
The extinction correction, which is
important in the blue bands, was set to $\tau_B=0.6$ with the
prescription given in \cite{1991ApJ...383L..37W}. As is shown in
Figs.~\ref{Fig:countsJ} and \ref{Fig:countsH} these two models, that
fit the B (Fig.~\ref{Fig:countsB}) and Ks (Fig.~\ref{Fig:countsKS})
galaxy counts, overestimate the slope variation at J$\sim$19 in the
Alhambra counts, and at H$\sim$18 in data from other surveys.

\begin{figure}
\epsscale{1.0}
\plotone{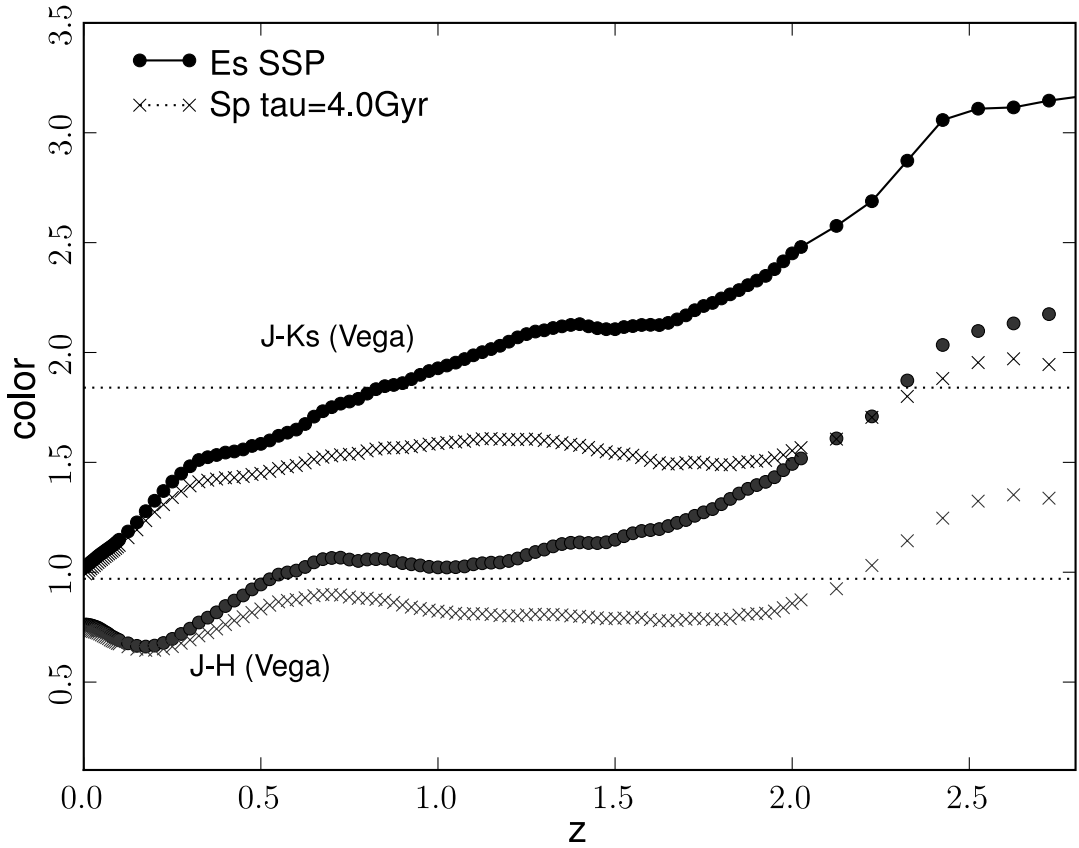}
\caption{\label{Fig:colorev} Color evolution with redshift of the NIR colors for an Elliptical and an early spiral (with the parameterizations given in Tab.~\ref{Tab:galparams}). 
The two upper lines correspond to the J-Ks color whereas the other two correspond to the J-H color. 
The horizontal dotted lines correspond to the J-H$\sim$0.97 and J-Ks$\sim$1.84 colors of the fitted point for the slope change.} 
\end{figure}

In order to explain the change of slope in the NIR galaxy counts, the
population of red Elliptical galaxies has to decrease with the
redshift. Although a model taking into account only the stellar
evolution with look back time fits the blue band counts (see
Fig.~\ref{Fig:countsB}), this model over-predicts the faint counts in
the NIR bands as can be seen from Fig.~\ref{Fig:countsKS}. Due to the
red color of the slope change only the Elliptical population
parameterized with a short burst of star-formation can play this
role. Fig.~\ref{Fig:colorev} shows the evolution with redshift of the
J-H and J-Ks colors for an Elliptical galaxy and an early spiral formed
at z=4 (with stellar populations according to the parameters in
Tab.~\ref{Tab:galparams}, and reddening in the spectra applied
directly from the \cite{2003MNRAS.344.1000B} code using $\tau_V=0.8$
and $\mu=0.5$). The estimated colors of the slope change from the fits
given in Tab.~\ref{Tab:slopes} are J-H$\sim0.97\pm 0.03$ and
J-Ks$\sim1.84\pm 0.03$, providing evidence that the Elliptical
population at z$\sim$1 must be the responsible of the turn down of the
NIR counts.

\begin{figure}
\epsscale{1.0}
\plotone{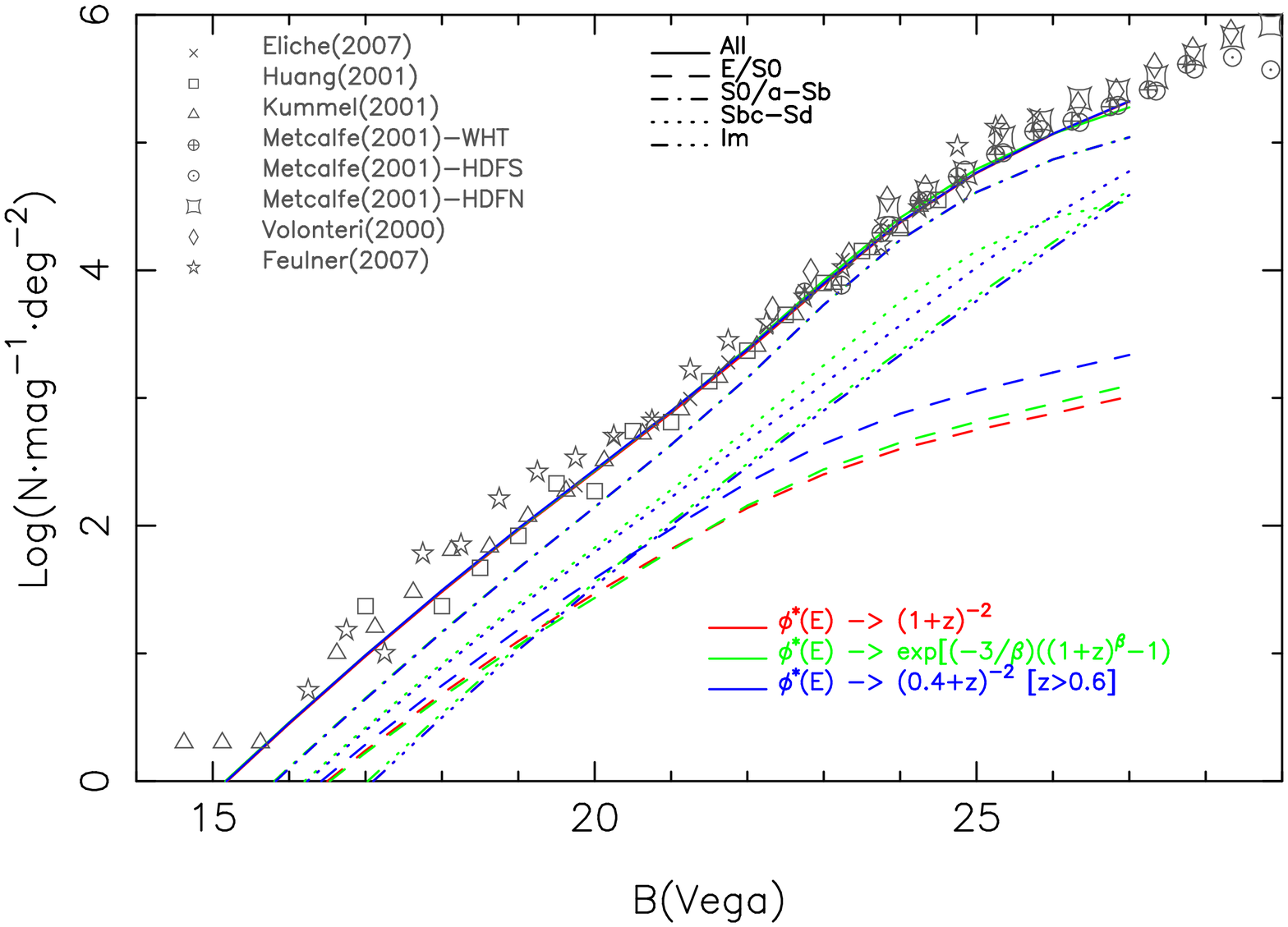}
\caption{\label{Fig:countsB2} Galaxy number count in B band taken from the literature. The lines correspond to the number counts predictions from the models described in the text.} 
\end{figure}

\begin{figure}
\epsscale{1.0}
\plotone{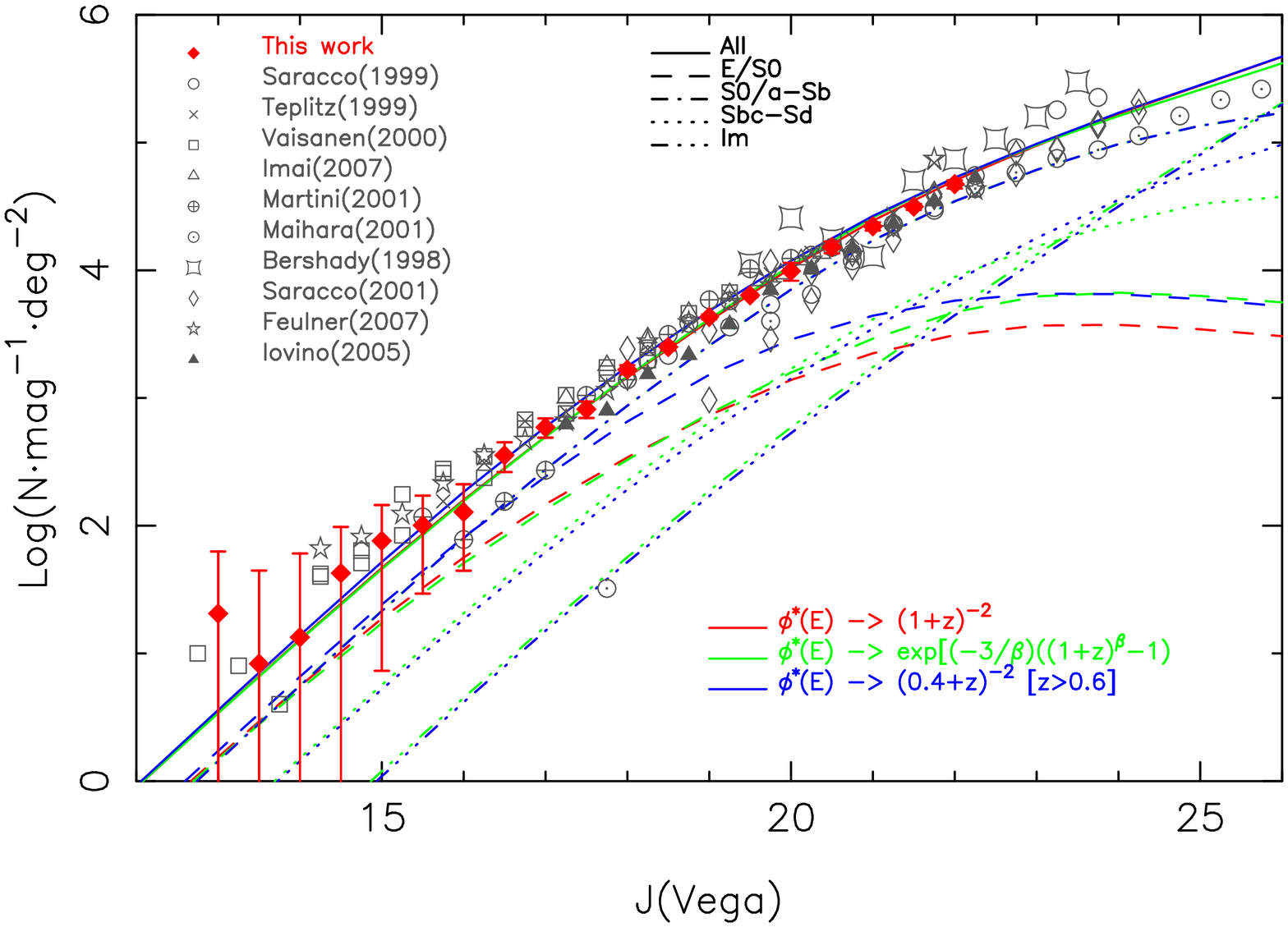}
\caption{\label{Fig:countsJ2} Galaxy number counts in the J filter compared with data from other surveys. The lines correspond to number counts models described in the text.} 
\end{figure}

\begin{figure}
\epsscale{1.0}
\plotone{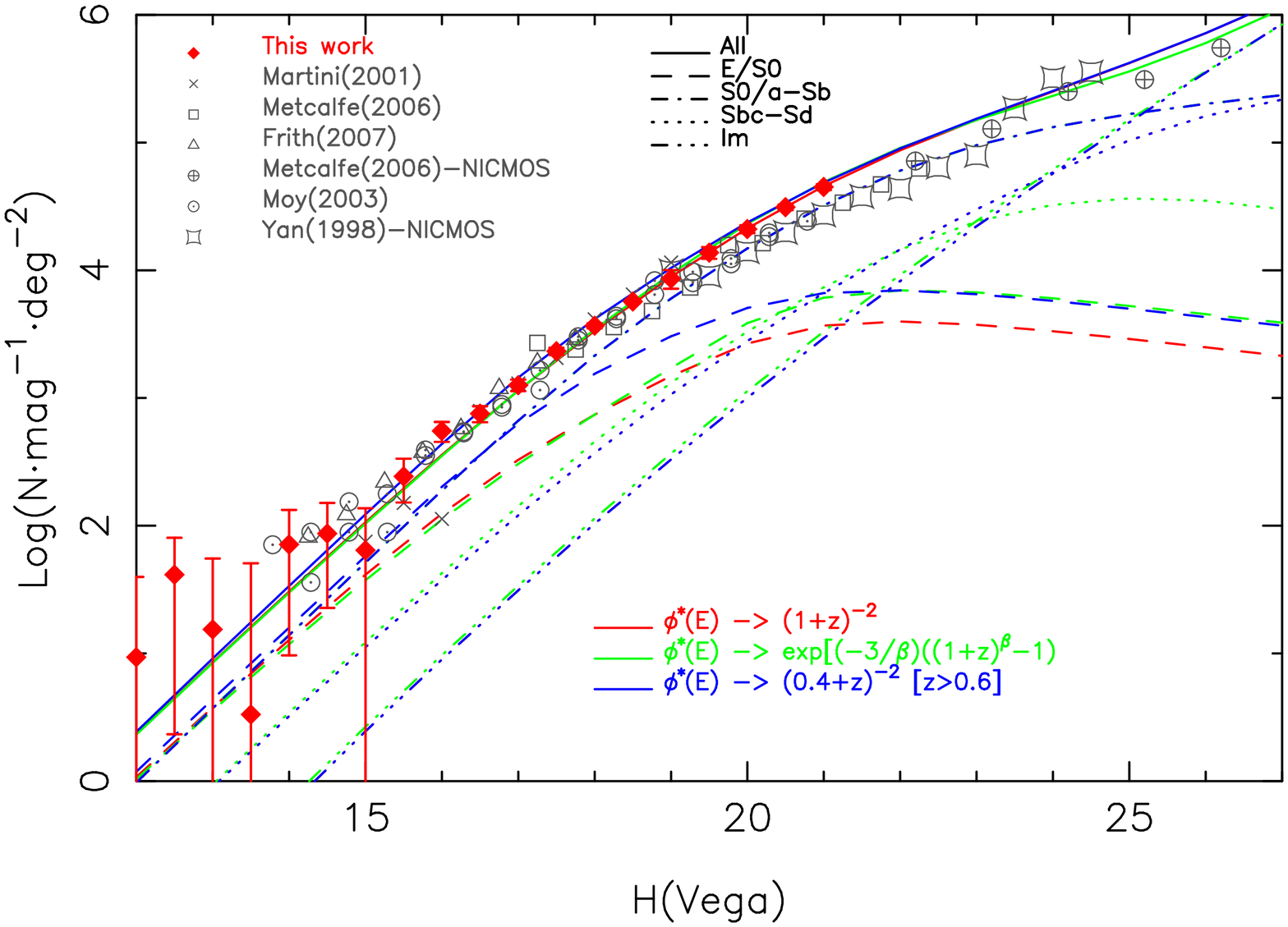}
\caption{\label{Fig:countsH2} Galaxy number counts in the H filter compared with data from other surveys and the number count models described in the text.} 
\end{figure}

\begin{figure}
\plotone{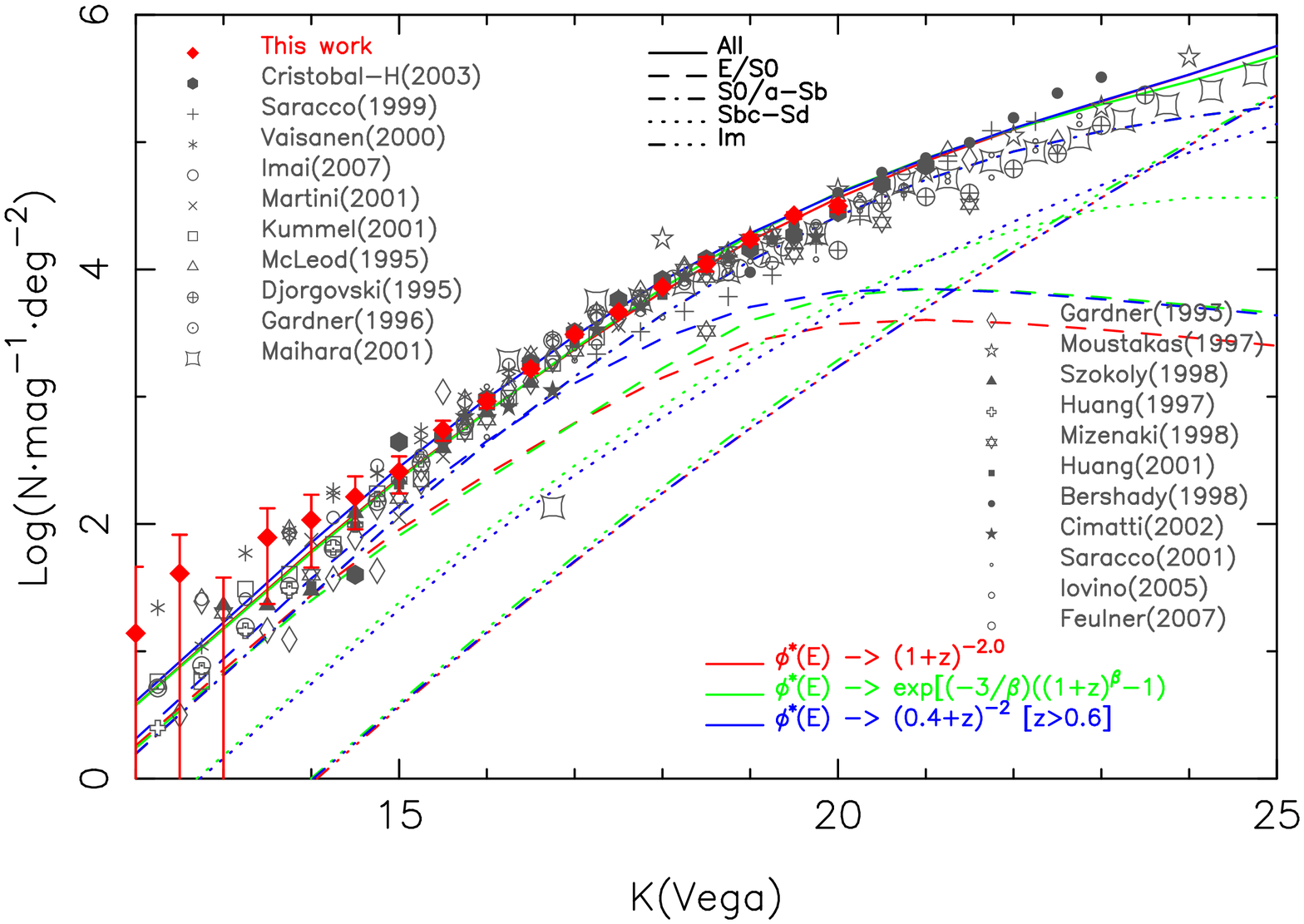}
\caption{\label{Fig:countsKS2} Galaxy number counts in the Ks filter compared with data from the literature and the galaxy counts models described in the text.}
\end{figure}

In the next model we introduce number-density evolution of the elliptical
population parameterized using $\phi^{\star}\propto (1+z)^{-2}$.
The formation redshifts is set to $z_f=4.0$ for all galaxy types, being in more agreement with 
the evolved red galaxies found at $z>2$ \citep{2003ApJ...587L..83V,2004ApJ...617..746D}.
The evolution in the early type densities was not accompanied by an
evolution in $M^{\star}$, arguing that a substantial number of
ellipticals formed in spiral-spiral mergers as expected for hierarchical galaxy
formation. For the early spiral galaxies no number-density evolution was
considered, and the density evolution parameterization in
\cite{2006ApJ...639..644E} $\phi^{\star}\propto (1+z)^{2}$ was used
for the two later type galaxies. The reddening in the spectra was applied
directly from the \cite{2003MNRAS.344.1000B} code using $\tau_V=0.8$
and $\mu=0.5$. This model, as can be seen in the
Figs.~\ref{Fig:countsB2}, \ref{Fig:countsJ2}, \ref{Fig:countsH2}, and
\ref{Fig:countsKS2} fits the galaxy counts in the optical and NIR
filters, reproducing the feature of the slope turn down in the three
NIR filters.

\begin{figure}
\epsscale{1.0}
\plotone{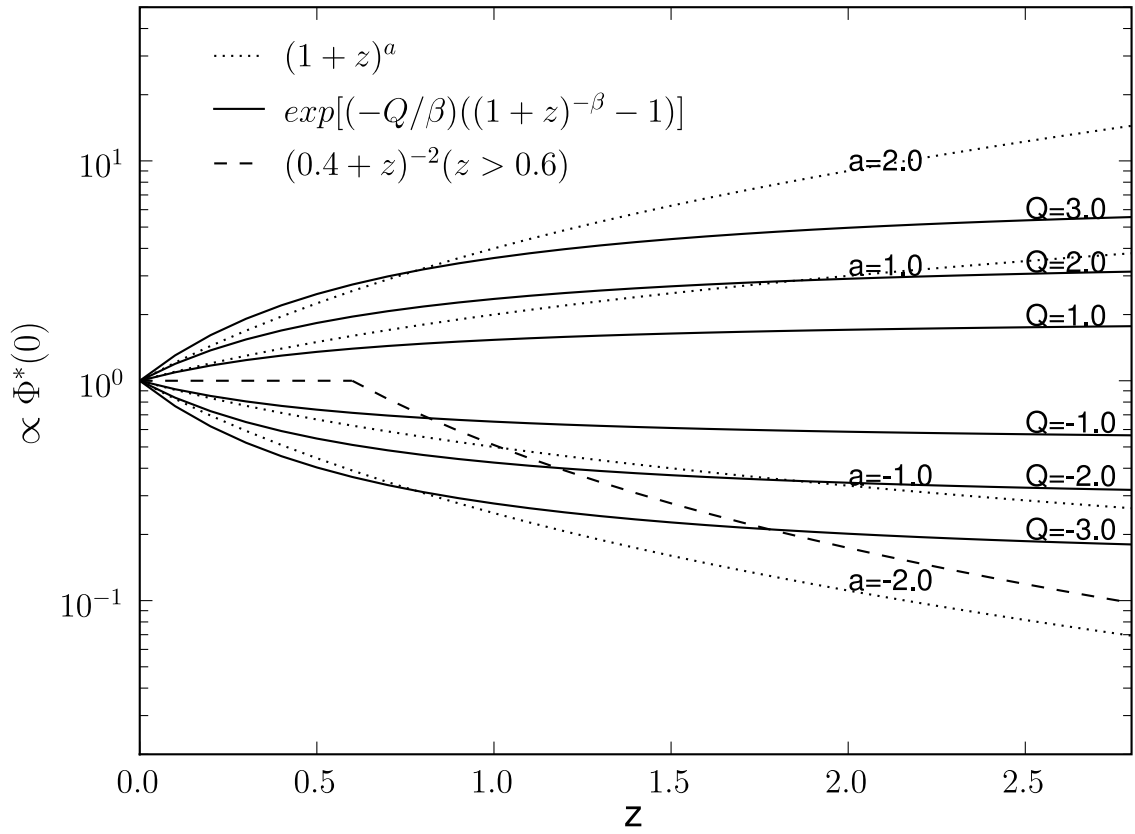}
\caption{\label{Fig:phievol} Functional form of the $\phi^*$ evolution parameterization.} 
\end{figure}

In order to avoid an unreasonable high number of late type galaxies at
high-z, the simple merger evolution $\phi^*\propto
\exp{[(-Q/\beta)((1+z)^{-\beta}-1)]}$ given in
\cite{1992Natur.355...55B} could be used with similar results. In this
case, we used Q=-3 to parameterize the decrease in number of E/S0
galaxies, and Q=1 to produce the required merger rate in the late
spirals and irregulars.  The number evolution given by these
parameterizations are displayed in Fig~\ref{Fig:phievol}, showing that
at $z\sim1$ the implied number density of E/S0 galaxies is only
$\sim1/4$ of the present day $\phi^*(0)$.  Density evolution in the
early type galaxies was observed in previous works studying the
type-dependent LF evolution
\citep{1996MNRAS.283L.117K, 2001A&A...367..788F, 2002MNRAS.333..633A,
2003A&A...401...73W,2005ApJ...622..116G,2006A&A...453..809I}. 
\cite{2003A&A...401...73W} found an increase in
$\phi^{\star}$ of an order of magnitude for the early type galaxies
from $z\sim1.2$ to $z=0$, that is over the $\phi^{\star} \propto
(1+z)^{-2}$ simulated here. However, they used the spectra of a present
day Sa type galaxy to separate the different galaxies, which leads to an
over-estimation of number-density evolution for the Early-type group.

In \cite{2007ApJ...669..184A} it is shown that evolution in the
fraction of the stellar mass locked in massive early-type galaxies is
produced in the interval $0.7<z<1.7$. A model in which $\phi^*$ for
the Elliptical galaxies is constant to $z\sim 0.6$ and then evolve as
$\phi^{\star} \propto (0.4+z)^{-2}$ for higher redshifts also produces
a good fit to the optical and NIR counts (see
Figs.~\ref{Fig:countsB2}, \ref{Fig:countsJ2}, \ref{Fig:countsH2}, and
\ref{Fig:countsKS2}). In this model the population of red elliptical
galaxies has doubled since $z=1$, in good agreement with the increase
of a factor of $\sim 2$ in the number evolution of red galaxies given
in \cite{2004ApJ...608..752B}.
 
Finally, we have implemented a simple recipe to simulate downsizing in
the elliptical population by maintaining $\phi^\star$ constant with
redshift for the LF of bright galaxies. The results are compatible
with our NIR galaxy counts and B-band counts from the literature in
the case that $\phi^\star$ is constant with redshift for
red-ellipticals brighter than $M^\star-0.7$ ($\sim-22.0$ in the Sloan
$r'$ band in AB system), decreasing the number densities for the bulk
of the ellipticals as $\phi^{\star}\propto (1+z)^{-2}$.

\section{Color analysis}

\begin{figure}
\epsscale{0.6}
\plotone{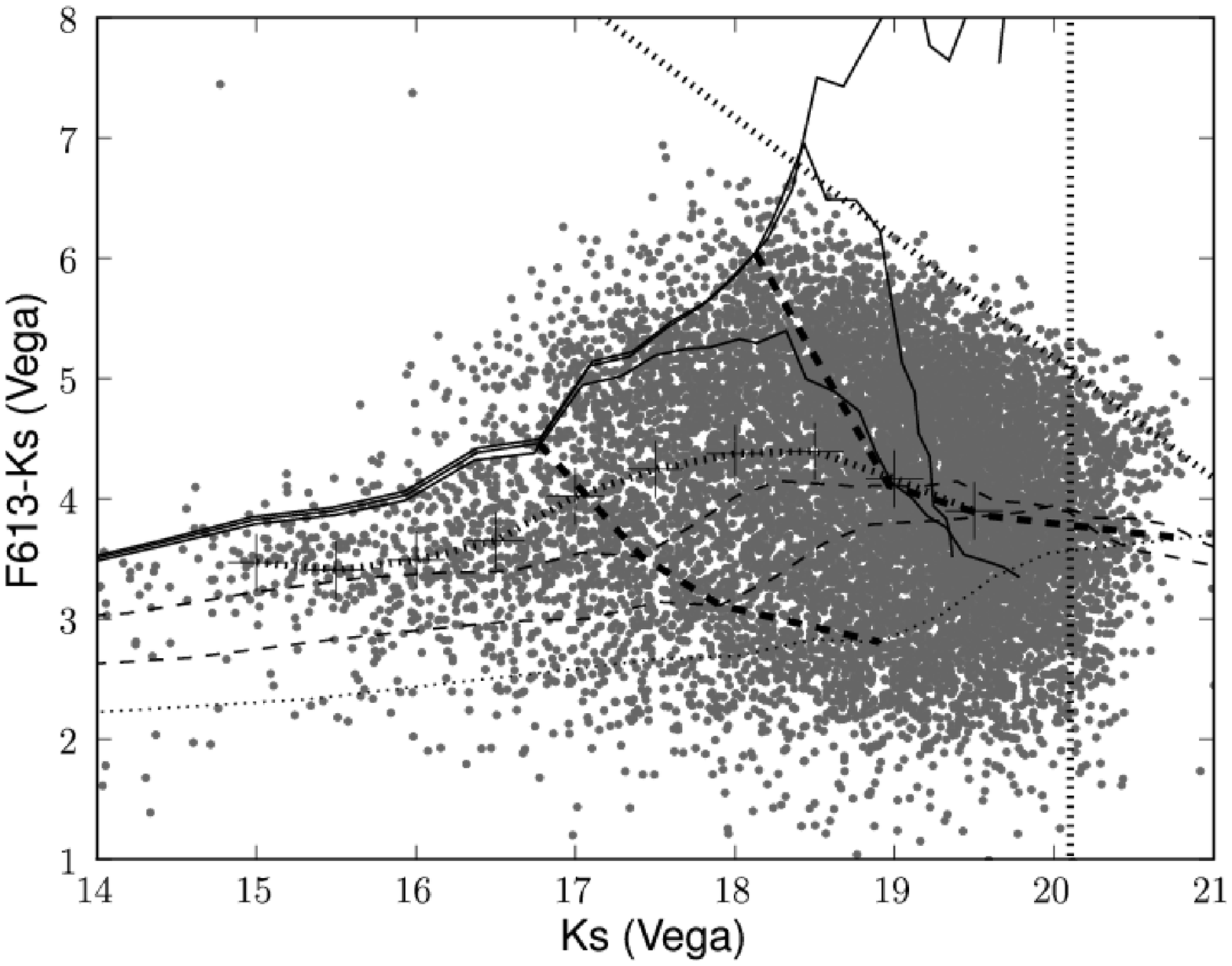}
\plotone{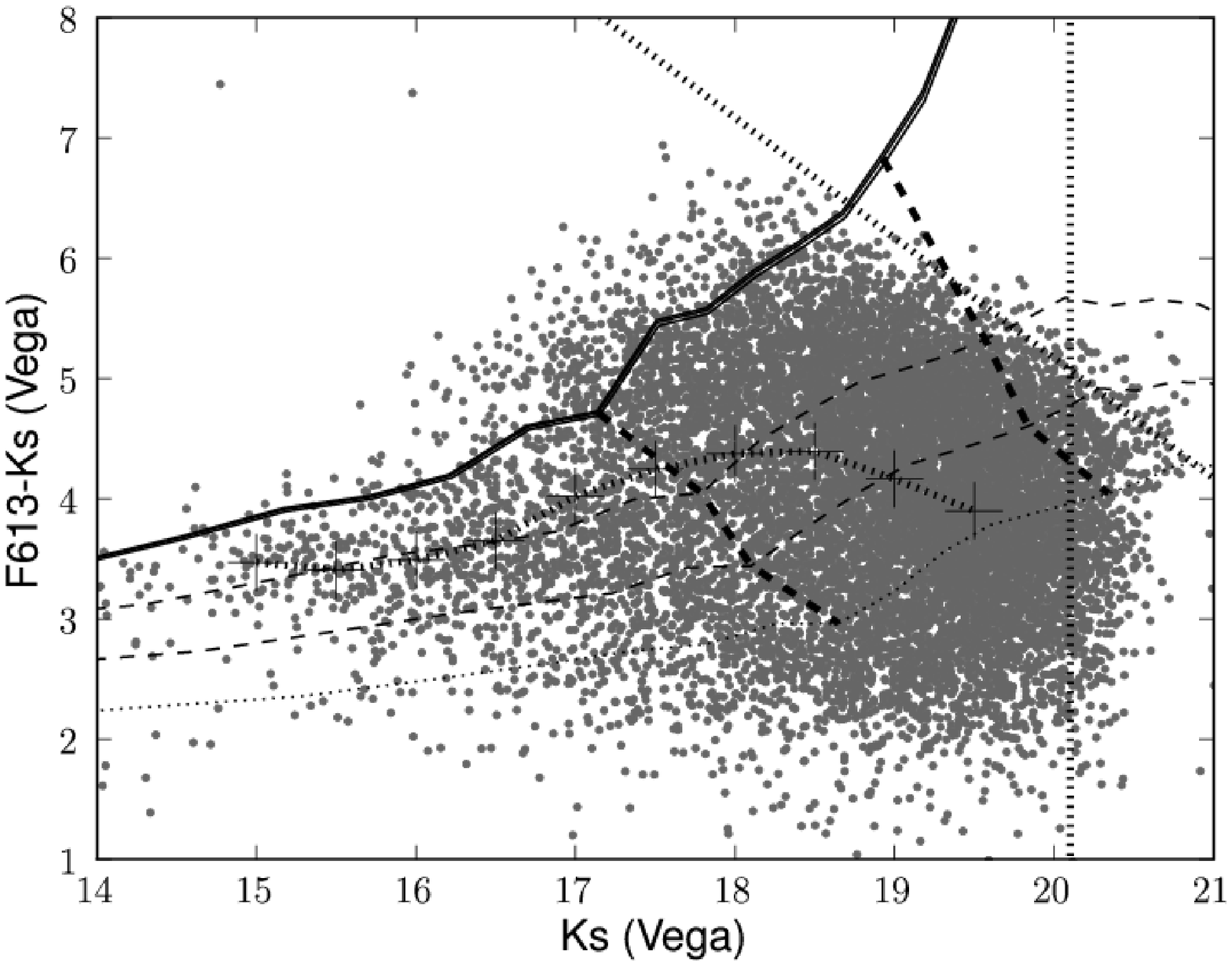}
\caption{\label{Fig:tracs} Color-magnitude diagram for the Alhambra data in the filter with effective
wavelength 6130$\AA$ (F613) and Ks. The tracks for the galaxy classes in Tab.~\ref{Tab:galparams} formed at zf=4 are also shown, for the passive evolution 
scenario {\it top panel)}, 
and the no-evolution case {\it bottom panel)}. Each
galaxy track is computed for the characteristic luminosity given in Tab.~\ref{Tab:lumfunc}. The solid tracks from top to bottom correspond to
a simulated E/S0 galaxy with single stellar population and e-folding timescales $\tau=0.5,1$Gyr. The dashed lines correspond to the early and late type spiral galaxies, and the dotted line
to the Irr galaxy. The vertical dotted line and the diagonal one are the $5\sigma$ detection levels. The two dashed lines
join the spectra points for z=0.5 and z=1.0. The median color for each Ks bin is marked with crosses.}
\end{figure}

More information about the evolution of the galaxy populations could be obtained from color
histograms. The separate number counts in each band at the magnitude ranges that we are sampling 
are less sensitive to the formation redshift (for values $zf>=4$) or the e-folding timescale of the 
star formation than 
color histograms. Figure ~\ref{Fig:tracs}  shows the color-magnitude diagram builded
with the ALHAMBRA data through the filter centered at 6130 $\AA$ (F613) and Ks.
The modelled evolutionary tracks for the 4 galaxy spectra considered in Tab.~\ref{Tab:galparams} are also displayed. Models with no evolution (top panel) and passive evolution (bottom panel
) have been considered. As can be appreciated the evolved spectra produce better match to the data than
the no evolved version, principally at the faint blue end which is better described by models considering passive evolution 
in the late Sp and Irr spectra.

\begin{figure}
\epsscale{0.5}
\plotone{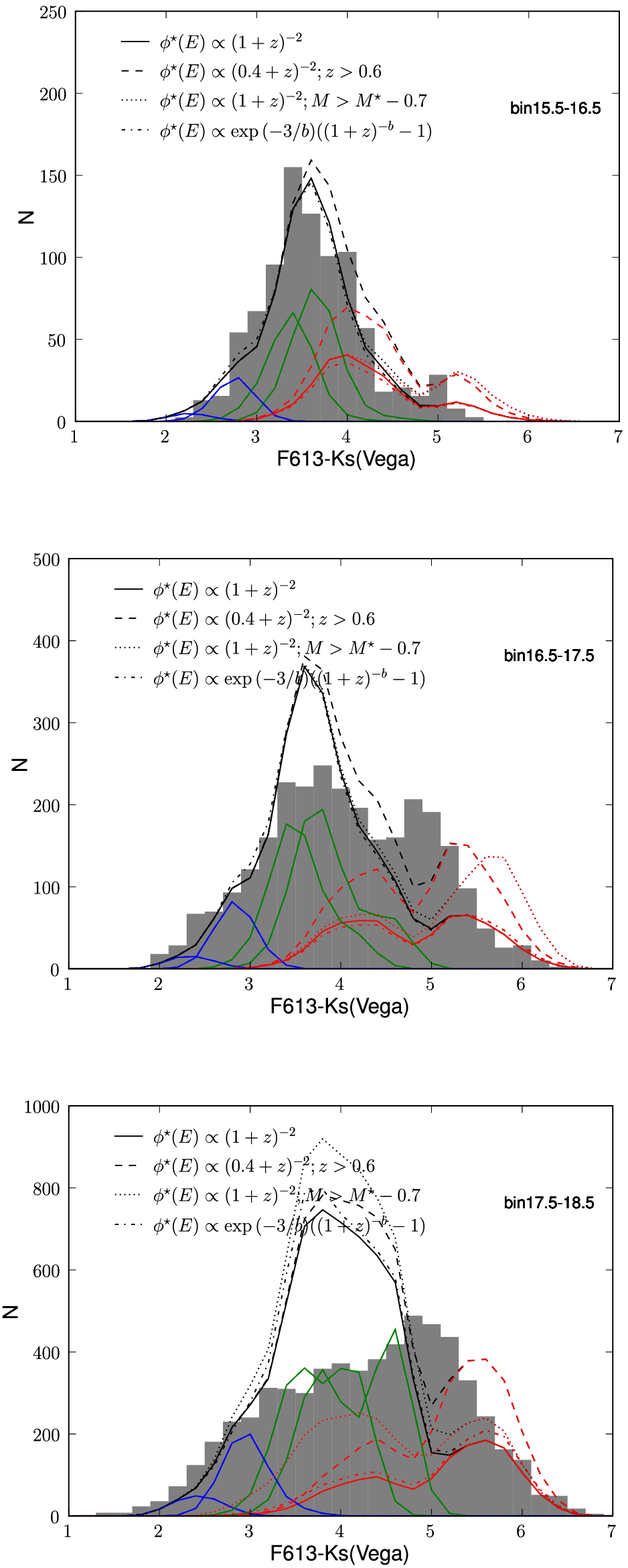}
\caption{\label{Fig:chisto}Color F613-Ks histogram (normalized for 1 square degree) for sources in different Ks bins. The histograms produced
by the models have been convolved with a Gaussian kernel $\sigma=0.2$ in order to reproduce the photometric errors. The histograms computed
for each population are shown in red for the E/S0, green for the early type spiral galaxies, and in blue for the late type spirals and Irr
galaxies.}
\end{figure}

In Fig.~\ref{Fig:chisto} it is shown the F613-Ks color histogram for different Ks magnitude bins. We have used an e-folding timescale $\tau=0.7$ Gyr to describe the Elliptical galaxies in those plots, this longer timescale produce better fits to the red end of the color histograms than an instantaneous star forming event which over-predict the number of red galaxies. Both models produce similar results when fitting the galaxy counts in individual optical and NIR bands. The simulated histograms
correspond to models where the population of E/S0 galaxies decrease with redshift, the population of spirals stay constant,
and the late type galaxies increase as as $\phi^{\star}\propto (1+z)^{2}$ conserving the luminosity densities.
The population of early spiral galaxies have been divided in two classes: one remain as in Tab.~\ref{Tab:galparams}, 
whereas the amount of extinction have been doubled for the other. This try to avoid the fact that
the discretization of the actual galaxy population in four classes tend to produce sharp histograms. 

\begin{figure}
\epsscale{0.5}
\plotone{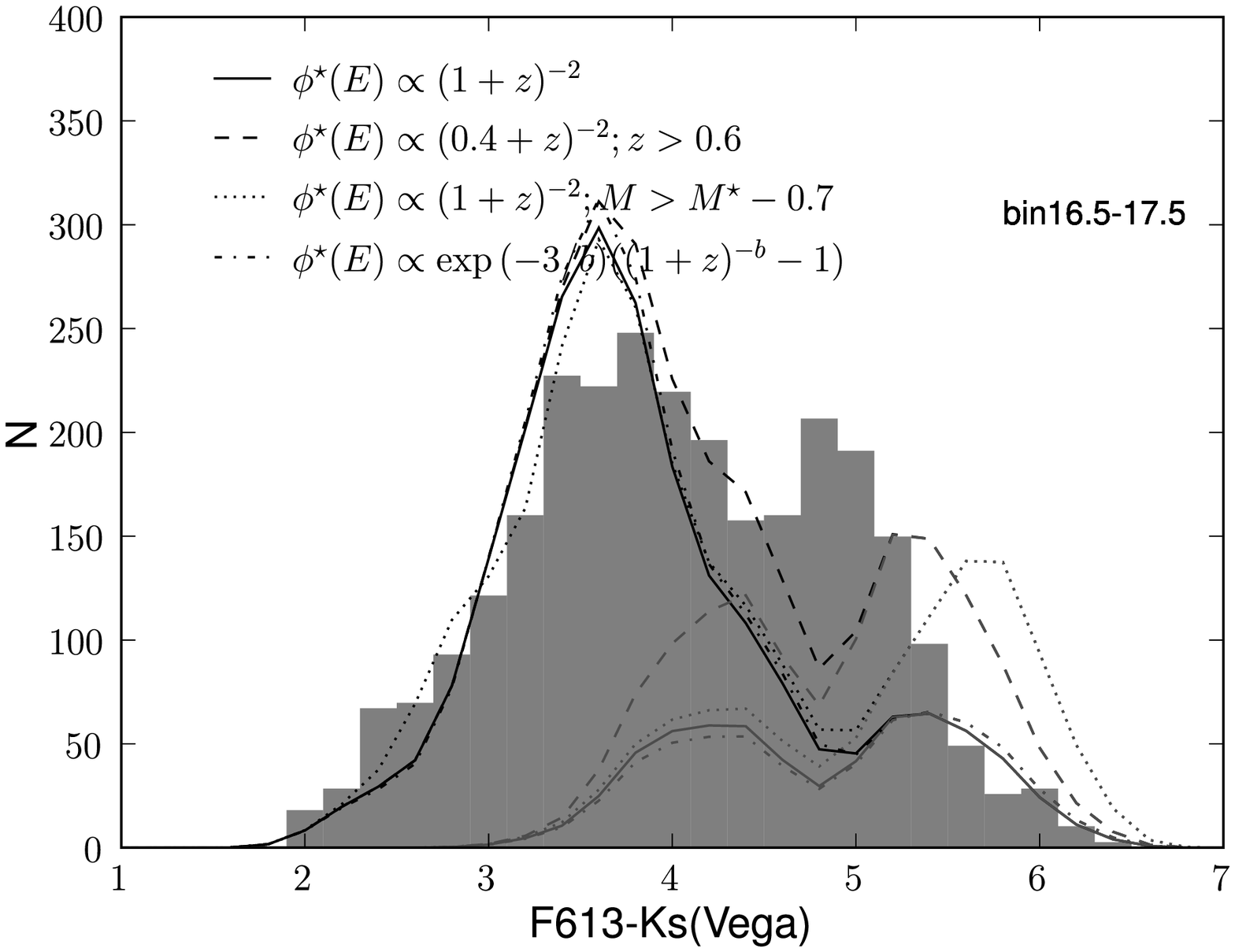}
\plotone{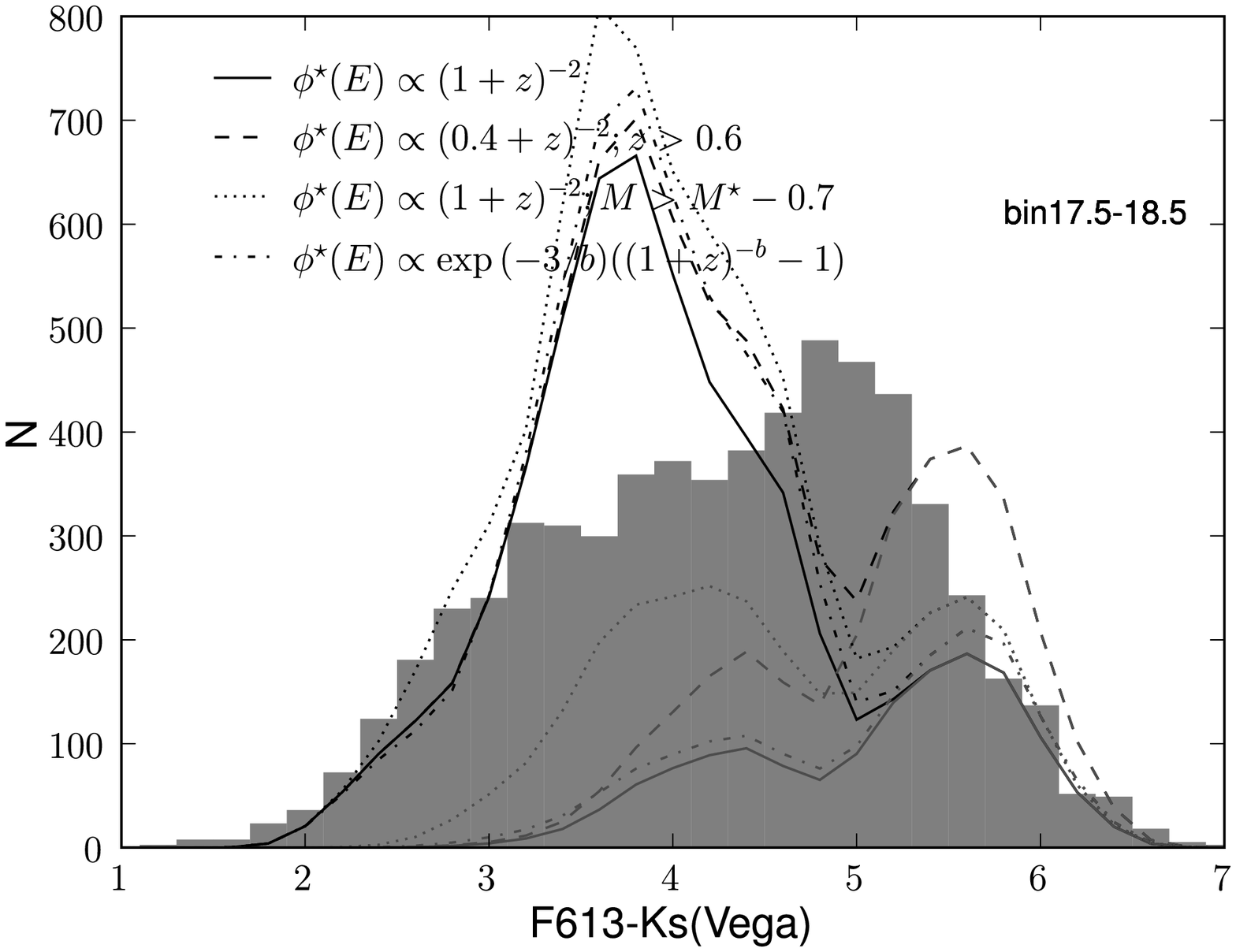}
\caption{Color F613-Ks histogram for sources in different Ks bins. The histograms have been normalized to 1 square degree.
The modelled histograms have been computed for 
models in which the number densities for the
early spirals decrease with redshift as  $\phi^{\star}\propto (1+z)^{-1}$, the late type spirals number density remain constant with
redshift, and number densities of Irr galaxies increase $\propto (1+z)^{3}$. The number of E/S0 galaxies decrease with $z$ as 
specified in the labels.
The histograms obtained
with this models have been convolved with a Gaussian kernel $\sigma=0.2$ in order to reproduce the photometric errors.
\label{Fig:chisto2}}
\end{figure}

\begin{figure}
\epsscale{1.0}

\plotone{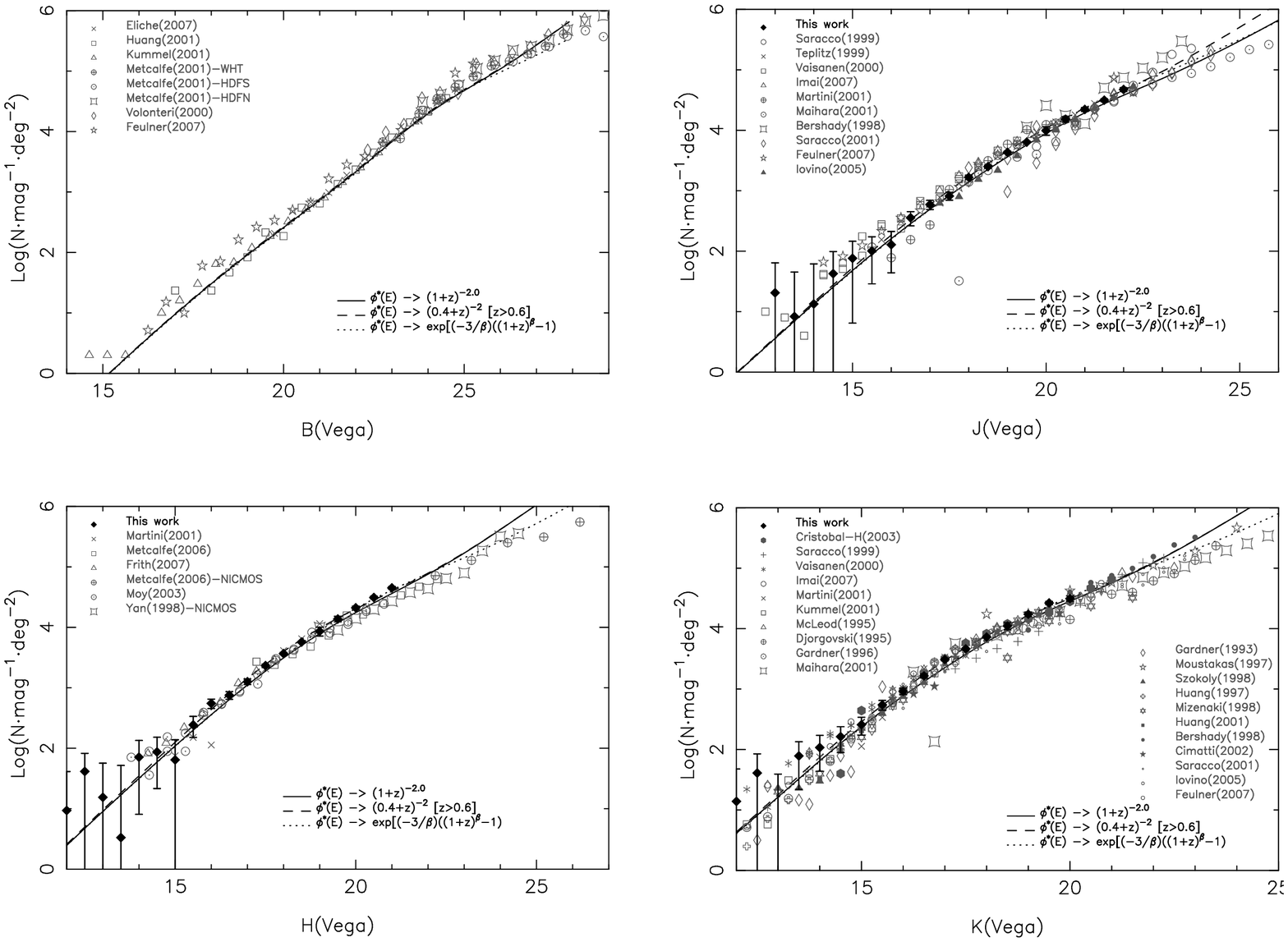}

\caption{Differential number counts in the B and NIR bands compared with a models in which the number densities for the
early spirals decrease with redshift as  $\phi^{\star}\propto (1+z)^{-1}$, the late type spirals number density remain constant with
redshift, and number densities of Irr galaxies increase $\propto (1+z)^{3}$.
\label{Fig:models2}}
\end{figure}

As can be appreciated the models reproduce the overall shape of the data for bright Ks magnitudes. 
Nevertheless, at faint Ks magnitudes the models predict higher values in the color range $3\le F613-Ks\le5$. 
As could be inferred from Fig.~\ref{Fig:chisto} to obtain a better match to the data the number densities of early spiral galaxies
formed in a shorter time-scale has to decrease with $z$, such fading of the spirals will lead to an under-prediction 
of the blue-band number counts unless that the
number of star-forming increases at higher redshift. 
In Fig.~\ref{Fig:chisto2} the F613-Ks histograms for the Ks bins:
[16.5,17.5], [17.5,18.5] show a better concordance with the observed data. Those histograms correspond to a models where
the number densities for the
early spirals decrease with $z$ as  $\phi^{\star}\propto (1+z)^{-1}$, the late type spirals number density remain constant with
redshift, and number densities of Irr galaxies increase $\propto (1+z)^{3}$. The luminosity density in not conserved within any
galaxy class. Similar results could be obtained using  $\phi^*\propto \exp{[(-Q/\beta)((1+z)^{-\beta}-1)]}$ 
\citep{1992Natur.355...55B}, with Q=-1, and Q=3 to describe the number evolution of early spiral and Irr galaxies. This number evolution formulation avoid a step increase of Irr galaxies at high redshift. However in Fig.~\ref{Fig:chisto2} a shortage of red galaxies is seen at about F613-Ks$\sim5$, this could be due to the fact that
we only use a discrete number of galaxy parameterizations, for example is well know the existence of dusty starburst with the 
same red colors than the passive extremely red objects \citep{2000A&A...361..535D}.
Covering a wider range in 
galaxy internal extinction or formation timescale will tend to smear the bi-modality present in the simulated histograms.
As could be seen in Fig.~\ref{Fig:models2}
the number counts produced by this model in B + NIR filters also produce good fits to the observed data points. In this models
the number counts at faint magnitudes will be dominated by the star-forming galaxies. The number count slope at faint magnitudes
will be $-0.4(\alpha +1)$ \citep{2003RMxAC..16..203B}, being $\alpha$ the slope of the dominant luminosity function for
$M<<M^\star$. With this parameterization the slope of the number counts tend to $0.36$ at the fainter end.

\clearpage

\section{Summary}

We have presented galaxy counts in the J,H, and Ks filters covering an
area of 0.45 square degrees and an average 50\% detection efficiency
depth of J$\sim22.4$, H$\sim21.3$ and Ks$\sim20.0$ (Vega system). The
depth reached, and the precision of the counts over a range of five
magnitudes makes the data valuable for examining the change of the
count slope reported in the Ks filter, and to extend this
examination to the J and H bands. We find that a change in slope occur
in each of the NIR bands in the range J=[18.5,19.5], H=[18.0,19.0]
and Ks=[17.0,18.0]. The NIR colors where the break in the galaxy counts are
found imply that this change is related to the population of red
galaxies at z$\sim$1.

We have compared our number counts results with predictions from a
wide range of number count models, concluding that in order to
reproduce the described changes in the NIR slopes, a decrease in bulk
of the population of red elliptical galaxies is needed. Good fits to
the B-band and NIR counts are obtained with a 
parameterization for the number evolution of the elliptical population
as $\phi^{\star}\propto (1+z)^{-2}$ with no accompanying evolution 
in $M^{\star}$, corresponding to evolution in which 
the majority of ellipticals formed in spiral-spiral
mergers. 

Performing a color analysis show that also the population of early spirals has to decrease
at higher redshift in order to describe the color distribution in r-Ks. Models using
the parameterization of \cite{1992Natur.355...55B} $\phi^*\propto \exp{[(-Q/\beta)((1+z)^{-\beta}-1)]}$, 
 with Q=-3, Q=-1, and Q=3 to describe the number evolution of ellipticals, early spiral and Irr galaxies,
 with no number density evolution for late-spiral systems, produce good fits to the observed 
 distribution, avoiding at the same time a high number of young systems at high $z$.
 A good match to the optical and NIR data is also obtained 
if the population of red-galaxies in the models remain constant to $z\sim0.6$ and
afterwards its number density decrease as $\phi^{\star} \propto (0.4+z)^{-2}$,
or if the number density of red-ellipticals is constant with redshift
for galaxies brighter than $M^\star-0.7$ ($\sim-22.0$ in the Sloan
$r'$ band in AB system), decreasing as $\phi^{\star}\propto (1+z)^{-2}$ for the bulk
of the ellipticals. 

Alhambra is processing the data obtained in 20 medium-band optical and 3 NIR filters
reaching high quality photometric redshift measurements ($\Delta z/(1+z)\le0.03$).
Also an accurate classification by Spectral Energy distribution will be acquired.  
Those data will allow for the study of the evolution of the
different galaxy types to $z\sim1$, which will complement the results given
in this article, 
disentangling
what populations contribute to the number counts at different redshift intervals. 
Also the
study of number counts for red galaxy populations, passive EROS  or BzK \citep{2004ApJ...617..746D} 
galaxies will constrain the
formation redshift and formation timescale for massive Elliptical galaxies.

The authors
acknowledge support from the Spanish Ministerio de Educaci\'on y
Ciencia through grants AYA2002-12685-E, AYA2003-08729-C02-01,
AYA2003-0128, AYA2007-67965-C03-01, AYA2004-20014-E, AYA2004-02703,
AYA2004-05395, AYA2005-06816, AYA2005-07789, AYA2006-14056, and from
the {\sl Junta de Andaluc\'{\i}a}, TIC114, TIC101 and {\sl Proyecto de
Excelencia} FQM-1392. NB, JALA, MC, and AFS acknowledge support from
the MEC {\sl Ram\'on y Cajal} Programme. NB acknowledges support from
the EU IRG-017288.  

This work has made use of software designed at
TERAPIX.

This publication makes use of data products from the Two Micron All Sky Survey, which is a joint project of the University of Massachusetts and the Infrared Processing and Analysis Center/California Institute of Technology, funded by the National Aeronautics and Space Administration and the National Science Foundation

    This publication makes use of data from the Sloan Digital Sky Survey. Funding for the Sloan Digital Sky Survey (SDSS) has been provided by the Alfred P. Sloan Foundation, the Participating Institutions, the National Aeronautics and Space Administration, the National Science Foundation, the U.S. Department of Energy, the Japanese Monbukagakusho, and the Max Planck Society. The SDSS Web site is http://www.sdss.org/.

    The SDSS is managed by the Astrophysical Research Consortium (ARC) for the Participating Institutions. The Participating Institutions are The University of Chicago, Fermilab, the Institute for Advanced Study, the Japan Participation Group, The Johns Hopkins University, Los Alamos National Laboratory, the Max-Planck-Institute for Astronomy (MPIA), the Max-Planck-Institute for Astrophysics (MPA), New Mexico State University, University of Pittsburgh, Princeton University, the United States Naval Observatory, and the University of Washington.

\appendix

\section{Individual image combination}\label{Ap:Combination}

As mentioned in the text, to combine the processed images we used
SWARP \citep{2002ASPC..281..228B}. With this software, individual
images were projected into subsections of the final frame using the
inverse mapping, in which each output pixel center was associated to a
position in the input image at which it is interpolated. With this
schema, the code corrected at the same time the geometrical
distortions in the individual images using the astrometric information
stored in the image headers.

\subsection{Estimating the relative transparency}

Using a filtered version of the SExtractor catalogs computed over the
sky subtracted images, an accurate estimate of the relative
transparency was computed by tracing the high S/N objects in all the
images. The relative transparency values were used inside SWARP to
scale the individual images to the same flux level in order to
uniformize the zero points in the outer dither areas, this allowed to
use 2MASS catalogs to calibrate the ALHAMBRA NIR photometry in the
final images.

\subsection{Astrometry calibration} 

When computing the resampled version of the individual frames, SWARP
uses the WCS information stored in the headers. In order to obtain a
better matching of the individual images, firstly the pipeline
calculated the external astrometrical solution for a reference
image. The individual images were then calibrated internally with
respect to it, thus obtaining the equatorial coordinates from the
reference image. In this paper the individual image with better
transparency in a given pointing was used as reference. However, to
get a better internal astrometry between the different filters, after
completing a pointing in the 23 ALHAMBRA filters, the images with
better FWHM and transparency in a set of selected optical filters, are
combined to produce a deep image that will be used afterwards as
reference.

\begin{figure}
\begin{center}
\epsscale{1.0}
\plotone{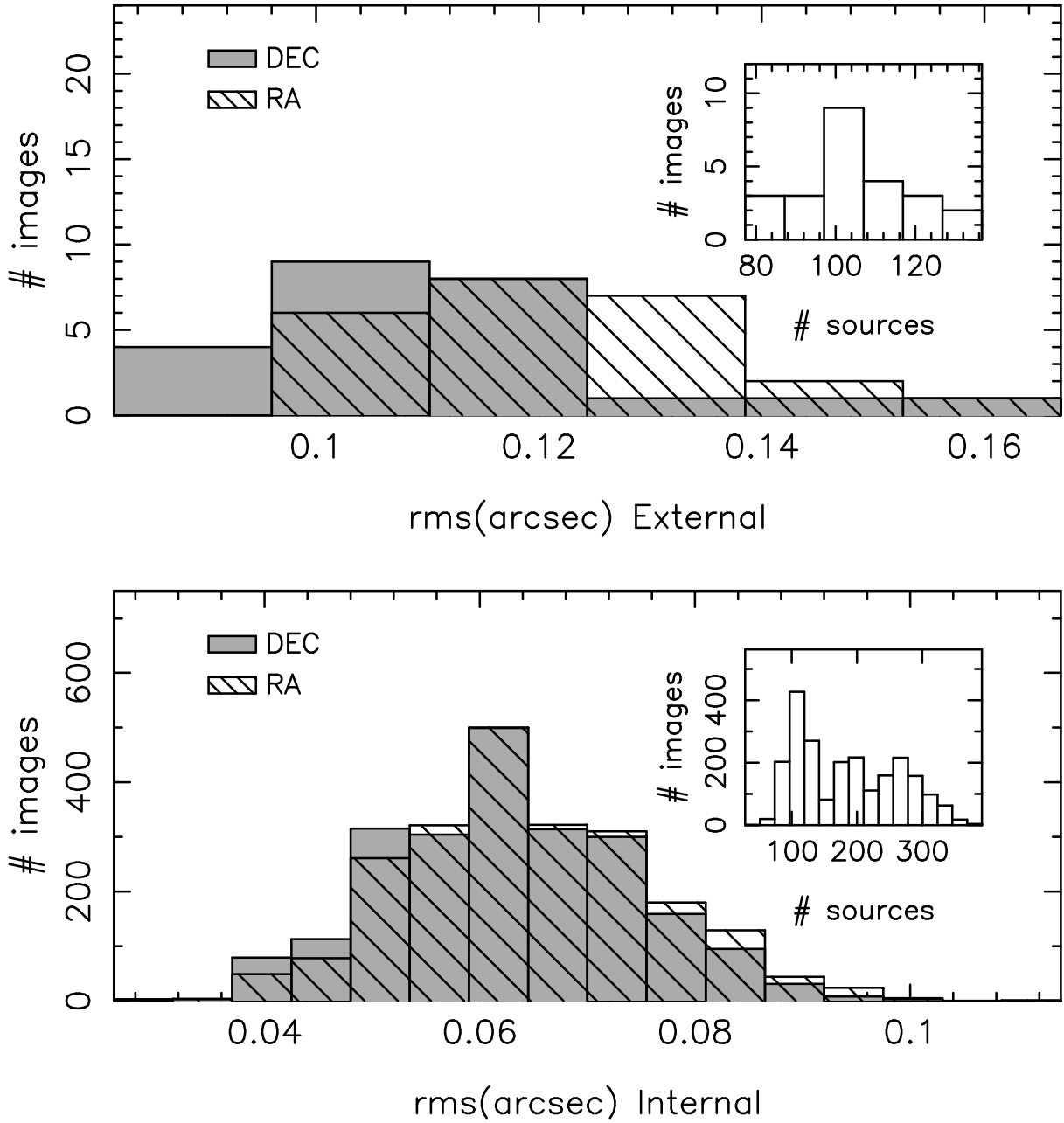}
\caption{Histograms of the astrometry solution rms for the ALH08 images. {\it (Top panel)} External USNO-B1.0. {\it (Bottom panel)} internal. The small
panels show the histograms of the number of objects that enter into the final astrometry fitting }
\label{Fig:wcs}
\end{center}
\end{figure}

We have determined that the USNO-B1.0 catalog
\citep{2003AJ....125..984M} provides an adequate number of objects to
perform a high quality external astrometric calibration. We used our
own code to match the sources with brighter apparent magnitudes in the
reference image with those in the USNO-B1.0. Once a meaningful number
of pairs was identified, the CCMAP IRAF task was used in two
iterations to acquire the required astrometric solution with a 2nd
order polynomial. A histogram of the external astrometric solution rms
and the number of objects used for the final images of ALH08 is shown
in Fig.~\ref{Fig:wcs}. The median external astrometry rms is
$0.12^{\prime\prime}\pm 0.01$ in RA and $0.11^{\prime\prime}\pm 0.01$
DEC.

Having calibrated a reference image, the rest of the individual images
were calibrated internally. The median internal rms in the astrometric
solution for the OMEGA2000 data used in this paper is
$0.06^{\prime\prime}$ in RA and DEC using a median of 160 objects, as
shown in Fig~\ref{Fig:wcs}.

\subsection{Image co-adding}

Swarp allows the user to choose among several interpolation functions for
inverse mapping. For selecting the more appropriate
kernel we analyzed the resulting final image FWHM and its
pixel-to-pixel correlation. Tab.~\ref{correlations} shows the
correlation values, between adjacent pixels and for pixel pairs
separated by 2 pixels, in the final images obtained using different
available interpolation functions. As can be seen in the table, the
bilinear function produce a higher correlation which translate into an
underestimation of the flux errors. Using the Lanczos-3 function the
FWHM of the final image was improved by $\sim0.05^{\prime\prime}$
compared with the nearest neighbor interpolation, whereas the
auto-correlation at 1 pixel remain acceptable $\sim0.16$ (when the
images are combined using the average). The Lanczos-4 function did not
decrease substantially nor the FWHM neither the correlation at 1
pixel, on the contrary it produced large artifacts at the bad pixels
and image borders, so finally we have decided to take the Lanczos-3
function.

\begin{deluxetable}{cccccc}
\tabletypesize{\small}
\tablewidth{0pt}
\tableheadfrac{0.01}
\tablecaption{Values for the correlation between adjacent pixels when co-adding the images using different interpolation functions\label{Tab:correlations}}
\tablehead{
\colhead{}&\colhead{Nearest} & \colhead{Bilinear} & \colhead{Lanczos2} & \colhead{Lanczos3} & \colhead{Lanczos4}}
\startdata
\multicolumn{6}{c}{Combining with median}\\
1pix & +0.017 & +0.186 & +0.112 & +0.074 & +0.068 \\
2pix & -0.035 & -0.013 & -0.072 & -0.059 & -0.101 \\
\\
\tableline
\\
\multicolumn{6}{c}{Combining with average}\\
1pix & +0.035 & +0.274 & +0.177 & +0.156 & +0.130 \\
2pix & -0.030 & -0.018 & -0.069 & -0.110 & -0.083 \\
\enddata
\label{correlations}
\end{deluxetable}

\end{document}